\documentclass[11pt,english,aps,preprint,superscriptaddress]{revtex4-1}
\pdfoutput=1
\usepackage{mathpazo}
\usepackage{helvet}

\usepackage[T1]{fontenc}
\usepackage[latin9]{inputenc}
\usepackage{dcolumn}
\usepackage[a4paper]{geometry}
\usepackage{babel}
\usepackage{float}
\usepackage{bm}
\usepackage{amstext}
\usepackage{graphicx}
\usepackage[unicode=true,pdfusetitle,
 bookmarks=true,bookmarksnumbered=false,bookmarksopen=false,
 breaklinks=true,pdfborder={0 0 0},backref=false,colorlinks=true]{hyperref}

\makeatletter

\providecommand{\tabularnewline}{\\}

\@ifundefined{textcolor}{}
{%
 \definecolor{BLACK}{gray}{0}
 \definecolor{WHITE}{gray}{1}
 \definecolor{RED}{rgb}{1,0,0}
 \definecolor{GREEN}{rgb}{0,1,0}
 \definecolor{BLUE}{rgb}{0,0,1}
 \definecolor{CYAN}{cmyk}{1,0,0,0}
 \definecolor{MAGENTA}{cmyk}{0,1,0,0}
 \definecolor{YELLOW}{cmyk}{0,0,1,0}
}

\usepackage{babel}
\usepackage{amstext}

\providecommand{\tabularnewline}{\\}

\@ifundefined{textcolor}{}{%
 \definecolor{BLACK}{gray}{0}
 \definecolor{WHITE}{gray}{1}
 \definecolor{RED}{rgb}{1,0,0}
 \definecolor{GREEN}{rgb}{0,1,0}
 \definecolor{BLUE}{rgb}{0,0,1}
 \definecolor{CYAN}{cmyk}{1,0,0,0}
 \definecolor{MAGENTA}{cmyk}{0,1,0,0}
 \definecolor{YELLOW}{cmyk}{0,0,1,0}
}

\@ifundefined{showcaptionsetup}{}{%
 \PassOptionsToPackage{caption=false}{subfig}}
\usepackage{subfig}
\makeatother

\begin{document}

\title{Looking for bSM physics using top-quark polarization and decay-lepton
kinematic asymmetries}

\author{Rohini M. Godbole}
\author{Gaurav Mendiratta}
\affiliation{Centre for High Energy Physics,
Indian Institute of Science, Bangalore 560 012, India}
\author{Saurabh Rindani}
\affiliation{Theoretical Physics
Division, Physical Research Laboratory, Navrangpura, Ahmedabad 380
009, India}
\begin{abstract}
We explore beyond Standard Model (bSM) physics signatures in the $l+jets$ channel of $t\overline{t}$
pair production process at the Tevatron and the LHC. We study the
effects of bSM physics scenarios on the top quark polarization and
on the kinematics of the decay leptons. To this end, we construct
asymmetries using the lepton energy and angular distributions. Further,
we find their correlations with the top polarization, net charge asymmetry
and top forward backward asymmetry. We show that when used together, these
observables can help discriminate effectively between SM
and different bSM scenarios which can lead to varying
degrees of top polarization at the Tevatron as well as the LHC. We
use two types of coloured mediator models to demonstrate the effectiveness
of proposed observables, an $s$-channel axigluon and a $u$-channel
diquark.

\end{abstract}

\date{\today}

\maketitle
\tableofcontents{} 

\section{Introduction}

Most experimental observations at particle accelerators fit the Standard
Model (SM) very well. However, there are some major puzzles to be
solved. One needs to have physics beyond the standard model (bSM)
to explain the presence of dark matter, explain quantitatively the
observed baryon asymmetry of the universe, and to explain the puzzle
of dark energy. Looking for signs of bSM, one finds that most of the
terrestrial experimental observations that are in tension with the SM results are
in the properties of third generation fermions. For example $B\to\tau\nu$
\cite{Lees:2012ju} , $h\to\mu\tau$ \cite{CMS-PAS-HIG-14-005} and
$b\overline{b}$ forward backward asymmetry (AFB) at LEP and Tevatron \cite{Abazov:2014ysa,ALEPH:2005ab,Murphy:2015cha}
show such deviations. One of these long standing puzzles is the
top-quark AFB measured by the D0 and
CDF detectors at the Tevatron collider in 2008 \cite{d02008afb,cdf2008afb}.
These observations by two independent collaborations were updated
with full data from the Tevatron and were consistent with each other
and in tension with the SM calculations until 2015. Recent experimental
results from $D0$ \cite{Abazov:2014cca} and theoretical calculations
\cite{Kidonakis:2015ona} point towards the possibility that the anomalous
nature of these observations may be a statistical phenomenon.

Due to its large mass which is close to the electroweak scale and the implied connection with electroweak symmetry breaking, the top quark is an important laboratory for various bSM
searches at colliders. In fact various proposals put forward to solve the different theoretical problems of the SM  often involve modifications in the top sector. Various extensions to the SM have also been proposed inspired by the possibly anomalous value of measured top-AFB
at the Tevatron. These bSM proposals involve explanation of the AFB in terms of processes involving a) $s$-channel resonances like the axigluon,
KK gluon, coloron \cite{DC:2007AfbAx,Antunano:2007da,Ferrario:2009ax,framptonshu2010,DC:2010_W'z'Axdiq,Dutta:2012ai,Fajfer:2012si,AguilarSaavedra:2011ug,Gresham:2012kv}
or b) t-channel exchange of particles with different spin and SM charges 
like the $Z'$ , diquarks etc \cite{Jung:2009_Z',Papaefstathiou:2011kd,Fajfer:2012si,Shu_Tait_2009diquark,DC:2010_W'z'Axdiq,Barger:2011W'Z',Rajaraman:2011rw,Patel:2011eh,Berger:2013ysa,AguilarSaavedra:2011ug}. Effective operator
approach has also been used in this context \cite{effLag_Ko,Franzosi:2015osa,Biswal:2012mr}.
Measurements of other related observables such as lepton angular asymmetries and $t\bar{t}$ invariant-mass dependence of the top
quark AFB are also compatible with the hypothesis of a heavy bSM particle,
see for example \cite{Shu:2011au,Westhoff:2013ixa}.

In this study, we will focus on the lepton + jets final state ($p\overline{p}/pp\to t\overline{t}\to b\nu l\overline{t}$) of the $t\overline{t}$ pair production process.
This channel has a larger cross-section
as compared to the dilepton + jets channel, and it has a much smaller
background compared to the all jets channel. For lighter quarks, hadronization
smears the information available about their spin and polarization.
The mass of the top quark is large enough that it decays into its
daughter particles before strong interactions can initiate the hadronization
process. Hence top-quark polarization
leaves a memory in the kinematic distribution of the decay products
and can be tracked \cite{Jezabek:1988topdecayleptondist,Godbole:2006ldistandtpol}.
We  study the correlations between various  kinematic asymmetries and  polarization to distinguish between different sources of these asymmetries
within an $s$-channel (axigluon) and a $t$-channel (diquark) extension
of the SM. For the Tevatron, the top pair production process is dominated
by $q\overline{q}$ collisions and at the LHC it is dominated by $gg$
collisions which means that new physics can manifest at differently at the two colliders.

A wide variety of observables have been studied in the literature to explore the top sector as a bSM portal \cite{Antunano:2007da,Fajfer:2012si,Rajaraman:2011rw,Bernreuther:2010ny,AguilarSaavedra:2012rx,Bernreuther:2012sx,AguilarSaavedra:2012xe,Hewett:2011wz}.
A brief review of some of these observables which have been experimentally
measured and are relevant to this work is presented in section
\ref{sec:Status-of-Experimental}. In section  \ref{sec:Flavor-Non-Universal-axigluon},\ref{sec:U-channel-Scalar-Model}
we describe the flavour non-universal axigluon
and diquark models which we use as templates for our analysis. Constraints on these models from top pair production cross-section and forward backward asymmetry at Tevatron, charge asymmetry, top quark pair production, dijet and four jet production cross-sections at LHC are discussed in section \ref{sec:dijet-constraints}. In section \ref{sec:Asymmetry-Definitions}
we construct the asymmetries which we use to explore the bSM models. In section \ref{sec:Correlations}
we present the correlations between various asymmetries and discuss
the role of top quark polarization and kinematics in discerning
the various regions of parameter space of the bSM models. We contrast
our results for the axigluon and diquark models and the resulting
conclusions can be generalized to other new physics scenarios. Our
results are presented for the Tevatron $\sqrt{s}=1.96$\textrm{ TeV}
and the LHC $\sqrt{s}=7$ TeV , $13$ TeV. We discuss the effects of transverse
polarization coming from the off-diagonal terms in the top-quark density
matrix in section \ref{sec:Off-Diagonal-Density-Matrix}
and then conclude in section \ref{sec:Conclusions}.

\section{Status of experimental results\label{sec:Status-of-Experimental}}

We begin by summarizing some of the experimental results from the
LHC and the Tevatron concerning the top quark and compare them with
the corresponding SM calculations from the literature.

The measured $t\bar{t}$ production cross-section for the Tevatron
at $\sqrt{s}=1.96$ TeV is $\sigma_{p\overline{p}\to t\overline{t}}^{Tevatron}=7.60\pm0.41$
pb \cite{Aaltonen:2013wca} and that for the LHC at $\sqrt{s}=7$
TeV is $\sigma_{pp\to t\overline{t}}^{LHC}=173.30\pm10.10$ pb \cite{CMS:2013sca,ATLAS:2012dpa}.
These agree with the calculated SM NNLO cross-sections $\sigma_{p\overline{p}\to t\overline{t}}^{Tevatron}=7.16{}_{-0.50}^{+0.54}$
pb \cite{SM_tevatron_crs} for the Tevatron, and $\sigma_{pp\to t\overline{t}}^{LHC\,7TeV}=177.30\pm10.63$
pb \cite{Czakon:2011xx} for the LHC, within $1\sigma$. The uncertainties
coming from the top-quark mass dependence of $t\overline{t}$ cross-section
\cite{Czakon:2013goa} have been included in the given LHC cross-sections.
In the calculations in following sections, we use a common K factor
for the bSM+SM to estimate the NNLO total cross-section. For the Tevatron,
the K factor is $K_{Tevatron}=1.39{}_{-0.10}^{+0.10}$ \cite{k_Tevatron}.
The K factor for the LHC is calculated using the NNLO cross-section
cited above and LO cross-section calculated using CTEQ6l parton distribution
functions (pdf) with factorization scale $Q=2m_{t}$. The errors in
the K factors represent pdf uncertainties, scale dependence and statistical
errors in the NNLO cross-section. For the LHC with $\sqrt{s}=7$ TeV,
$K_{LHC7}=2.20_{-0.15}^{+0.14}$.

The cross-sections impose a constraint on any new particle to have
small couplings with the top quark and/or have a sufficiently high
mass. It is interesting to note that we can still find a range of
couplings of the bSM large enough to explain the measured anomalous
top-quark and lepton asymmetries reported at the Tevatron and remain
compatible with the measurements at the LHC.
The AFB of the top quark in the $t\overline{t}$
centre-of-mass (CM) frame is defined as 
\begin{eqnarray}
A_{Forward\, Backward} & = & \frac{N_{F}(y_{t}-y_{\bar{t}}>0)-N_{B}(y_{t}-y_{\bar{t}}<0)}{N_{F}(y_{t}-y_{\bar{t}}>0)+N_{B}(y_{t}-y_{\bar{t}}<0)}\\
 & = & \frac{N(\cos\theta_{t}>0)-N(\cos\theta_{t}<0)}{N(\cos\theta_{t}>0)+N(\cos\theta_{t}<0)},
\end{eqnarray}
where $y_{t}$, $y_{\bar{t}}$ are respectively the rapidities of
the $t$ and the $\bar{t}$ and $\theta_{t}$
and $\theta_{\bar{t}}$ are their respective polar angles measured
with respect to the beam direction.

The CDF measurement of the $t\overline{t}$ CM frame AFB with the
full data set is $A_{FB}^{t\bar{t}}=0.164\pm0.045$ \cite{FBA_mttbar2013}.
The corresponding SM result is 0 at tree level in QCD. At NLO in QCD,
the value predicted is $0.0589_{-0.0140}^{+0.0270}$ (the errors only
represent scale variation) which upon including NLO electroweak corrections
becomes, $0.0734_{-0.0058}^{+0.0068}$ \cite{Czakon:2014xsa}. Recently,
AFB has been calculated at NNLO to be $0.0749_{-0.0086}^{+0.0049}$
in pure QCD and $0.095\pm0.007$ including EW corrections \cite{Czakon:2014xsa}
and including effective $N^{3}LO$ QCD, $A_{FB}^{t\overline{t},SM}=0.100\pm0.006$
\cite{Kidonakis:2015ona}. D0 has come out recently with a measurement
$A_{FB}=0.106\pm0.03$ \cite{Abazov:2014cca} which agree with
the theoretical results. However for the purpose
of this study, we use the CDF measurement which is still in tension
with the SM and with the D0 measurement.

Since the LHC is a $pp$ collider, its symmetric initial state makes
the forward and backward regions trivially symmetric. For the LHC,
instead of top quark AFB, a charge asymmetry (AC) is defined in
the lab frame as,
\begin{equation}
A_{C}=\frac{N(\Delta|y_{t}|>0)-N(\Delta|y_{t}|<0)}{N(\Delta|y_{t}|>0)+N(\Delta|y_{t}|<0)}\label{eq:charge_assym_defn}
\end{equation}
where $\Delta|y_{t}|=|y_{t}|-|y_{\bar{t}}|$. The AC
at the LHC is much smaller than the AFB at the Tevatron both in the
case of the SM and of the bSM models aimed at explaining the Tevatron's
anomalous AFB. The measured value of AC with CMS and ATLAS combination is $A_{c}=0.005\pm0.009$
\cite{CMS:2014jua}. The theoretical results for the SM values of
the AC \cite{SM_LHC_Ac} (QED+EW+NLO QCD) are given
in table \ref{tab:Charge-asymmetry} for different energies at the
LHC.

\begin{table}[th]
\begin{ruledtabular}
\begin{tabular}{|c|c|}
\hline 
$\sqrt{s}$\textrm{(TeV)} & $A_{C}$\tabularnewline
\hline 
\hline 
7 & 0.0115(6)\tabularnewline
\hline 
12 & 0.0068(3)\tabularnewline
\hline 
13 (From-Fit See Appendix \ref{sec:Charge-Asymmetry-AtLHC13}) & 0.0063\tabularnewline
\hline 
14 & 0.0059(3)\tabularnewline
\hline 
\end{tabular}\caption{\label{tab:Charge-asymmetry}Charge asymmetry in the lab frame at
the LHC, as defined in eqn (\ref{eq:charge_assym_defn}).}
\end{ruledtabular}
\end{table}

Measurements have also been made for a number of other observables including $M_{t\bar{t}}$, rapidity-dependent top
AFB \cite{FBA_mttbar2013}, lepton and di-lepton asymmetries  \cite{cdf_Alpm,Abazov:2014oea}, some of which show a deviation from
the standard model \cite{Bernreuther:2012sx} of up-to 1-3 $\sigma$ . Some
CDF results are shown in table \ref{tab:CDF-asymmetries} and D0
results \cite{Abazov:2014oea} in table \ref{tab:D0-lepton-asymmetries}.
\begin{table}[th]
\begin{ruledtabular}
\begin{tabular}{|c|c|c|}
\hline 
Asymmetry & Experimental Value & SM calculation\tabularnewline
\hline 
\hline 
$A_{FB}^{l}\left(or\, A_{\theta_{l}}\right)$ & $0.090_{-0.026}^{+0.028}$ & $0.038\pm0.003$\tabularnewline
\hline 
$A_{t\overline{t}\_FB}^{M_{t\bar{t}}>450GeV}$ & $0.295\pm0.058\pm0.031$ & $0.100\pm0.030$\tabularnewline
\hline 
$A_{t\overline{t}\_FB}^{M_{t\bar{t}}<450GeV}$ & $0.084\pm0.046\pm0.026$ & $0.047\pm0.014$\tabularnewline
\hline 
$A_{FB}^{l^{+}l^{-}}$ & $0.094\pm0.024_{-0.017}^{+0.022}$ & $0.036\pm0.002$\tabularnewline
\hline 
\end{tabular}\caption{\label{tab:CDF-asymmetries}CDF lepton and $M_{t\overline{t}}$ dependent
top level asymmetries \cite{FBA_mttbar2013,cdf_Alpm,Bernreuther:2012sx}}
\end{ruledtabular}
\end{table}

\begin{table}[th]
\begin{ruledtabular}
\begin{tabular}{|c|c|c|}
\hline 
Asymmetry & Experimental Value & SM calculation\tabularnewline
\hline 
\hline 
$A_{FB}^{l}\left(or\, A_{\theta_{l}}\right)$(extrapolated) & $0.047\pm0.027$ & $0.038\pm0.003$\tabularnewline
\hline 
$A_{FB}^{l}$$\left(|y_{l}|<1.5\right)$ & $0.042_{-0.030}^{+0.029}$ & $0.02$\tabularnewline
\hline 
\end{tabular}\caption{\label{tab:D0-lepton-asymmetries}D0 lepton asymmetries  \cite{Abazov:2014oea} }
\end{ruledtabular}
\end{table}
$t\overline{t}$ spin correlations have been measured using decay
particle double distributions in polar and azimuthal angles at the
Tevatron \cite{Aaltonen:2010nz,Peters:2012wg} and the LHC \cite{Aad:2014pwa,Chatrchyan:2013wua}.
The polarization of the top quark, as defined in eqn (\ref{eq:polarization}),
has also been observed at CMS, for the LHC 7 TeV run to be\textrm{
$0.01\pm0.04$} \cite{Chatrchyan:2013wua} compared to the corresponding
SM prediction from MC@NLO \cite{Frixione:2002ik} $0.000\pm0.002$. The ATLAS collaboration
also observed the polarization at 7 TeV beam energy, assuming CP conserving
$t\overline{t}$ production and decay process, to be $0.035\pm0.040$
\cite{Aad:2013ksa}, in agreement with the SM prediction.

\section{Flavour non-universal axigluon model\label{sec:Flavor-Non-Universal-axigluon}}

An axigluon is a massive, coloured ( $SU(3)_c$ adj), vector boson. Models of axigluon
which have only axial couplings with the quarks have been suggested in the literature
in many GUT like theories as chiral extensions of the QCD \cite{Frampton:1987dn,Frampton:1992flv_nonunivAx}.
Contribution to AFB for such a particle was studied even before the possible anomalous AFB was
observed at the Tevatron in 2008 \cite{DC:2007AfbAx}. For this flavour universal, axially interacting massive gluon with coupling $g_s$, the top quark AFB becomes
negative for masses above $m_{A}\sim500$ GeV. Upon the observation
of a positive AFB by Tevatron in 2008, this model was found to be incompatible
in the mass parameter regions allowed by the di-jet constraints from
Tevatron. The AFB turns back positive if the assumption of universality
of the interaction of axigluon with the quark families is dropped
\cite{Ferrario:2009ax}. In our study here, we have used a
more general, flavour non-universal axigluon with axial vector + vector couplings
\cite{framptonshu2010}. This model is obtained by breaking a larger
symmetry group of $SU(3)_{A}\times SU(3)_{B}$ to the QCD colour
group $SU(3)_{C}$ and a $SU(3)_{C'}$. The axial-vector coupling
of the axigluon to the first and second generation quarks is negative
of that for the third generation and the vector couplings are the
same for all three generations. The couplings of the axigluon with quarks
are described by the Lagrangian 
\begin{equation}
\mathcal{L}=\bar{\psi}\gamma^{\mu}T^{a}(g_{V}+g_{A}\gamma^{5})\psi A_{\mu}^{a},
\end{equation}
where $T^{a}$ are the Gell-Mann matrices. The couplings are parametrized
by $g_{V}=-\frac{g_{s}}{\tan(2\theta_{A})}$ , $g_{A}=\frac{g_{s}}{\sin(2\theta_{A})}$,
for the third generation of quarks. The parameters in this model are
$\theta_{A}$ and $m_{A}$. We vary the value of the coupling in the
range $\theta_{A}\in[0,\frac{\pi}{4}]$ which corresponds to varying
the axial and vector couplings from a large value at small $\theta_{A}$
to $g_{V}=0,g_{A}=g_{s}$ for $\theta_{A}=\frac{\pi}{4}$. A mass range
of $m_{A}\in[1,3]$\textrm{ TeV} is scanned.

 The decay width of axigluon and the density matrices for top-pair production mediated by an axigluon are given in Appendix \ref{sub:axigluon-Density-Matrices}.
For an s-channel resonance, the terms in the $t\overline{t}$ pair
production amplitude which are proportional to the linear power of
$\cos\theta$ (where $\theta$ is the top-quark polar angle) contribute towards the AFB. The
helicity dependent analysis of the top-quark decay distributions can
give additional information about the bSM couplings. We will show
in this study that this information can be accessed at the experiments
from correlations among top polarization, top-quark and decay-particle
asymmetries.

We first discuss constraints coming from $t\bar{t}$ production cross
section measurements, and top-quark level forward-backward and charge
asymmetries measured at the Tevatron and the LHC (as appropriate).

\subsection{Constraints on the axigluon model}\label{sec:const-axigluon}

\begin{figure}[h]
\subfloat[Cross-section at the Tevatron with $\sqrt{s}=1.96$\textrm{ TeV}]{\includegraphics[scale=0.155]{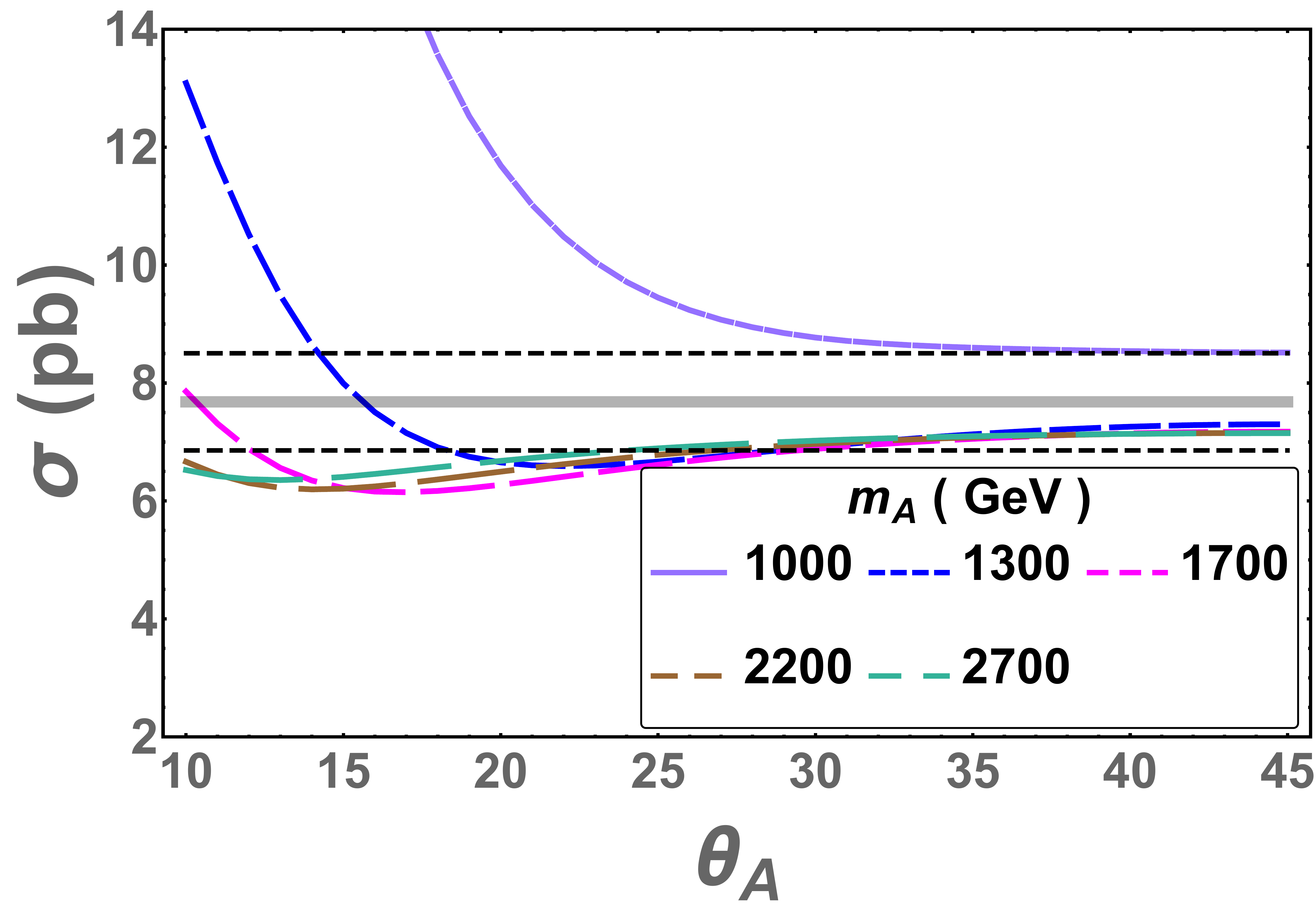}

}\hfill{}\subfloat[AFB at the Tevatron with $\sqrt{s}=1.96$\textrm{ TeV}]{\includegraphics[scale=0.155]{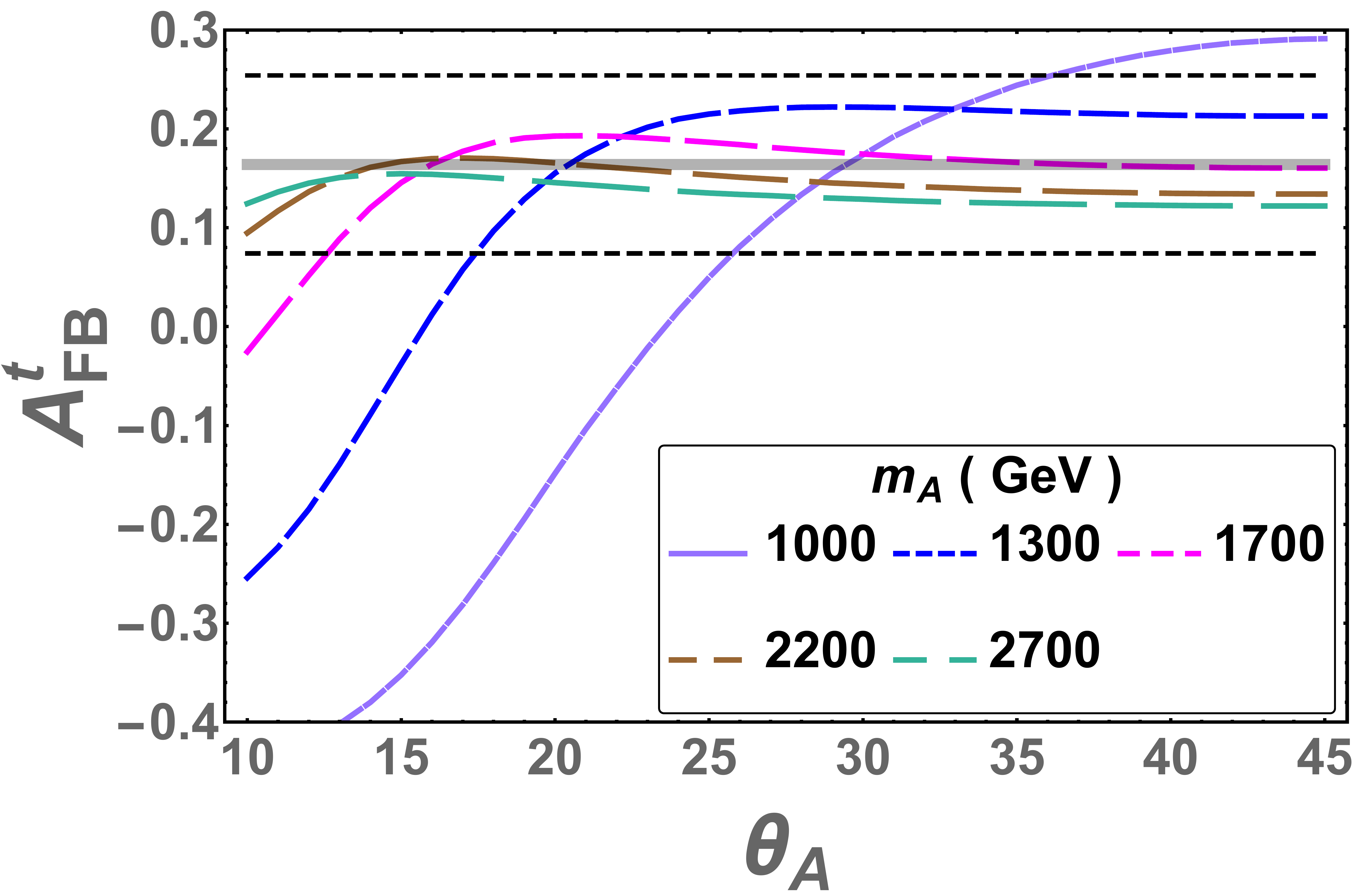}

}

\subfloat[Cross-section at the LHC with $\sqrt{s}=7$\textrm{ TeV}]{\includegraphics[scale=0.155]{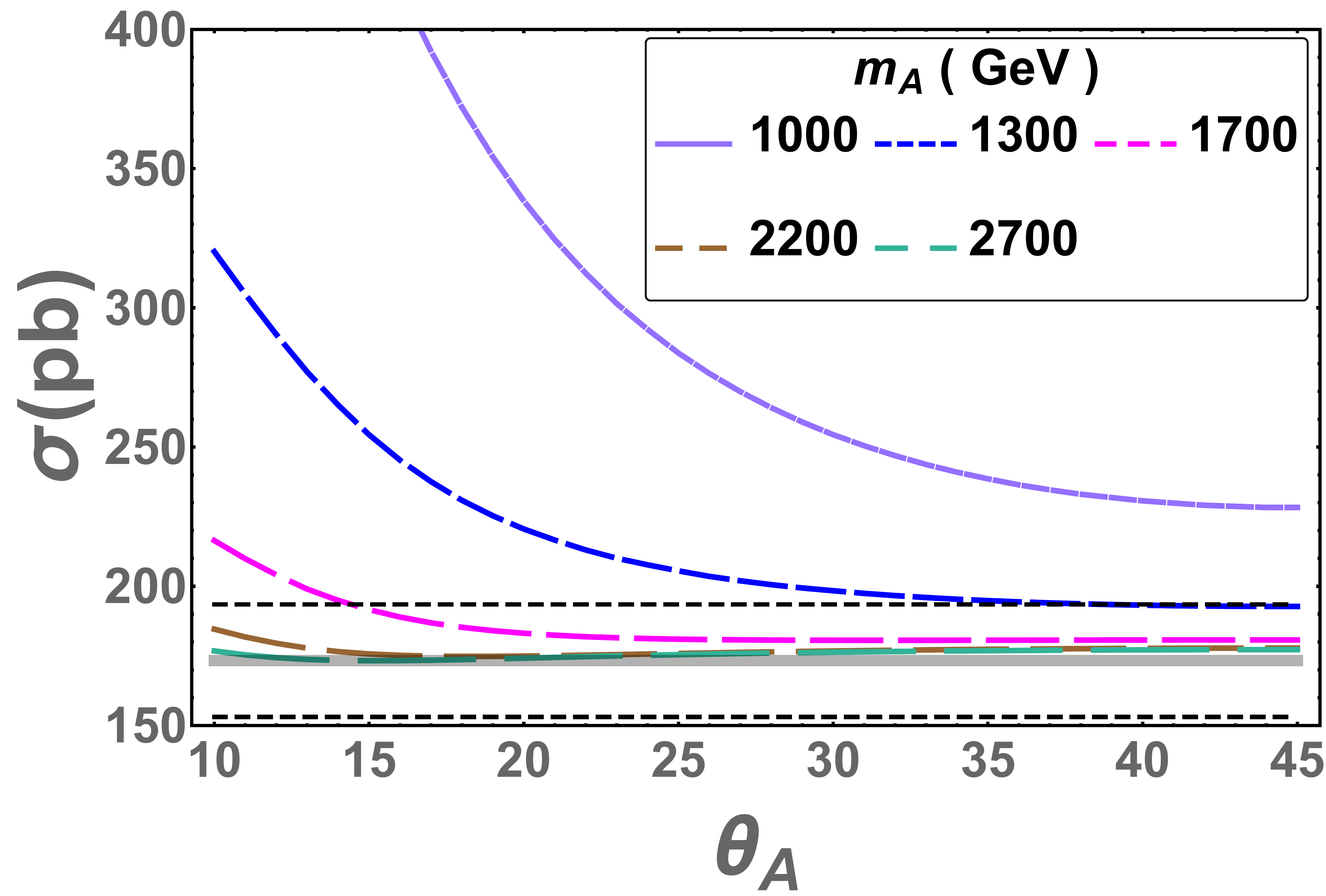}

}\hfill{}\subfloat[$A_{C}$ at the LHC with $\sqrt{s}=7$\textrm{ TeV}]{\includegraphics[scale=0.155]{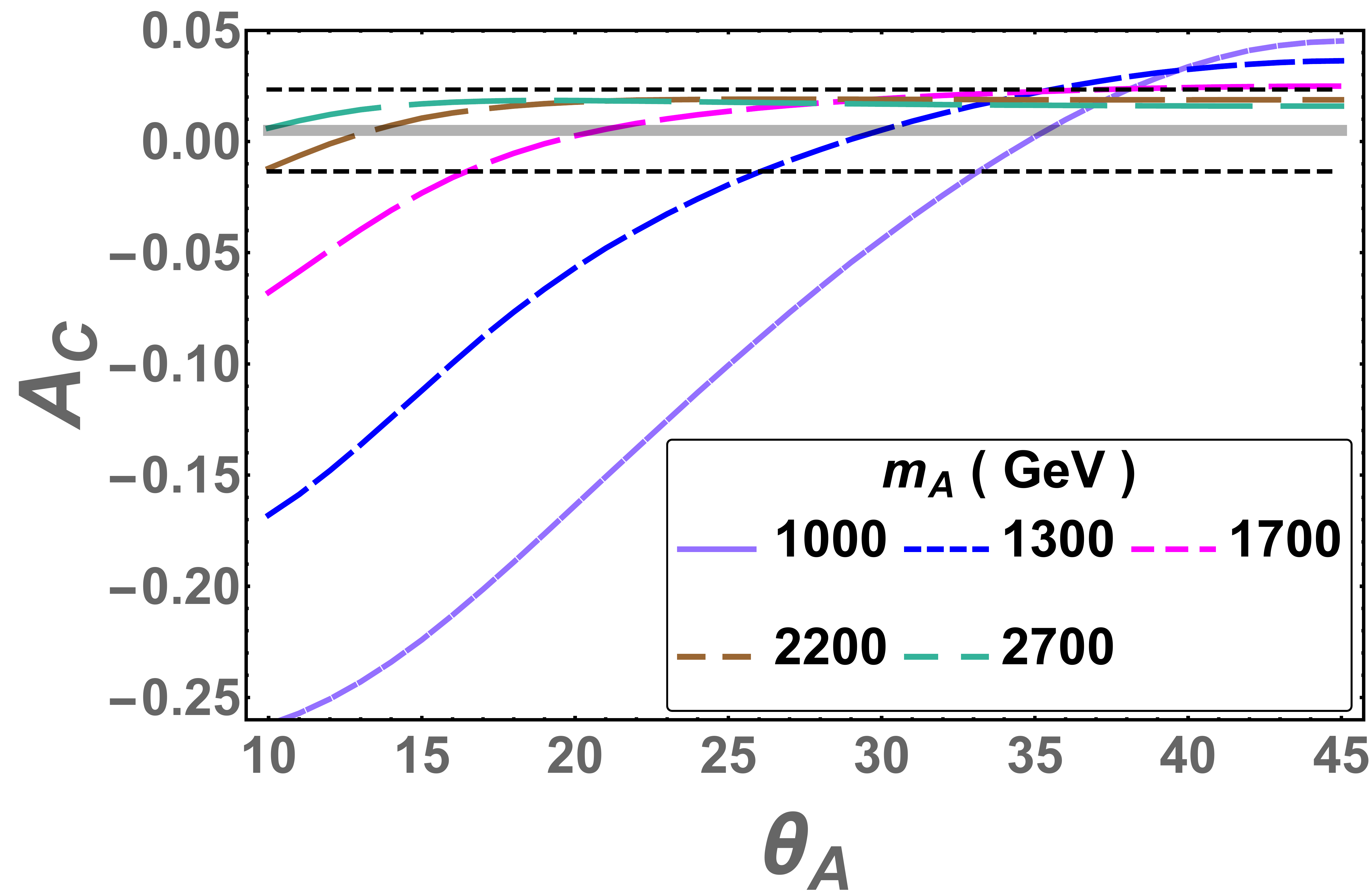}

}

\caption{\label{fig:crs,AFB,Ac_Ax}Observables at the top quark level at the Tevatron and the LHC as a function of $\theta_{A}$ for various values of masses for the axigluon. The experimentally measured
values are marked in grey and the respective 2$\sigma$ errors in
dotted black lines. As the lines go from solid to dashed with larger
gaps, the mass of the axigluon rises from 1 TeV to 2.7 TeV.\hfill{}.}
\end{figure}

We calculate the differential cross-section of the process ($pp$)$p\bar{p}\to t\bar{t}\to l\nu b\bar{t}$
at the Tevatron with $\sqrt{s}=1.96$\textrm{ TeV} and at the LHC
with $\sqrt{s}=7$\textrm{ TeV} and $\sqrt{s}=13$\textrm{ TeV} for
the SM + bSM with CTEQ6l \cite{CTEQ6l1:Pumplin2002} parton distribution
functions with factorization scale fixed at $Q=2m_{t}=345$ GeV, the
top quark mass is taken to be $m_{t}=172.5$\textrm{ TeV} and $\alpha_{s}\left(m_{t}\right)=0.108$.

The cross-section calculated for the Tevatron, LHC and the AC
and AFB of the $t\overline{t}$ at those
experiments in the axigluon model are shown in figure \ref{fig:crs,AFB,Ac_Ax}.
We constrain the model parameter space by limiting the predicted observables
$\sigma_{p\overline{p}\to t\overline{t}}$, $\sigma_{pp\to t\overline{t}}$,
$A_{FB}$ and $A_{C}$ to within $2\sigma$ of the experimental values.
As the values of $\theta_{A}$ and $m_{A}$ grow larger, the couplings
reduce, the mass of the mediating particle rises and the bSM contributions
to the observables reduce. At large values of $\theta_{A}$, the figures
correspond to an axigluon model with only an axial coupling
with the top quark and no resulting top polarization. For a lower
mass range, constraints from the LHC allow only larger $\theta_{A}$
and hence smaller coupling values, at the same time, interference
with SM gives a constraint at the Tevatron which allows some region
in the large coupling range as well. $A_{C}$ gives a complimentary
constraint and rules out large values of $\theta_{A}$ (couplings
close to $g_{s}$) for a smaller mass of the axigluon. The result
is that for the low masses of the axigluon, a range of couplings corresponding
to $\theta_{A}\sim(25\textdegree-35\textdegree)$ and masses $m_{A}\sim(1300-1900)$
GeV are allowed. Masses above these values are allowed for almost
all parameter space with the only constraints coming from the Tevatron
cross-section.

CMS results constrain the mass of an additional
massive spin-1 colour octet of particles (eg. Kaluza Klein-gluon) which couple to gluons and quarks to above $3.5$ TeV,
which excludes the parameter region favoured by
the experimental results from the  $t\overline{t}$ process mentioned above \cite{Khachatryan:2015sja}. The constraints can be evaded if the assumption of equal couplings
of axigluons to light quarks and the top quark is relaxed. In this
case, the values of coupling $g_{V},g_{A}$ we use can be split into
$g_{V}^{q},g_{A}^{q}$ and $g_{V}^{t},g_{A}^{t}$ where, the couplings
with quarks would be constrained strongly from the axigluon direct
production bounds. In the limit that the vector and axial couplings
are equal or any one of the vector or axial coupling is small, our
results can be recast into the modified model by using $g_{v/a}^{2}=g_{v/a}^{q}g_{v/a}^{t}$.
A more generalized version of such an axigluon model has already been
discussed in the literature \cite{Aguilar-Saavedra:2014nja} along
with constraints on the model from lepton and top quark asymmetries
at the Tevatron and the LHC.

The axigluon model can be constrained from B physics \cite{Chivukula:2010Axnotallowed}
results however, given the somewhat large hadronic uncertainties in some of the variables along with the possibility of relaxing these constraints in various modified axigluon models and/or by constructing UV
completions, for the purpose of this study, we do not take these constraints into account.

\section{U-channel scalar exchange model\label{sec:U-channel-Scalar-Model}}

In a second class of bSM models, AFB is explained due to contributions of a $t$- or $u$-channel
exchange of new particles between the top quark-antiquark pair. The
corresponding mediators do not show resonance behaviour
and are elusive in the bump-hunting type analyses in $t\overline{t}$
pair production though they do contribute significantly to the angular
distributions. We consider here a scalar particle called diquark, which, similar to a squark
with R parity violation, transforms as a triplet under $SU(3)_{c}$
and has a charge of $-\frac{4}{3}$. The corresponding coupling is
given by the lagrangian below,
\begin{eqnarray}
\mathcal{L} & = &\overline{t^{c}}T^{a}(y_{s}+y_{p}\gamma^{5})u\phi_{a}+h.c.\\
t^{c} & = & -i\gamma^{2}t^{*}\nonumber \\
 & = & -i(\overline{t}\gamma^{0}\gamma^{2})^{T}\label{eq: t^c}
\end{eqnarray}
We assume a right-handed coupling of the scalar with the up type
quarks with $y=y_{s}=y_{p}$. This ensures that flavour constraints
and proton stability bounds are avoided. The density matrices for
top pair production in the diquark model are given in Appendix \ref{sub:diquark-Density-Matrices}.
All calculations in this work are performed at tree level. The NLO
contributions become important to study the effects on invariant mass
distributions which we have not included here. These calculations
are under progress for both axigluon and diquark models.

\subsection{Constraints on the coloured scalar model}
\begin{figure}[h]
\subfloat[Cross-section at the Tevatron with $\sqrt{s}=1.96$\textrm{ TeV}]{\includegraphics[scale=0.205]{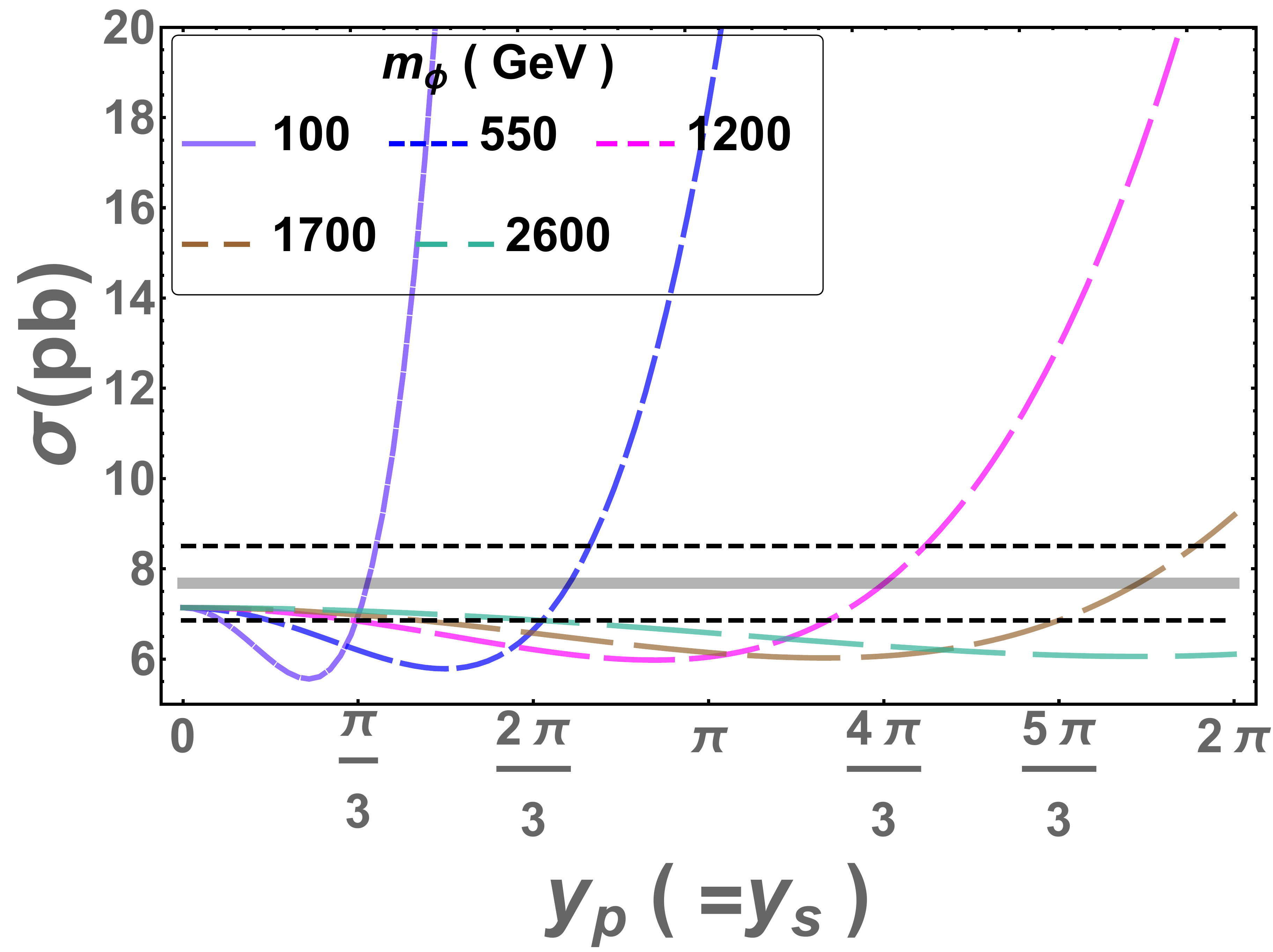}

}\hfill{}\subfloat[AFB at the Tevatron with $\sqrt{s}=1.96$\textrm{ TeV}]{\includegraphics[scale=0.21]{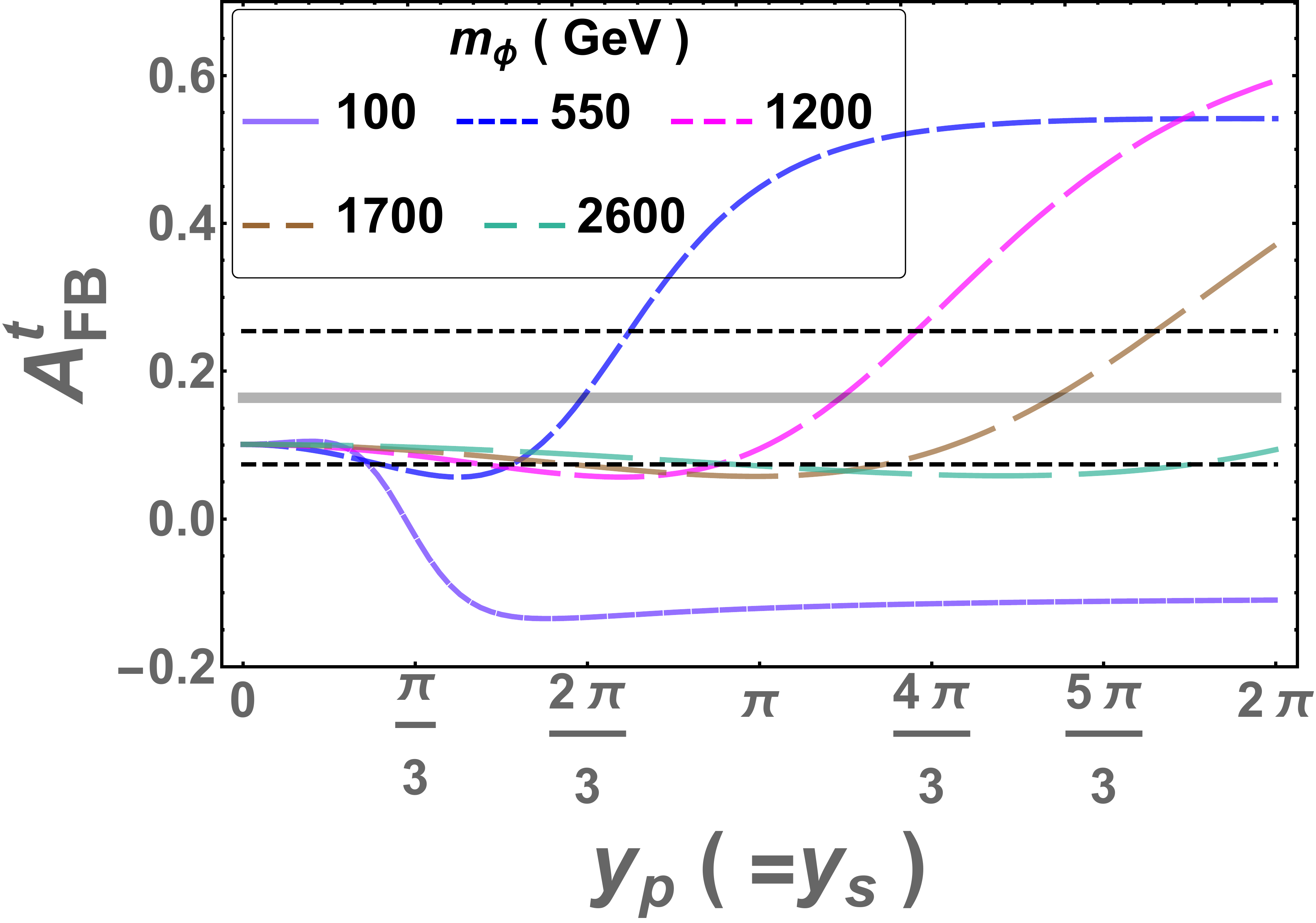}

}

\subfloat[Cross-section at the LHC with $\sqrt{s}=7$\textrm{ TeV}]{\includegraphics[scale=0.225]{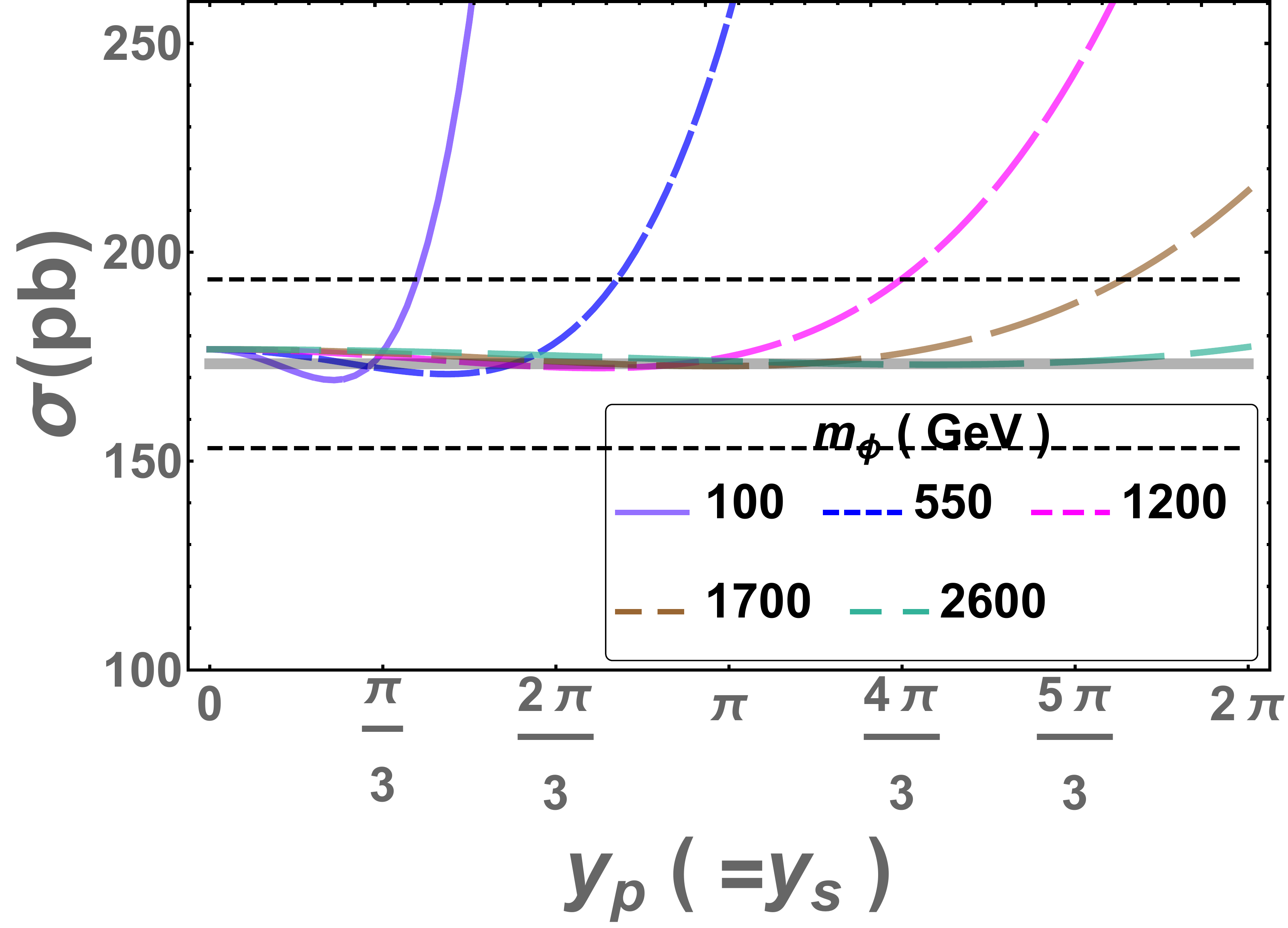}

}\hfill{}\subfloat[$A_{C}$ at the LHC with $\sqrt{s}=7$\textrm{ TeV}]{\includegraphics[scale=0.235]{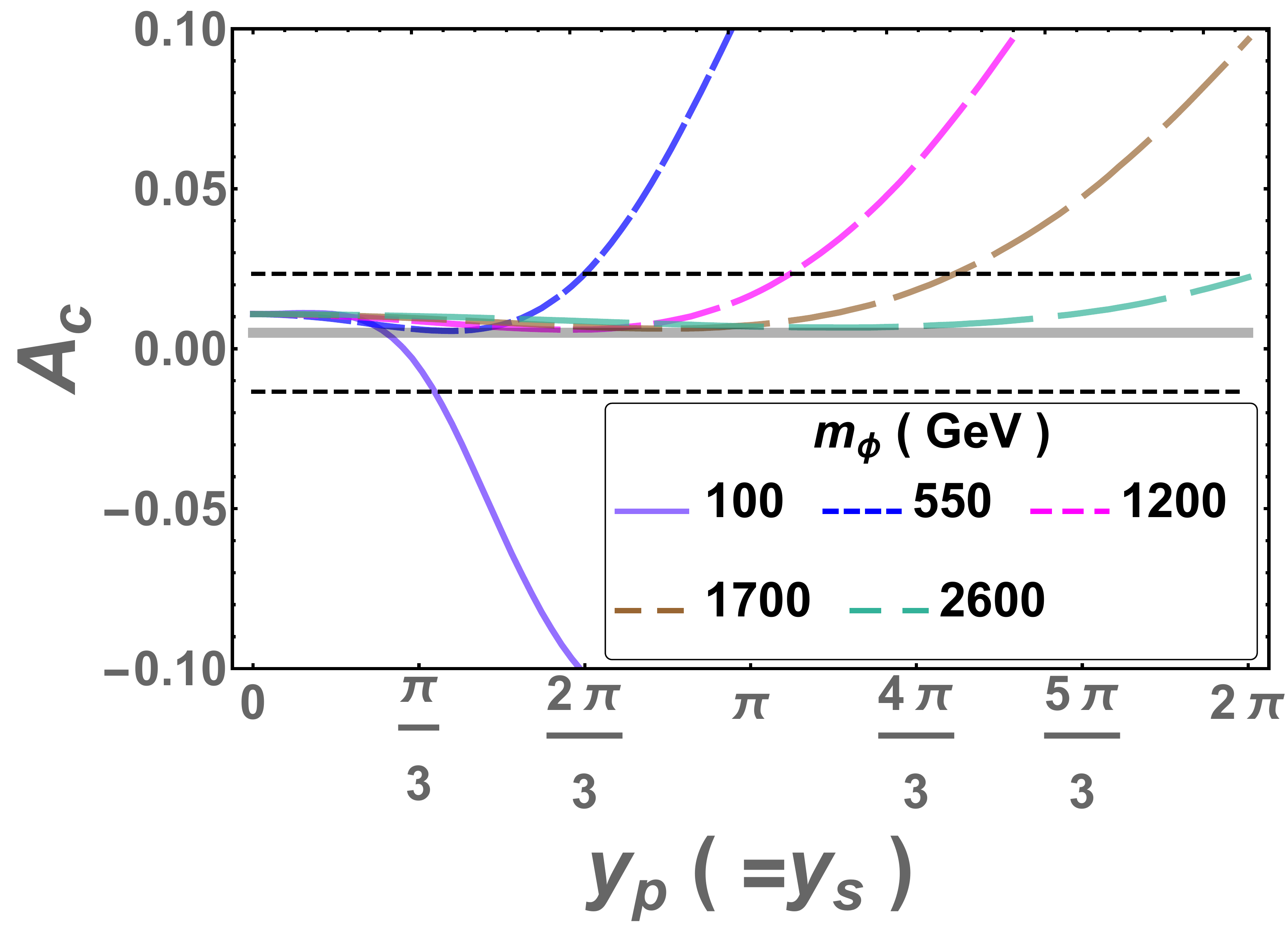}

}

\caption{\label{fig:crs,AFB,Ac_diq} Observables at the top-quark level at the Tevatron and the LHC (7 TeV) as a function of the Yukawa coupling for various values of the diquark masses. The experimentally
measured values are marked in grey and the respective 2$\sigma$ errors
in dotted black lines. The line spacing changes from solid to a dashed
line with wider spaces as mass values rise from 100-2600 GeV.\hfill{}.}
\end{figure}
As in the case of an $s$-channel resonance in the previous section,
the constraints are obtained from the measurements of the top pair
production cross-section at the Tevatron and the LHC (7 TeV), the
AFB and AC. We explore a parameter space of $m_{\phi}\in[100,3000]$
GeV and $y\in[0,\,2\pi]$ chosen so as to explore all the values of the coupling within
the perturbative limit. As the value of the coupling rises, contribution
of the bSM to all the observables becomes larger. For lighter diquarks, negative values of AFB and AC are predicted
for large values of the coupling, though this mass range is ruled
out by independent constraints from diquark pair production \cite{CMS:2013gqa}. In figure \ref{fig:crs,AFB,Ac_diq} we can notice from the
top left panel that, as in the case of the axigluon, the Tevatron
cross-section provides the constraints in the parameter space of lower
masses and couplings. In the next panel, the AFB
measured at the Tevatron disallows lighter scalars and also constraints
a part of coupling values for larger masses. The LHC cross-section
constraints large coupling regions which give larger contribution
and the cut-off coupling increases as mass of the scalar becomes heavier.
The AC also allows larger coupling parameter space for
higher masses of the scalar.

The constraints from pair production of the coloured scalar from gluon
fusion at the LHC are weak (\textasciitilde{}300 GeV) as reinterpreted
from corresponding constraints on squarks \cite{Chatrchyan:2013izb}.
There are further constraints on lower mass scalars from atomic parity violation
\cite{Gresham:2012wc}. Constraints from $uu\to tt$ can be avoided
by adding flavour symmetries (see for example \cite{Grinstein:2011yv}).

\section{Constraints from dijet production at LHC}\label{sec:dijet-constraints}
\begin{figure}[th]
\subfloat[\label{fig:param-space-ax}Constraints on axigluon parameter space]{\includegraphics[scale=0.2]{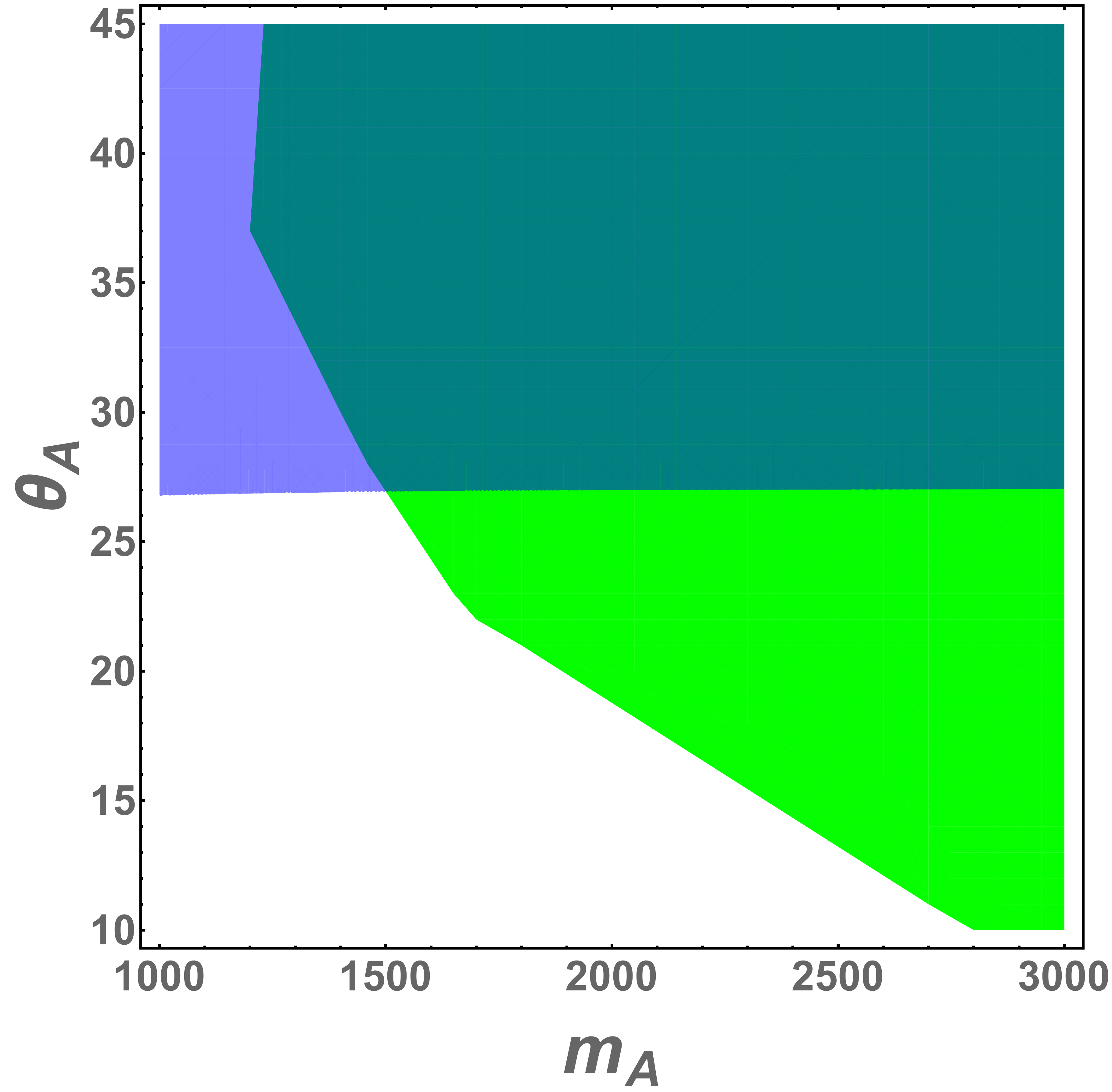}}\hfill{}
\subfloat[\label{fig:param-space-dq}Constraints on diquark parameter space]{\includegraphics[scale=0.2]{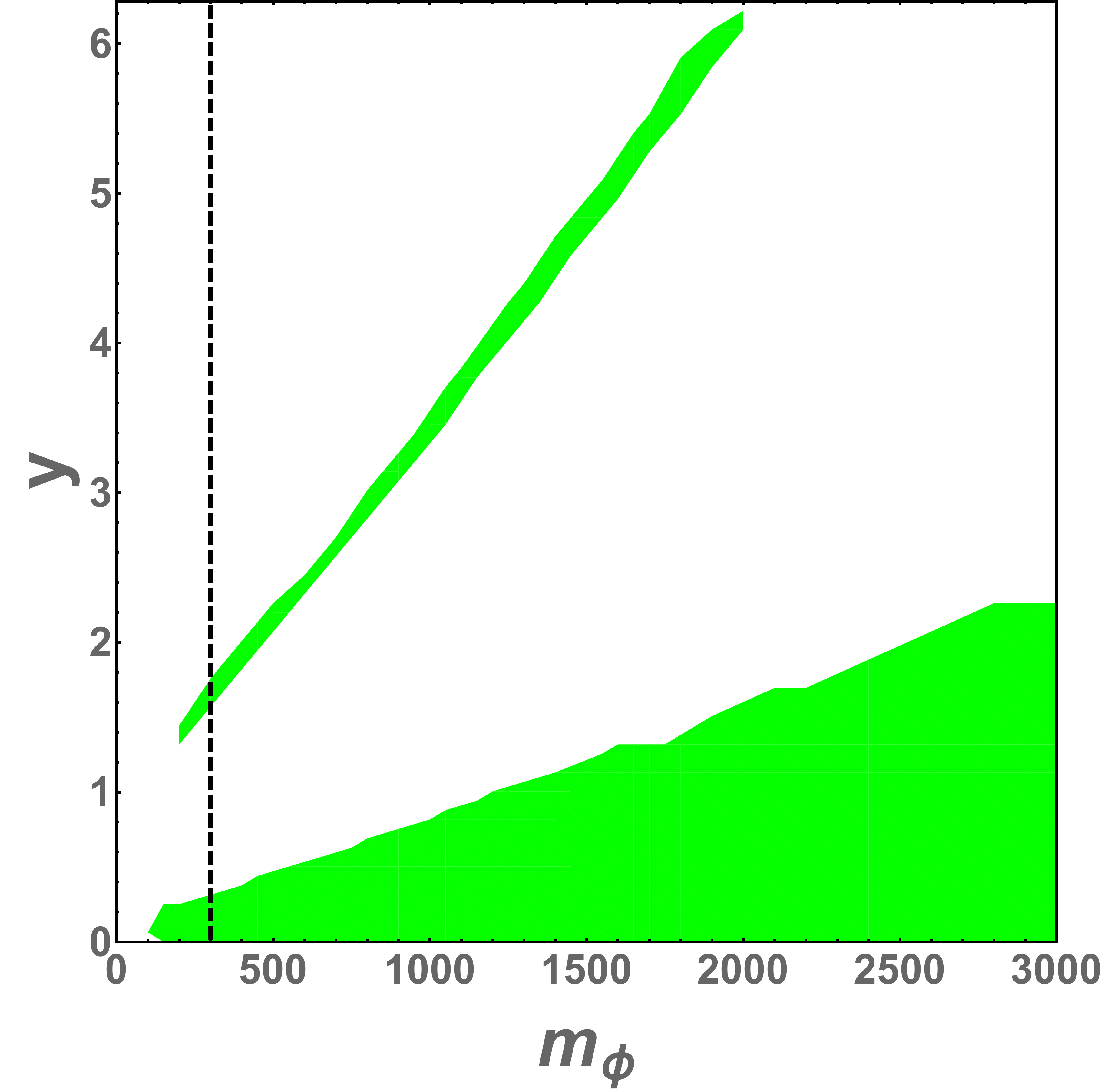}}
\caption{\label{fig:param-space}Allowed parameter space for axigluon and diquark models are depicted as Green (lighter) coloured regions. Figure \ref{fig:param-space-ax} shows the constraints from dijet searches as the blue (darker) shaded area. In figure \ref{fig:param-space-dq} the dotted line represents the bound from pair production of bSM particles at LHC.}
\end{figure}
The coloured scalar and vector bSM models get constrained from searches for direct production of bSM particles and subsequent decay to di-jet and four jet final states ( $ q\bar{q}\to A\to 2j$ , $gg\to\phi\phi^\dagger\to 4j$). Earlier constraints on axigluon model were obtained from searches of narrow resonances from dijet spectrum at 8 TeV LHC and were extended to the case of a wider width axigluon model with $\frac{\Gamma_A}{m_A}=0.3$ where the axigluon has only axial-vector couplings \cite{Diaz:2013tfa} (also see \cite{Choudhury:2011cg} for bSM particle off-shell effects in dijet searches). We reinterpret these constraints to the case of the axigluon model in this study which has both axial-vector and vector couplings with the quarks and find the excluded parameter range where $\frac{\Gamma_A}{m_A}<0.3$. Figure \ref{fig:param-space-ax} shows the parameter space allowed for the axigluon model. The following constraints are put on the model parameters to obtain the allowed values : reinterpreted searches for bSM resonances in dijet production, $t\bar{t}$ cross-section and top charge asymmetry measurements at 7 TeV LHC and cross-section, top forward backward asymmetry measurements at the Tevatron as discussed in section \ref{sec:const-axigluon}. The coupling values corresponding to $\theta_A>27^\circ$ are ruled out for axigluon masses up to $\sim 4$ TeVs as these narrow, resonant particles would have been detected in the dijet searches. The allowed values of couplings correspond to $\theta_A\sim10^\circ-27^\circ$ for the mass range between 1.5 TeV to 3 TeV. Note that the constraints from dijet searches may be relaxed if the magnitude of coupling of axigluon is different for the third generation of quarks as compared to the first and the second generations.

The diquark mass is bound from below to $m_\phi>\sim300$ GeV from pair production of diquarks via gluon fusion at the LHC \cite{Chatrchyan:2013izb}. The direct production bounds along with the constraints obtained from top quark pair production cross-section and top charge, forward backward asymmetry measurements at LHC and Tevatron (see section \ref{sec:U-channel-Scalar-Model}) are shown in figure \ref{fig:param-space-dq}. A narrow strip of parameter space is allowed when couplings are large due to destructive interference effects. Besides this region, the rest of the allowed diquark parameter space follows the expected behaviour of small coupling values $<0.2$ for lower masses and for a diquark of mass 3 TeV, $y_s$ as large as 2 is allowed.

\section{Polarization of the top quark and decay-lepton distributions\label{sec:Asymmetry-Definitions}}

The decay kinematics of leptons embeds the information regarding top
quark production dynamics, kinematics and polarization \cite{Papaefstathiou:2011kd}. Different
lepton observables embed these effects in different ways and so provide
 a number of probes which are all correlated with the top
quark kinematics and polarization. For a detailed analysis of top
quark decay see \cite{Jezabek:1988topdecayleptondist,Godbole:2006ldistandtpol}.
In this section we discuss distributions of the lepton polar angle, azimuthal angle
and energy in SM decay of top quark and construct asymmetries based
on these distributions to probe top quark bSM interactions.

A proper treatment of the decay distributions of the top quark requires
the spin density matrix formulation, which preserves correlations
between the spin states in the production and in the decay.

The spin density matrix for $t$ in the production of a $t\overline{t}$
pair with the spin of $\overline{t}$ summed over, can be expressed
as 
\begin{equation}
\rho_{t\overline{t}\, production}\left(\lambda_{t},\lambda_{t}^{\prime}\right)=\sum_{\lambda_{\bar{t}}}\mathcal{M}_{production}\left(\lambda_{t},\lambda_{\bar{t}}\right)\mathcal{M}_{production}^{\star}\left(\lambda_{t}^{\prime},\lambda_{\bar{t}}\right)\label{eq:top-density-matrix}
\end{equation}
The density matrix gets SM contributions $\rho_{SM}^{gg}\left(\lambda_{t},\lambda_{t}^{\prime}\right)$
and $\rho_{SM}^{q\bar{q}}\left(\lambda_{t},\lambda_{t}^{\prime}\right)$
respectively from gluon-gluon and quark-anti-quark initial states,
a contribution $\rho_{bSM}\left(\lambda_{t},\lambda_{t}^{\prime}\right)$
from the bSM model, and a contribution $\rho_{interference}\left(\lambda_{t},\lambda_{t}^{\prime}\right)$
from the interference between the SM amplitude and the bSM amplitude:
\begin{equation}
\rho\left(\lambda_{t},\lambda_{t}^{\prime}\right)=\rho_{SM}^{gg}\left(\lambda_{t},\lambda_{t}^{\prime}\right)+\rho_{SM}^{q\overline{q}}\left(\lambda_{t},\lambda_{t}^{\prime}\right)+\rho_{bSM}\left(\lambda_{t},\lambda_{t}^{\prime}\right)+\rho_{Interference}\left(\lambda_{t},\lambda_{t}^{\prime}\right)
\end{equation}

The spin density matrix for the decay of the top quark is given by
\begin{equation}
\Gamma_{top\, decay}\left(\lambda_{t},\lambda_{t}^{\prime}\right)=\mathcal{M}_{decay}\left(\lambda_{t}\right)\mathcal{M}_{decay}^{\star}\left(\lambda_{t}^{\prime}\right),
\end{equation}
with the spins of the decay products summed over.

\begin{figure}[t]
\subfloat[{\footnotesize{\label{fig:Lepton-polar-angle-rest}Lepton polar
angle distribution in top quark rest frame.}}]{\includegraphics[scale=0.26]{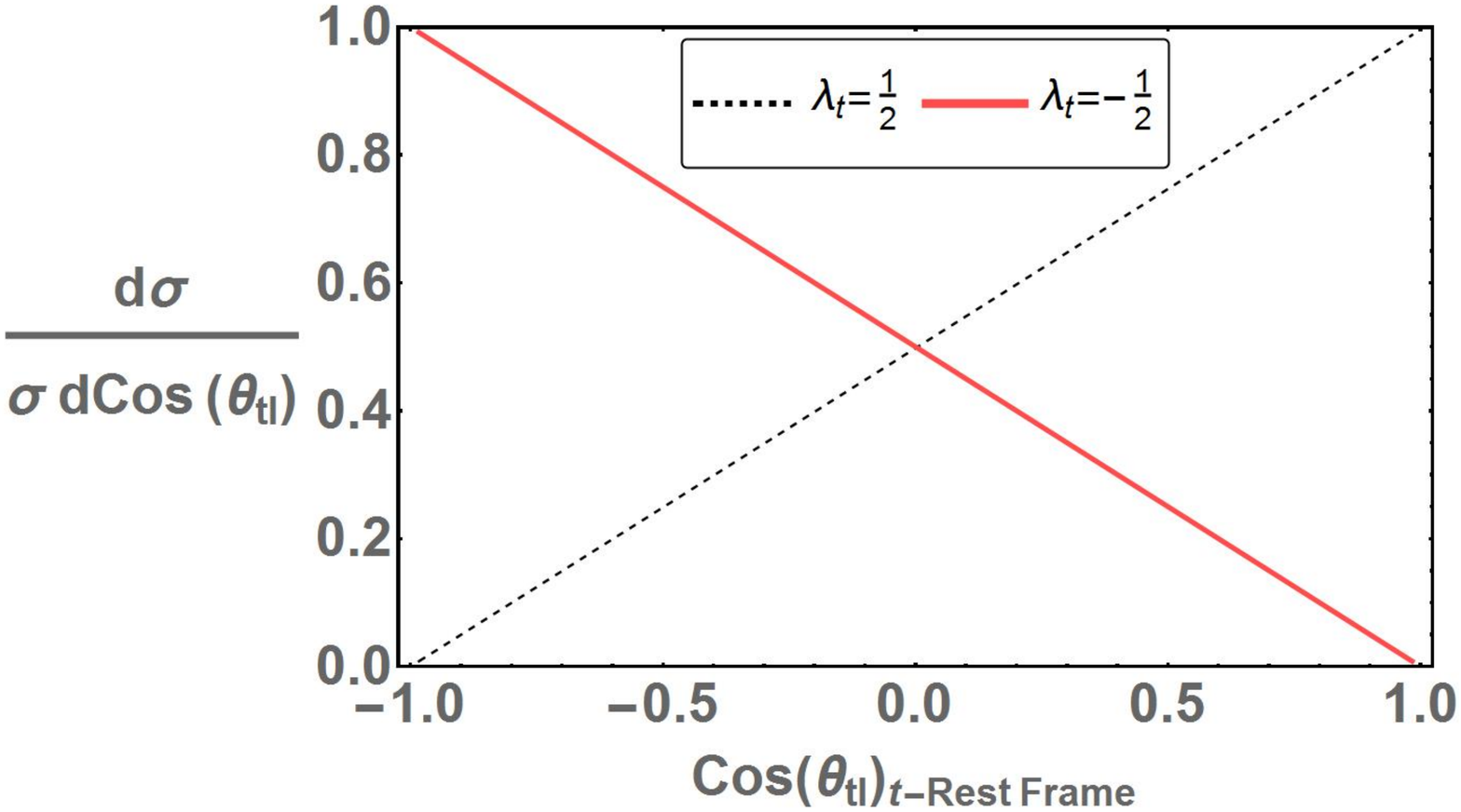}

}\hfill{}\subfloat[{\footnotesize{\label{fig:Lepton-polar-angle-lab}Lepton polar
angle distribution in the lab frame.}}]{\includegraphics[scale=0.26]{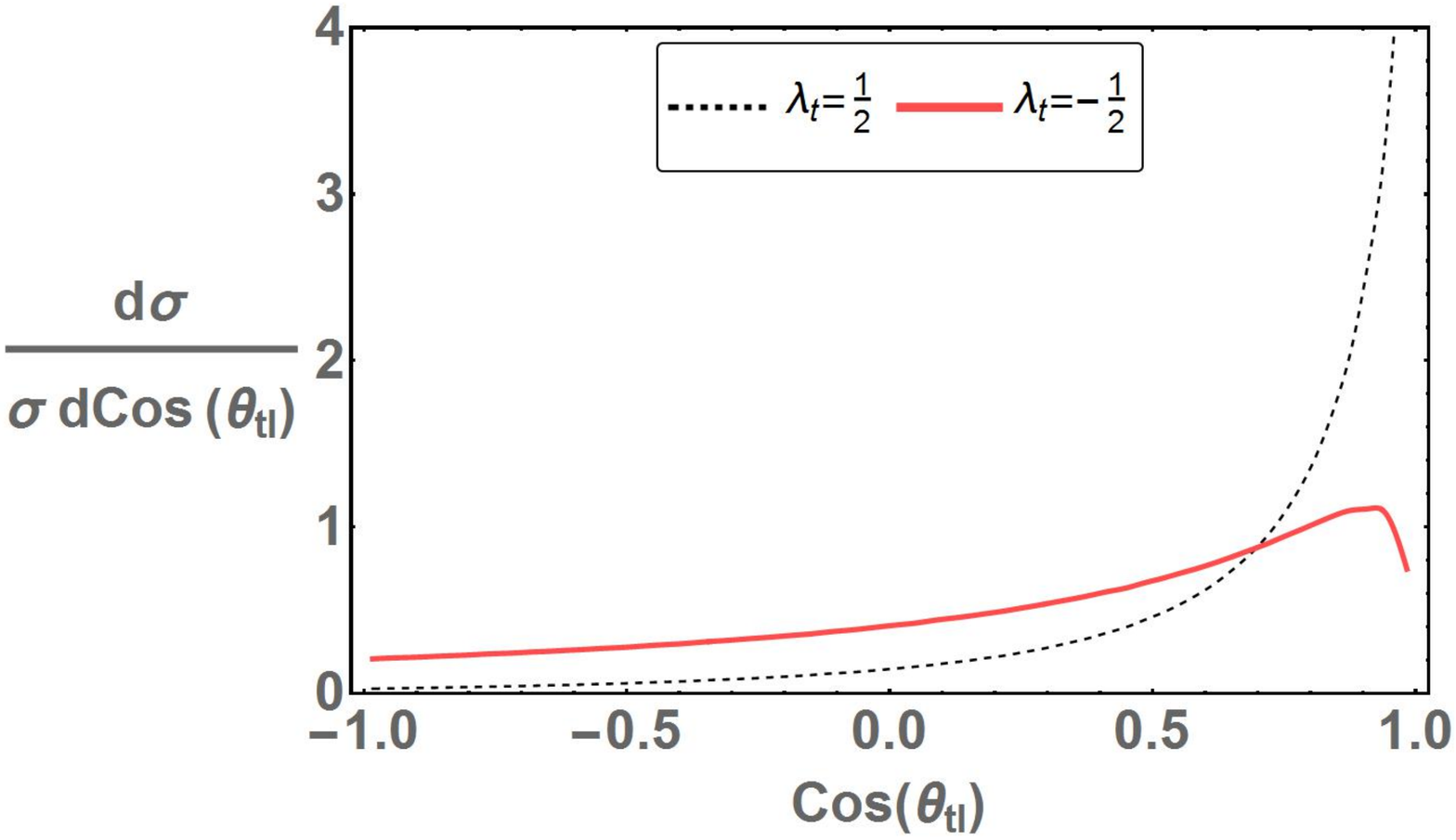}

}

\subfloat[{\footnotesize{\label{fig:Lepton-Azimuthal-angle-dist}Lepton Azimuthal
angle distribution in the lab frame }}]{\includegraphics[scale=0.26]{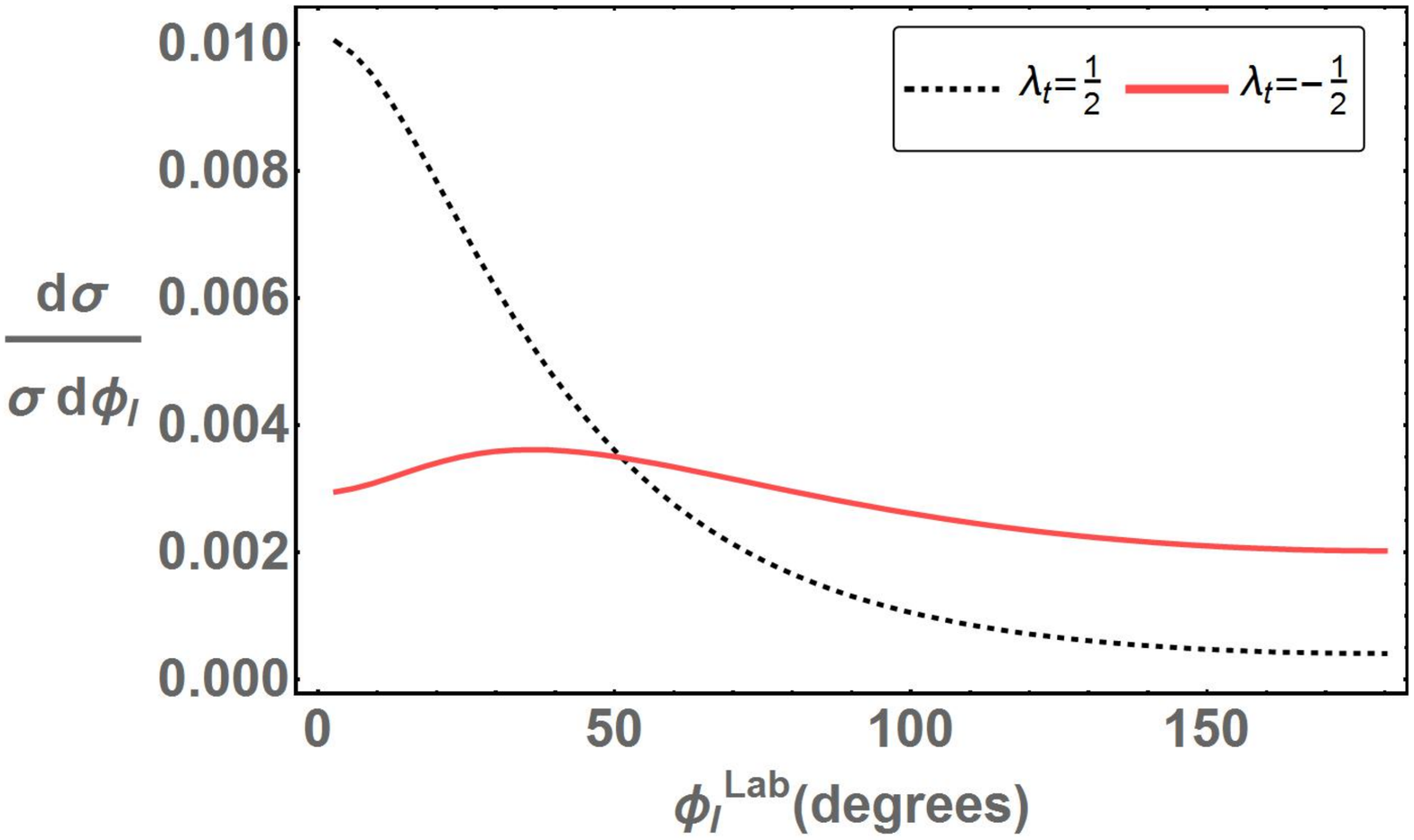}

}\hfill{}\subfloat[{\footnotesize{\label{fig:Lepton-energy-distribution}Lepton energy
distribution in the lab frame}}]{\includegraphics[scale=0.26]{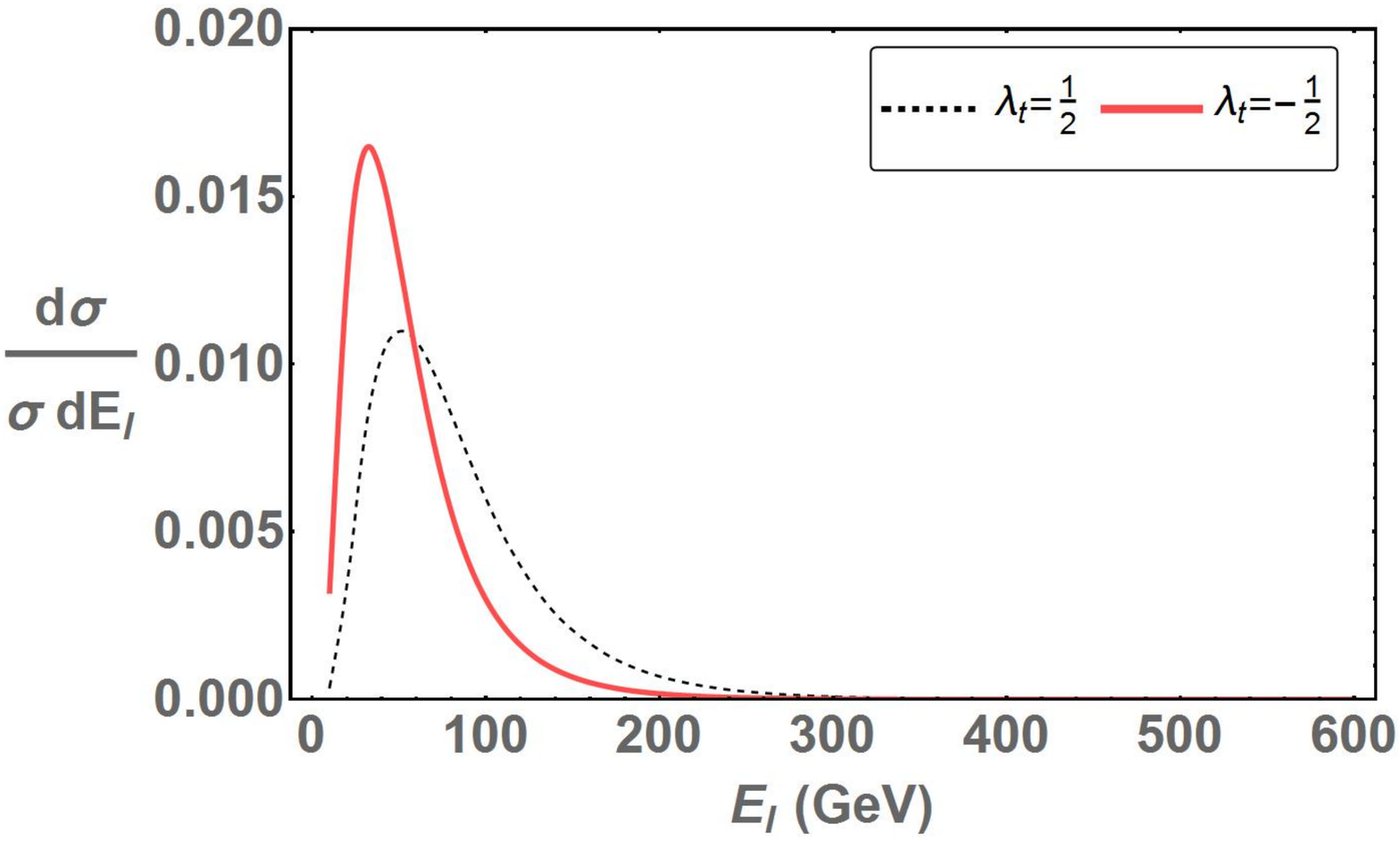}}

\caption{\label{fig:Lepton-distributions} The tree-level lepton polar and
azimuthal distributions for $p\overline{p}\to t\overline{t}\to l+jets$
with $\sqrt{s}=1.96$\textrm{ TeV}. In the above plots, the average boost of the $t\bar{t}$ pairs is 0.34.}
\end{figure}

The squared amplitude for the combined process of production and decay
is given by 
\begin{eqnarray}
\overline{|\mathcal{M}|^{2}} & = & \frac{\pi\delta\left(p_{t}^{2}-m_{t}^{2}\right)}{\Gamma_{t}m_{t}}\sum_{\lambda_{t},\lambda_{t}^{\prime}}\rho\left(\lambda_{t},\lambda_{t}^{\prime}\right)\Gamma\left(\lambda_{t},\lambda_{t}^{\prime}\right)\label{eq:total_amp sq}
\end{eqnarray}
This expression assumes a narrow-width approximation for the top quark.
Top decay is assumed to progress through SM processes.
In the rest frame of top quark, the differential decay distribution
of the top quark is given by 
\begin{equation}
\frac{d\Gamma_{t}}{\Gamma d\cos(\theta)}=\frac{1+A_{p}k_{f}\cos\theta}{2}\label{eq:top rest theta_l}
\end{equation}
where $\theta$ is the angle between top-quark spin direction and
the momentum of the decay product $f$. For $N(\lambda_t)$ number of top quarks with helicity $\lambda_t$, polarization $A_{p}$ is defined as, 
\begin{equation}
A_{P}=\frac{N(\lambda_{t}=+)-N(\lambda_{t}=-)}{N(\lambda_{t}=+)+N(\lambda_{t}=-)},\label{eq:polarization}
\end{equation}
and the coefficient $k_{f}$ is called the top-spin analysing power
of the decay particle $f$. For the case of leptons as the final state
particles, the factor $k_{f}=1$ at tree level in the SM. When the top quark is boosted
in the direction of its spin quantization axis, eqn (\ref{eq:top rest theta_l})
gets modified to 
\begin{equation}
\frac{\mbox{d}\Gamma^{Boosted}}{\Gamma\mbox{d}\cos(\theta_{tl})}=\frac{\left(1-\beta^{2}\right)(1+\lambda_{t}\cos\theta_{tl}-\beta(\cos\theta_{tl}+\lambda_{t}))}{2(1-\beta\cos\theta_{tl})^{3}},\label{eq:top_theta_l
boosted}
\end{equation}
where $\theta_{tl}$ is defined as the angle between lepton and top quark momenta in the boosted frame.
Lepton kinematic distributions for the tree-level SM differential
cross-section for the process $p\overline{p}\to t\overline{t}\to l+jets$
with $\sqrt{s}=1.96$ TeV are presented in the figure \ref{fig:Lepton-distributions}.
The energy and azimuthal lepton distributions are uncorrelated with
the polarization in the rest frame of the top quark, though correlate
with the polarization in the boosted frame. Higher-order corrections
to the production and decay processes have been calculated for the
SM and the distributions are found to be qualitatively unchanged \cite{Melnikov:2009dn,Bernreuther:2014dla}.
Due to this reason, we expect that the effect of higher-order corrections
to the asymmetries constructed from decay-lepton distributions to
be relatively small as the corrections partially cancel out within
the difference and ratio taken to derive the asymmetry. 

A lepton polar-angle asymmetry with respect to the top-quark direction
can be defined by 
\begin{eqnarray}
A_{FB}^{tl} & = & \frac{\sigma(\cos(\theta_{tl})>0)-\sigma(\cos(\theta_{tl})<0)}{\sigma(\cos(\theta_{tl})>0)+\sigma(\cos(\theta_{tl})<0)}\label{eq: definition top-lepton polar asymmetry}
\end{eqnarray}
$A_{FB}^{tl}$ has been measured at both LHC and Tevatron in the lab and $t\bar{t}$ center of momentum frame, albeit with large statistical errors and different results from CDF and D0 \cite{cdf_Alpm}. Integrating eqn (\ref{eq:top rest theta_l}) the top rest-frame
lepton asymmetry can be related to the polarization of the top quark,
\begin{eqnarray}
A_{P}=\mbox{\ensuremath{\frac{1}{2}A_{FB}^{tl,t-rest}}}.\label{eq:tl-polar-asymmetry and pol}
\end{eqnarray}
In QCD, $A_{FB}^{tl,rest}=A_{p}=0$, though in the boosted frame, the lepton polar asymmetry with respect to the top quark is large even in a tree-level SM calculation.

In the lab frame where the top quarks and leptons are
boosted and the cross-section convoluted with the pdf, the correlations
between various angles and energies become more complicated. The lepton
polar angle with respect to the proton beam is a convenient observable
which does not require top-quark rest frame or momenta reconstruction. The lepton polar
asymmetry $A_{FB}^{l}$ in the lab frame is also 0 at tree level in SM
QCD. $A_{FB}^{l}$ is identically 0 at the LHC due to the symmetric
nature of the initial state. This asymmetry according to our analysis
correlates the best with the off-diagonal elements of the top quark density matrix for the $l+jets$ process considered (see
Section \ref{sec:Off-Diagonal-Density-Matrix}) for both axigluon and
diquark models. An analytic study of the lepton polar angle and its correlation with
top AFB and polarization has also been made by Berger et al. \cite{Berger:2012_assym_corr}.
They relate the lepton and the top quark level polar asymmetries and
subsequently use this relation to distinguish between a sequential
axigluon and a W' type model \cite{Berger2013_AlandPol}.

In the top-quark rest frame, other lepton kinematic variables : azimuthal angle
and its energy have no dependence on the helicity of the top quark
and hence the integrated asymmetries are uncorrelated with the polarization. It has been noted in the literature that the lepton azimuthal distributions
correlate with the polarization of the top quark in a boosted frame
\cite{Godbole:2006ldistandtpol,Godbole:2010phidist_tpol_LHC,Godbole:2011_AlAphiAp,Barger:2011philcorr}.
Sums and differences of azimuthal decay angle in top pair production
process have also been used in the literature to study the polarization
and spin correlations of top quark in detail \cite{Baumgart:2012ay}.
For a detailed analysis of analytic relation between polarization
of a heavy particle and decay particle azimuthal asymmetry, see \cite{Boudjema:2009fz}.
We reproduce the azimuthal distribution in the lab frame for the SM
$t\overline{t}$ pair production process at the Tevatron in figure \ref{fig:Lepton-Azimuthal-angle-dist}.
The azimuthal distribution can be measured at both Tevatron and LHC
and requires only partial reconstruction of top quark rest frame. 
The azimuthal angle is defined by assuming that the top quark lies
in the x-z plane with proton (beam) direction as z-axis. From this distribution,
an azimuthal asymmetry about a point $\phi_{0}$ can be defined as
\begin{eqnarray}
A_{\phi}^{l} & = & \frac{\sigma(\pi>\phi_{l}>\phi_{0})-\sigma(\phi_{l}<\phi_{0})}{\sigma(\pi>\phi_{l}>\phi_{0})+\sigma(\phi_{l}<\phi_{0})}\label{eq: definition lepton azimuthal asymmetry}
\end{eqnarray}
A natural choice for the value of $\phi_{0}$ would be the point of
intersection of the distributions corresponding to left and right
helicity top quarks. For SM, this point is about $\phi=50\textdegree$
for both Tevatron and the LHC. The SM point would correspond to 0
polarization and would maximize correlation with bSM contribution.
We assume a value of $\phi_{0}$ a bit lower at $40\textdegree$.
Since the positive helicity top quark have larger differential cross-section
in this region, this choice enhances correlations of the lepton level
asymmetry for larger positive (or smaller negative) values of polarization.
The standard model tree-level results for this asymmetry at the Tevatron
and the LHC respectively are given in table \ref{tab:SM-Asymmetries} . In the lab frame, due to
the boost and rotation from the direction of the top quark, $A_{\phi}$
is sensitive to both the polarization and the parity breaking or t-channel
structure of the top quark coupling.
\begin{table}[t]
\begin{ruledtabular}
\subfloat[Tevatron $\sqrt{s}=1.96$\textrm{ TeV}]{%
\begin{tabular}{|c|c|c|}
\hline 
Asymmetry & Q = $m_{t}$ & Q = $2m_{t}$ \tabularnewline
\hline 
\hline 
$A_{FB}^{tl}$ & $0.645$ & $0.642$\tabularnewline
\hline 
$A_{\phi}^{l}$ & $-0.113$ & $-0.116$\tabularnewline
\hline 
$A_{E_{l}}^{l}$ & $0.381$ & $0.397$\tabularnewline
\hline 
\end{tabular}

}\hfill{}\subfloat[LHC $\sqrt{s}=7$\textrm{ TeV}]{%
\begin{tabular}{|c|c|c|}
\hline 
Asymmetry & Q = $m_{t}$ & Q = $2m_{t}$ \tabularnewline
\hline 
\hline 
$A_{FB}^{tl}$ & $0.748$ & $0.748$\tabularnewline
\hline 
$A_{\phi}^{l}$ & $-0.075$ & $-0.077$\tabularnewline
\hline 
$A_{E_{l}}^{l}$ & $0.138$ & $0.146$\tabularnewline
\hline 
\end{tabular}

}\hfill{}\subfloat[LHC $\sqrt{s}=13$\textrm{ TeV}]{%
\begin{tabular}{|c|c|c|}
\hline 
Asymmetry & Q = $m_{t}$ & Q = $2m_{t}$ \tabularnewline
\hline 
\hline 
$A_{FB}^{tl}$ & $0.789$ & $0.788$\tabularnewline
\hline 
$A_{\phi}^{l}$ & $-0.041$ & $-0.044$\tabularnewline
\hline 
$A_{E_{l}}^{l}$ & $0.036$ & $0.038$\tabularnewline
\hline 
\end{tabular}

}

\caption{\label{tab:SM-Asymmetries}Scale dependence of SM values of various
asymmetries tree level.}
\end{ruledtabular}
\end{table}
Another observable which can be constructed from the decay-lepton
kinematics is the lepton energy asymmetry about a chosen energy $E_{0}$:
\begin{equation}
A_{E_{l}}^{l}=\frac{\sigma(E_{l}>E_{0})-\sigma(E_{l}<E_{0})}{\sigma(E_{l}>E_{0})+\sigma(E_{l}<E_{0})}\label{eq: lepton energy asymmetry definition}
\end{equation}
No reconstruction of the top-quark rest frame is needed to measure $E_{l}$.
Just like the azimuthal case, this asymmetry can be measured both at
the LHC and the Tevatron. The lepton energy distribution is sensitive
to the polarization of top quark \cite{Godbole:2006ldistandtpol},
as shown in figure \ref{fig:Lepton-energy-distribution}. Similar asymmetries
based on the energy of decay particles or the ratios of these energies
have been used in the literature to study bSM physics \cite{Shelton:2008topPol,Godbole:2011_AlAphiAp,Carmona:2014gra,Prasath:2014mfa}.
We define the lepton energy
asymmetry about a value of $E_{0}=80$ GeV, to act as a better discriminator
between bSM and SM. Ideally, the point of intersection of the positive
and the negative top-polarization curves should form the best correlation
with the top polarization, though this point varies with the energy
and the invariant mass of the initial state. Standard model values
of asymmetries mentioned in this section are given in table \ref{tab:SM-Asymmetries}.
It would be interesting to use SM distributions at NLO to decide the
reference points $E_{0},\phi_{0}$, but since in the end we construct
asymmetries, we expect that the qualitative behaviour of our results
would not change.

In the recent past, polarization measurements have been made by collaborations
both at the Tevatron and the LHC. The polarization at the Tevatron
points towards a small positive value and that at the LHC to small
negative values. This is consistent with the small coupling and large
mass regimes of both the models studied here.

In the next section, we use correlations among top charge and forward-backward
asymmetries, decay lepton angular and energy asymmetries, and polarization
to uncover specific properties of bSM particles which can be inferred
from the Tevatron and the LHC data.

\section{Correlations\label{sec:Correlations}}

The parameter space of $m_{A}\in[1000,3000]$\textrm{ GeV} and $\theta_{A}\in[10,45]$ are explored
for the axigluon model and $m_{\phi}\in[100,3000]$\textrm{ GeV} and
$y_{s}\in[0,2\pi]$ for the coloured scalar. The figures in the section \ref{sec:asym_correlations}
show parameter space allowed by the constraints mentioned in sections
\ref{sec:Flavor-Non-Universal-axigluon}, \ref{sec:U-channel-Scalar-Model}.

\subsection{Correlations between charge and forward-backward asymmetries}

\begin{figure}[th]
\subfloat[\label{fig:AfbAc corr axigluon}The red markers represent axigluon
model with only axial interactions($g_{V}=0,g_{A}^{t}=g_{s}$). The
size of the plus marks represent a mass range from 1000-3000 GeV]{\includegraphics[scale=0.21]{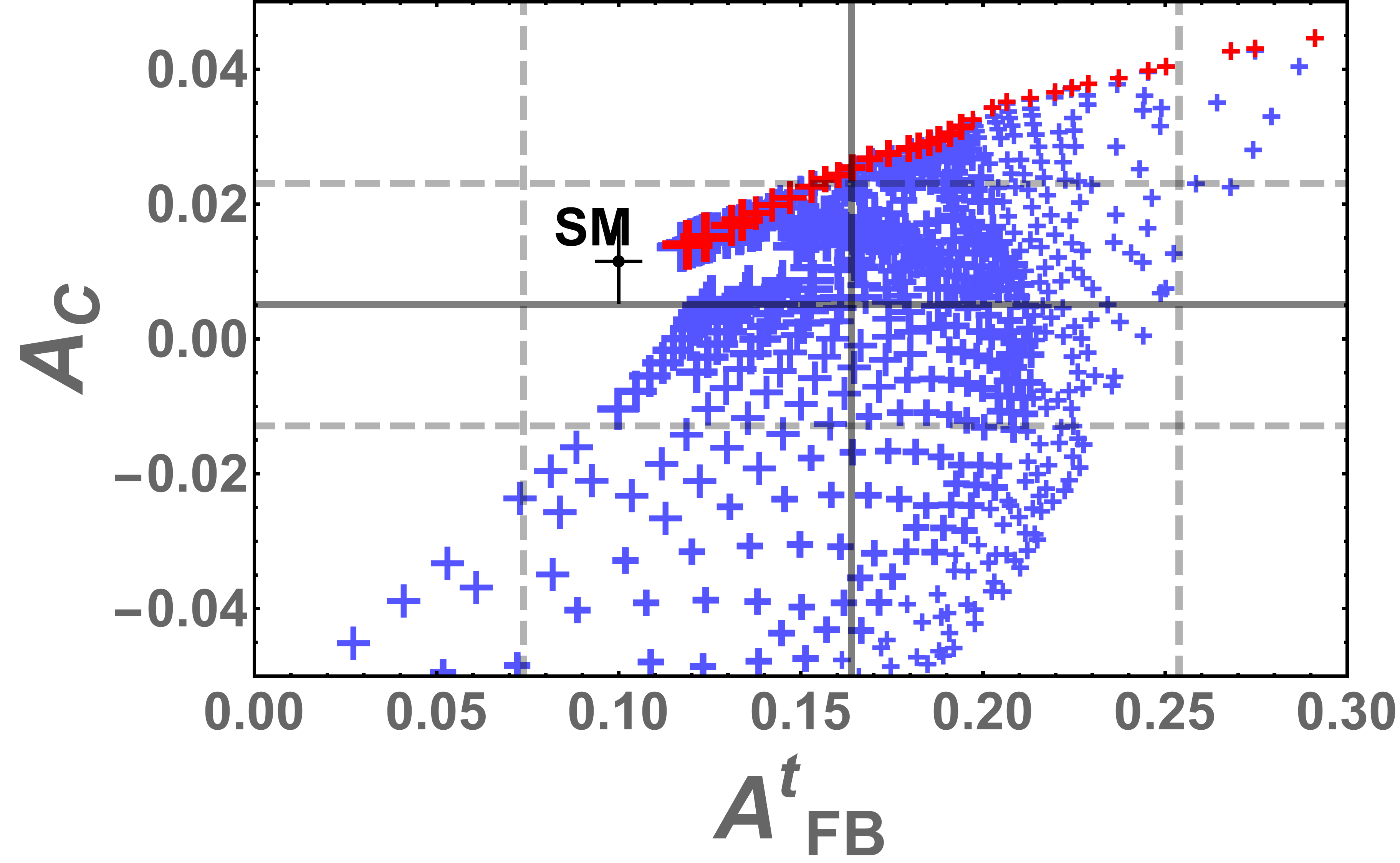}

}\hfill{}\subfloat[diquark model with right handed couplings to u,t quarks. The size
of the plus marks represent a mass range from 100-3000 GeV]{\includegraphics[scale=0.21]{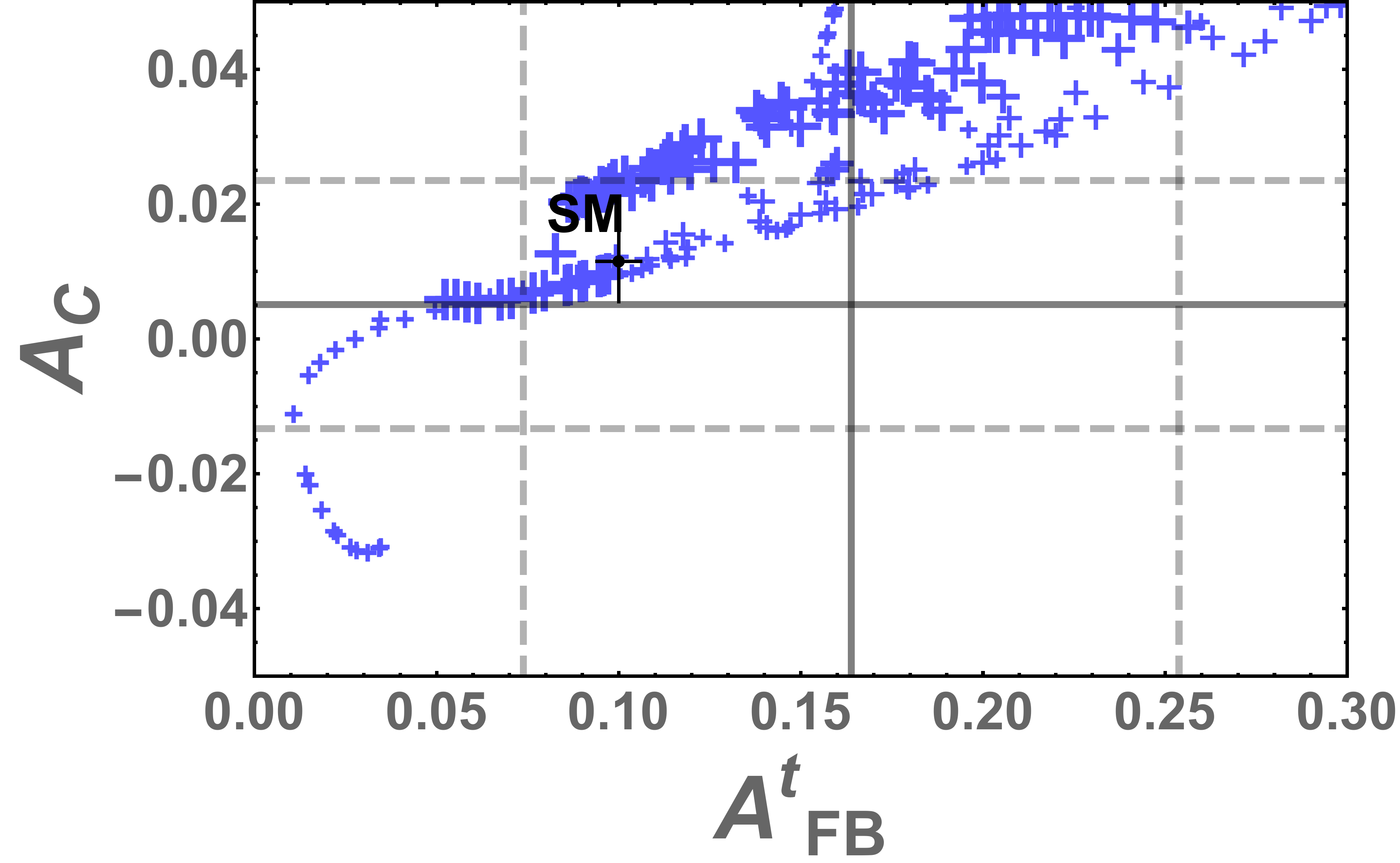}

}

\caption{\label{fig:Corr Afb and Ac}Correlation between top-quark asymmetries
$A_{FB}^{t}$ vs $A_{C}^{t}$ at the Tevatron and the LHC ($\sqrt{s}=7 $\textrm{TeV}). The grey solid and dashed lines represent the observed values of the respective asymmetries and their $2 \sigma $ errors.}
\end{figure}

The correlation between the AC at 7 TeV LHC and the
AFB at the Tevatron have been used in the literature constrain various
bSM models(see for example \cite{DC:2010_W'z'Axdiq}). These constraints are model dependent and the asymmetries are not in general tightly correlated \cite{Drobnak:2012cz}. We show similar correlations
in figure \ref{fig:Corr Afb and Ac} where we plot $A_{C}$ vs $A_{FB}$,
using the relation 
\begin{equation}
A_{C/FB}=A_{C/FB}^{SM\_NLO}+A_{C/FB}^{bSM}\label{eq:top asymmetry addition}
\end{equation}
This relation is valid as long as the bSM physics corrects the SM
cross-section of the $t\overline{t}$ pair production process by a
small amount.

The pure axial axigluon which leads to unpolarized top quark is disfavoured
as it does not have a parameter space where it can explain both $A_{FB}$
and $A_{C}$ experimental values. In the
diquark model, the coupling to right handed quarks is sampled from
0 to $2\pi$ where large mass or small couplings lead to a better
agreement with SM NLO values of asymmetries.

\subsection{Correlations among lepton and top asymmetries}\label{sec:asym_correlations}

In this section, we study the correlations among top polarization,
top asymmetries and decay lepton asymmetries. We show that
combined, they form sensitive discriminators between models with different
dynamics. The top-quark and decay-lepton asymmetries are calculated
at various points in the parameter space allowed by the experimental
constraints discussed in section \ref{sec:Flavor-Non-Universal-axigluon} and \ref{sec:U-channel-Scalar-Model}.
The expected polarization of the top quark, for corresponding points in the parameter space, is represented in colour contrast form inside the graphs and clear trends for the polarization can be observed. In all the following
figures, top-quark asymmetries represented on the $x$-axis are calculated
as shown in eqn (\ref{eq:top asymmetry addition}) and the lepton
asymmetries shown on the $y$-axis are calculated including SM+bSM
contributions at tree level.

\subsubsection{Asymmetry correlations for the Tevatron}
\begin{figure}[t]
\subfloat[\label{fig:Lepton-polar-asymmetry-tev}Lepton polar asymmetry Tevatron
$\sqrt{s}=1.96$ TeV.]{\includegraphics[scale=0.21]{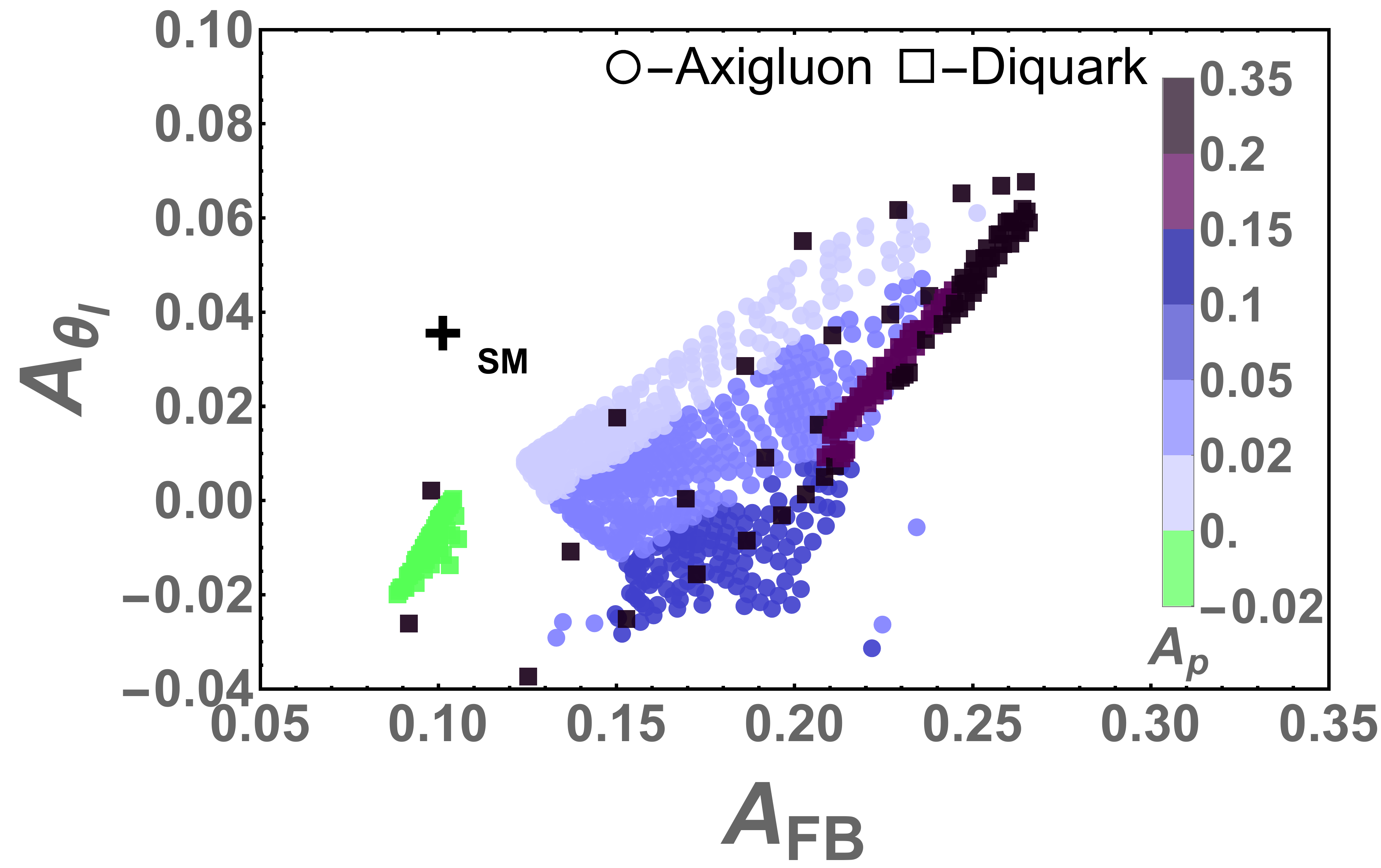}

}\hfill{}\subfloat[\label{fig:Top-Lepton-polar-asymmetry-tev}Lepton polar asymmetry
Tevatron $\sqrt{s}=1.96$ TeV.]{\includegraphics[scale=0.21]{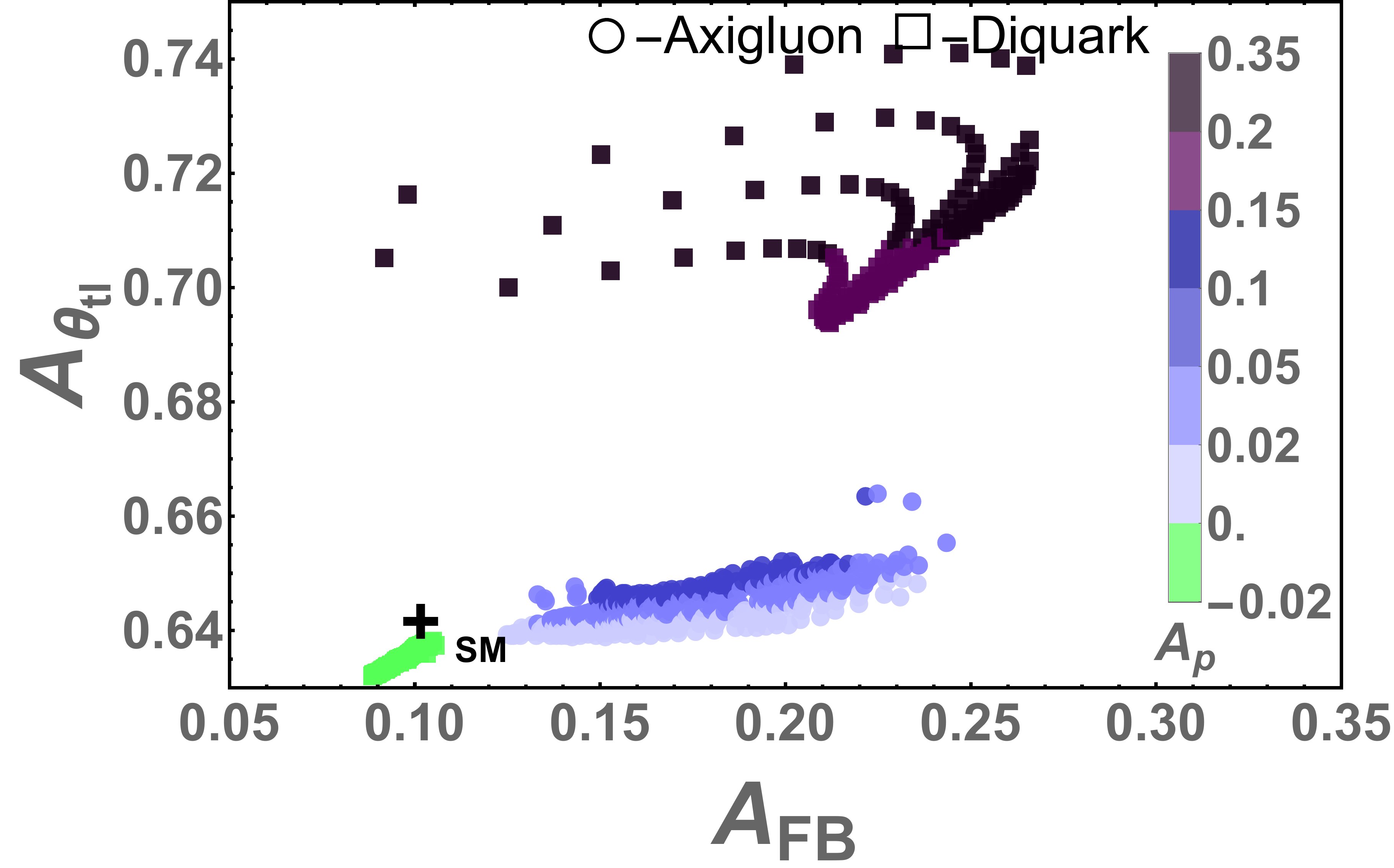}

}

\caption{Correlations between top AFB and lab frame $\theta_{l},\theta_{tl}$ asymmetries at the Tevatron $\sqrt{s}=1.96$\textrm{
TeV}}
\end{figure}

\begin{figure}[t]
\subfloat[\label{fig:Lepton-Azimuthal-asymmetry-tevatron}Lepton Azimuthal asymmetry
about $\phi_{0}$ for the Tevatron $\sqrt{s}=1.96$ TeV.]{\includegraphics[scale=0.21]{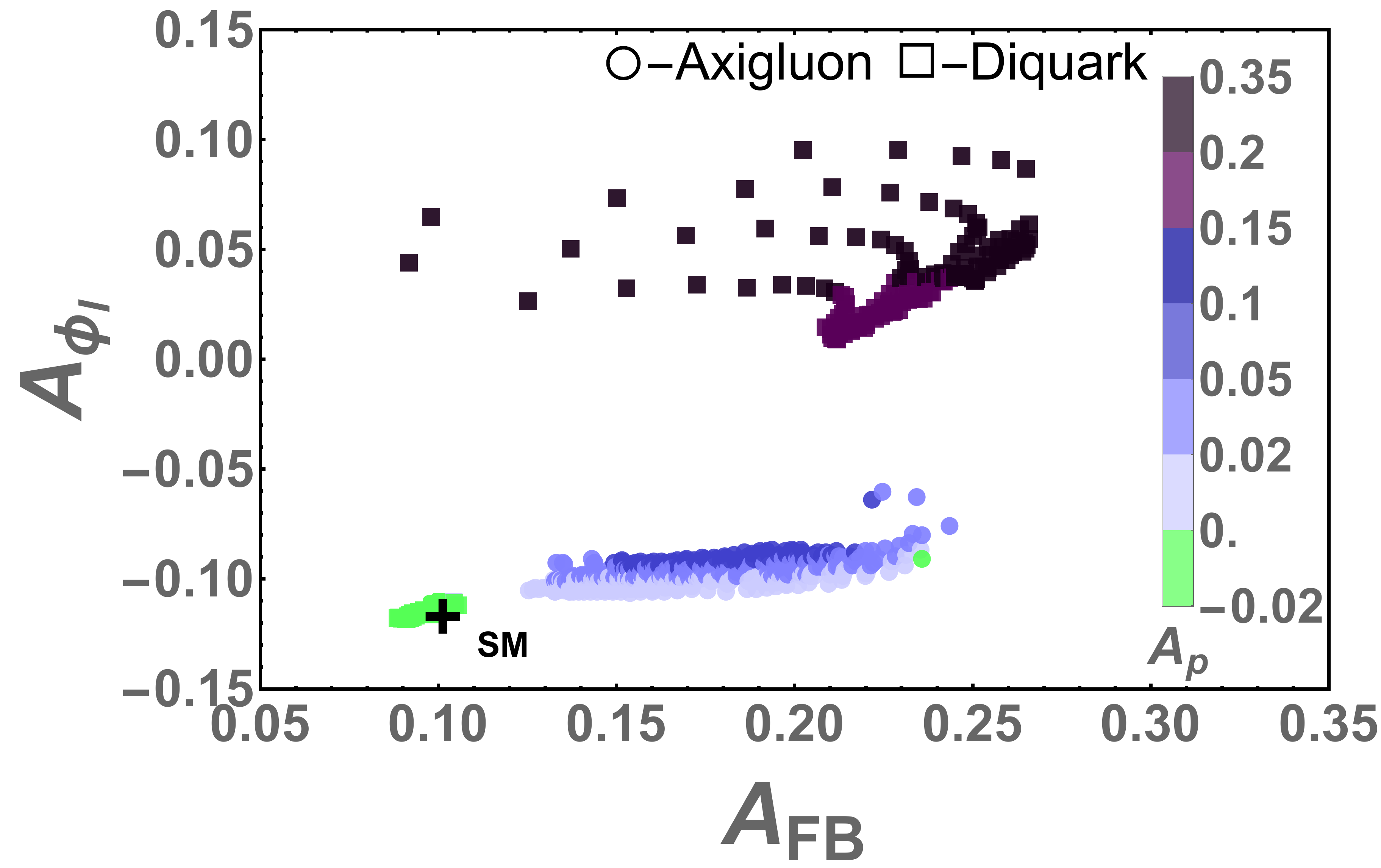}

}\hfill{}\subfloat[\label{fig:Lepton-energy-asymmetry-tev}Lepton energy asymmetry about
energy $E_{0}$ for the Tevatron $\sqrt{s}=1.96$ TeV.]{\includegraphics[scale=0.21]{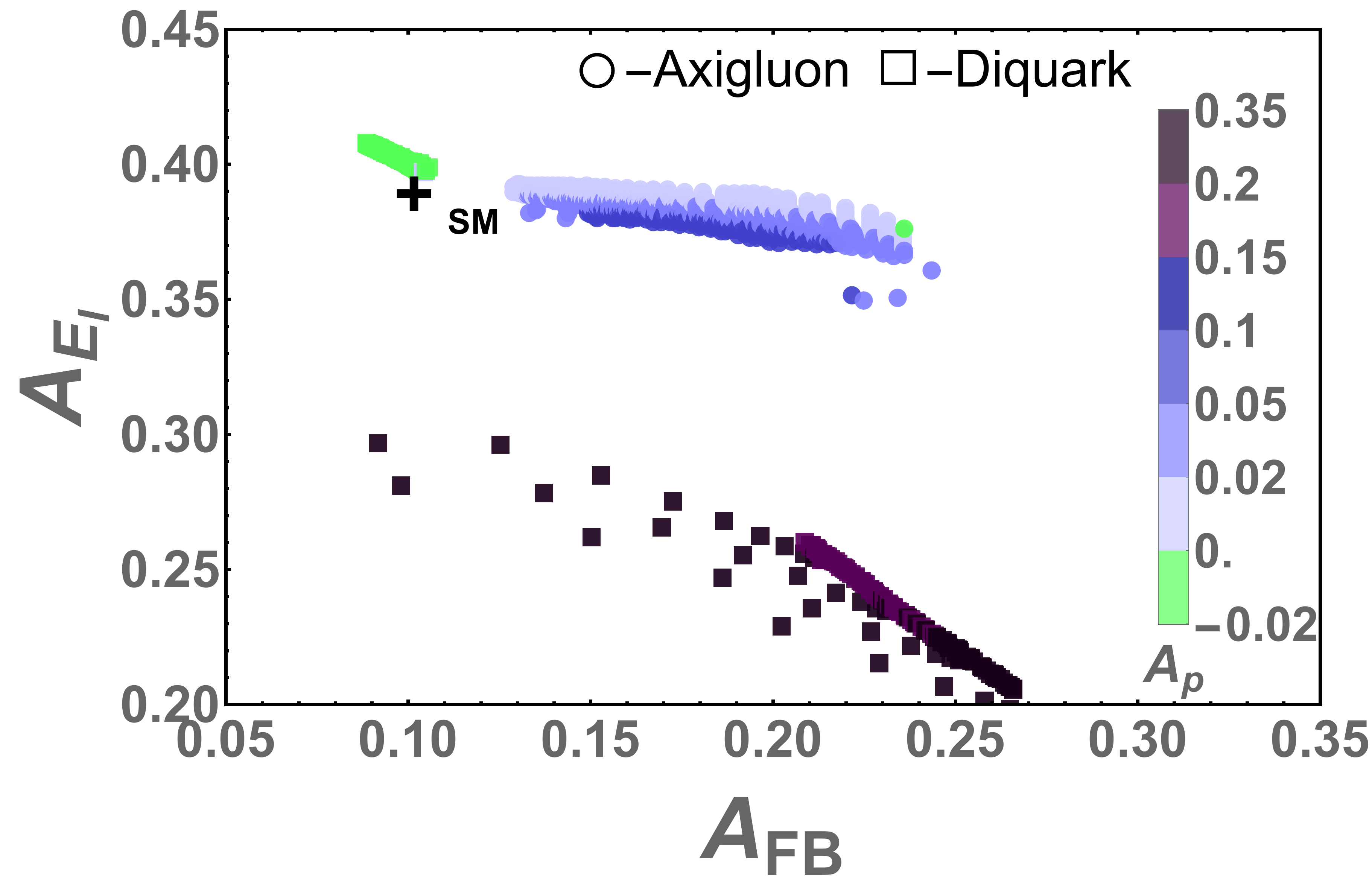}

}

\caption{Correlations between top AFB and the lepton energy and azimuthal asymmetries at the Tevatron $\sqrt{s}=1.96$\textrm{
TeV}}
\end{figure}

The correlations of the lepton-level asymmetries with the top AFB at the Tevatron are shown in figures \ref{fig:Lepton-polar-asymmetry-tev},
\ref{fig:Top-Lepton-polar-asymmetry-tev}, \ref{fig:Lepton-Azimuthal-asymmetry-tevatron}
and \ref{fig:Lepton-energy-asymmetry-tev}. For the case of axigluon as its mass is increased, the polarization rises until $m_{A}\sim1650$ GeV and
then drops again for even larger masses. The diquark model predicts negative
polarization for a significant portion of parameter space, turning
positive only for large couplings. The large mass region for the diquark
also favours large (negative) values of azimuthal asymmetry, smaller
lepton polar asymmetry and larger lepton energy asymmetry. Lepton
polar asymmetry correlation with top AFB shows large overlap between
the two models. The observed value for the lab-frame lepton polar
asymmetry in tables \ref{tab:CDF-asymmetries},\ref{tab:D0-lepton-asymmetries}
points towards a positive polarization between $A_{\theta_{l}}=0.2\mbox{ to }1.1$.
In this region (see figure \ref{fig:Lepton-polar-asymmetry-tev}),
a large positive value for polarization is favoured for the diquark
model and a large contribution to top AFB. The axigluon model is compatible
with both the observed value of $A_{\theta_{l}}$ and a small contribution
towards longitudinal polarization for a significant part of its parameter
space. Figure \ref{fig:Top-Lepton-polar-asymmetry-tev} shows the
asymmetry in the lepton polar angle with respect the top direction,
$A_{\theta_{tl}}$, which is equal to twice the top polarization,
eqn (\ref{eq:tl-polar-asymmetry and pol}), when calculated in top-quark rest
frame. It receives contribution from bSM physics via the boost of
the parent top quark. In the lab frame, large deviations of
$A_{\theta_{tl}}$ from the SM value correlate with large contribution
to the top-quark polarization from the bSM.
The values of asymmetries grow closer to the corresponding SM values
with increase in mass and reduction in the bSM coupling strength.
The azimuthal asymmetry and lepton polar asymmetry with respect to
the top-momentum direction ($A_{\theta_{tl}}$) and lepton energy
asymmetry in figures \ref{fig:Top-Lepton-polar-asymmetry-tev}, \ref{fig:Lepton-Azimuthal-asymmetry-tevatron}
and \ref{fig:Lepton-energy-asymmetry-tev} discriminate well between
the $s$-channel and $u$-channel exchange models though the parameter
spaces within the model are clumped together. When combined with polarization, all correlations
enhance their discriminating power especially to distinguish between
$s$-channel and $u$-channel models as they predict opposite signs
of polarization for a large portion of parameter space.

\subsubsection{Asymmetry correlations for the LHC }

\begin{figure}[t]
\subfloat[\label{fig:Lepton-Azimuthal-asymmetry-lhc}Lepton Azimuthal asymmetry
about $\phi_{0}=40$ degrees for the LHC $\sqrt{s}=7$ TeV]{\includegraphics[scale=0.21]{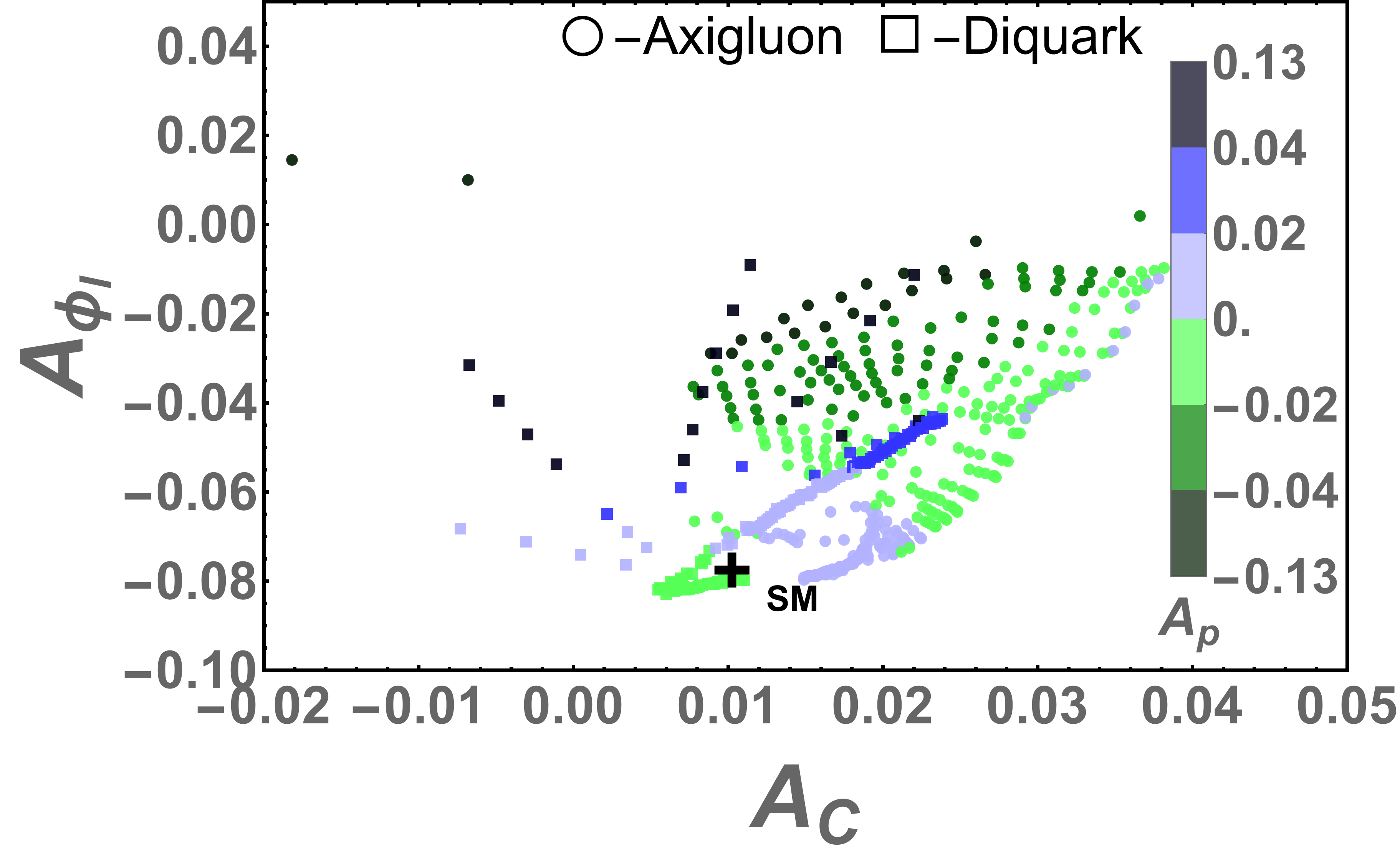}

}\hfill{}\subfloat[\label{fig:Lepton-polar-asymmetry-lhc}Lepton polar asymmetry(w.r.t
top quark direction) for the LHC $\sqrt{s}=7$ TeV]{\includegraphics[scale=0.2]{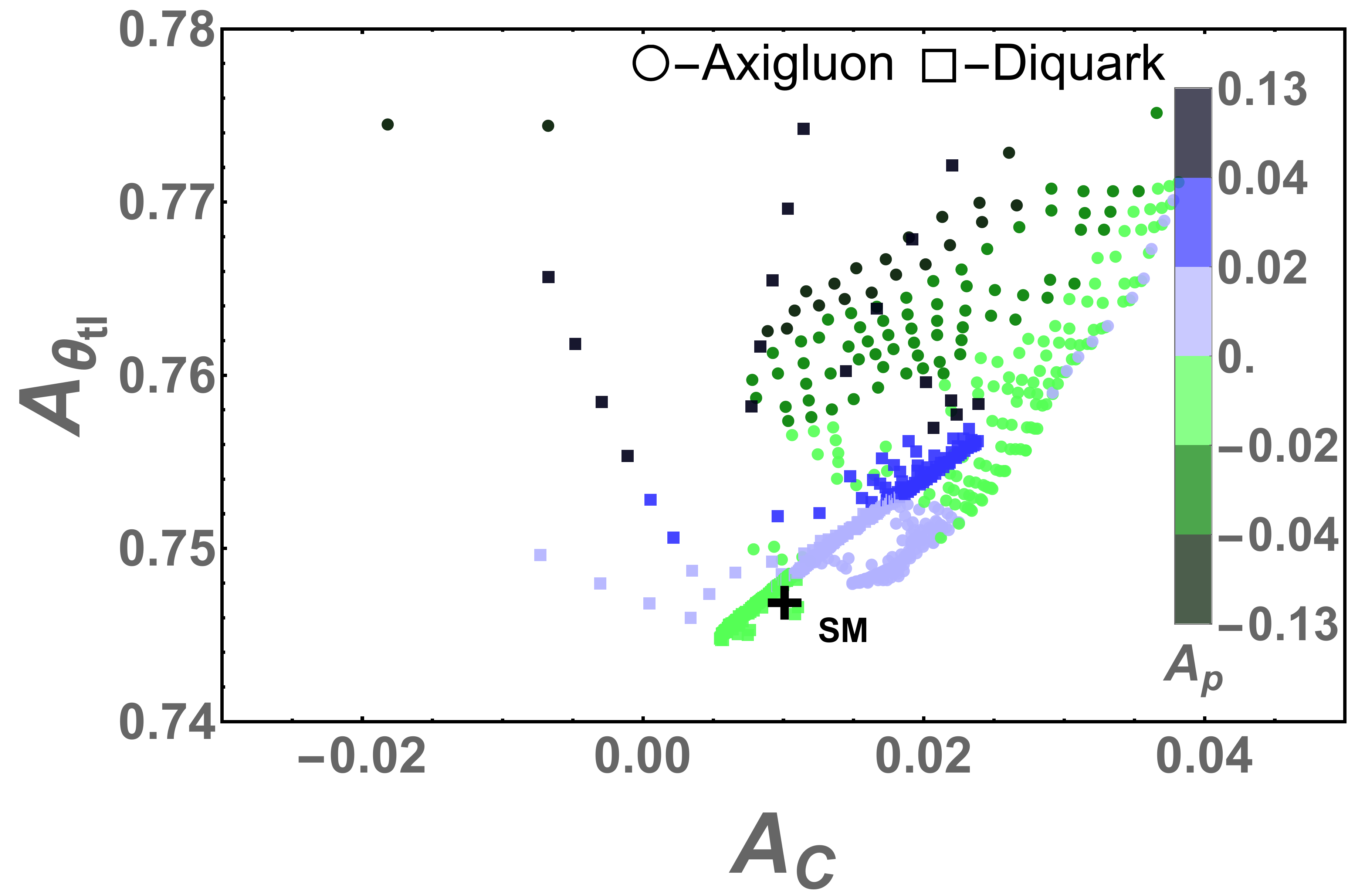}

}

\subfloat[\label{fig:Lepton-Energy-asymmetry-lhc}Lepton Energy asymmetry about
$E_{0}=80$ GeV for the LHC $\sqrt{s}=7$ TeV]{\includegraphics[scale=0.21]{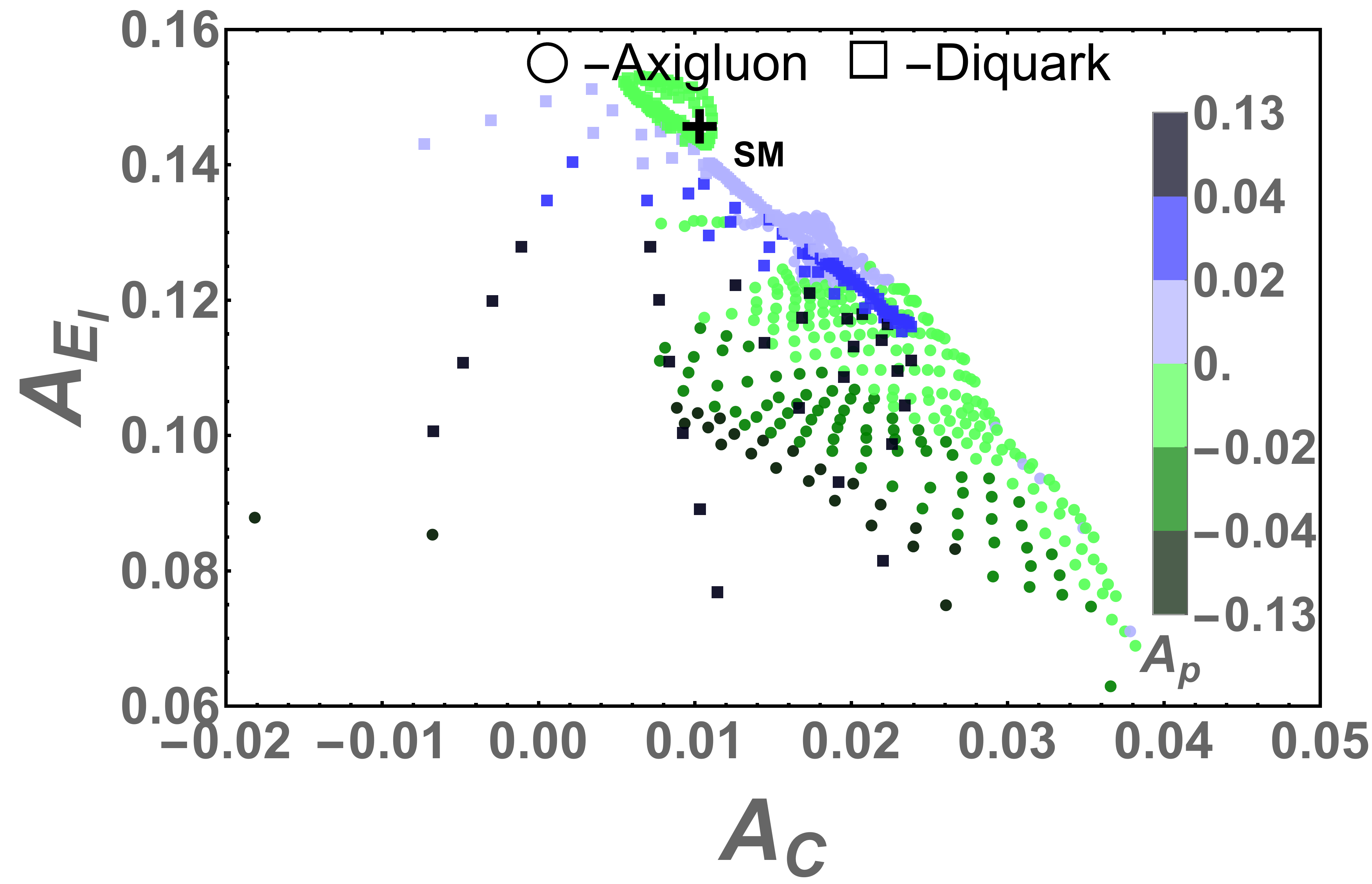}

}

\caption{Correlations between lepton and top kinematic asymmetries at the LHC-7
TeV.}
\end{figure}

\begin{figure}[t]
\subfloat[\label{fig:Lepton-Azimuthal-asymmetry-lhc13}Lepton Azimuthal asymmetry
about $\phi_{0}=40$ degrees for the LHC $\sqrt{s}=13$ TeV]{\includegraphics[scale=0.21]{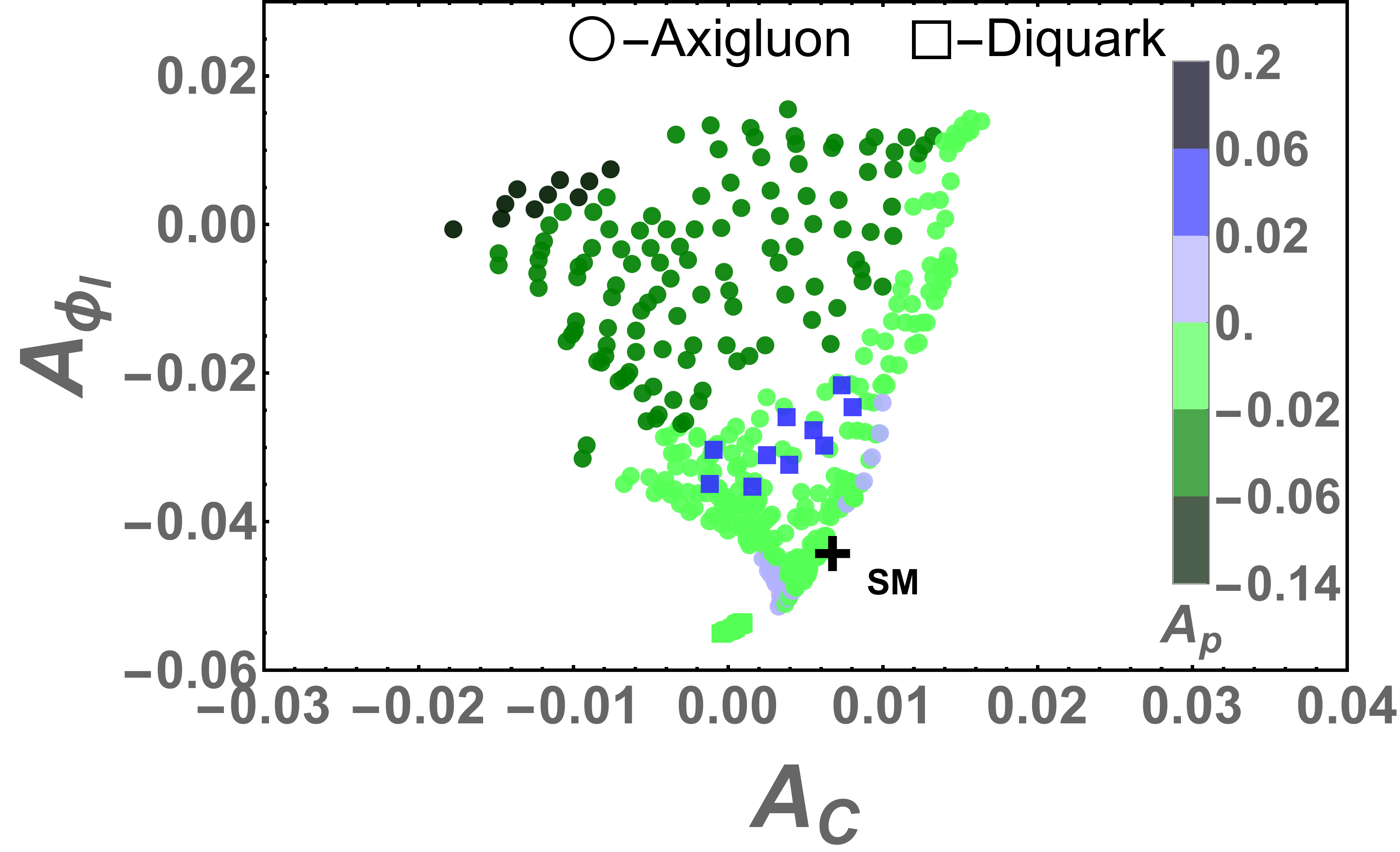}

}\hfill{}\subfloat[\label{fig:Lepton-polar-asymmetry-lhc13}Lepton polar asymmetry(w.r.t
top quark direction) for the LHC $\sqrt{s}=13$ TeV]{\includegraphics[scale=0.21]{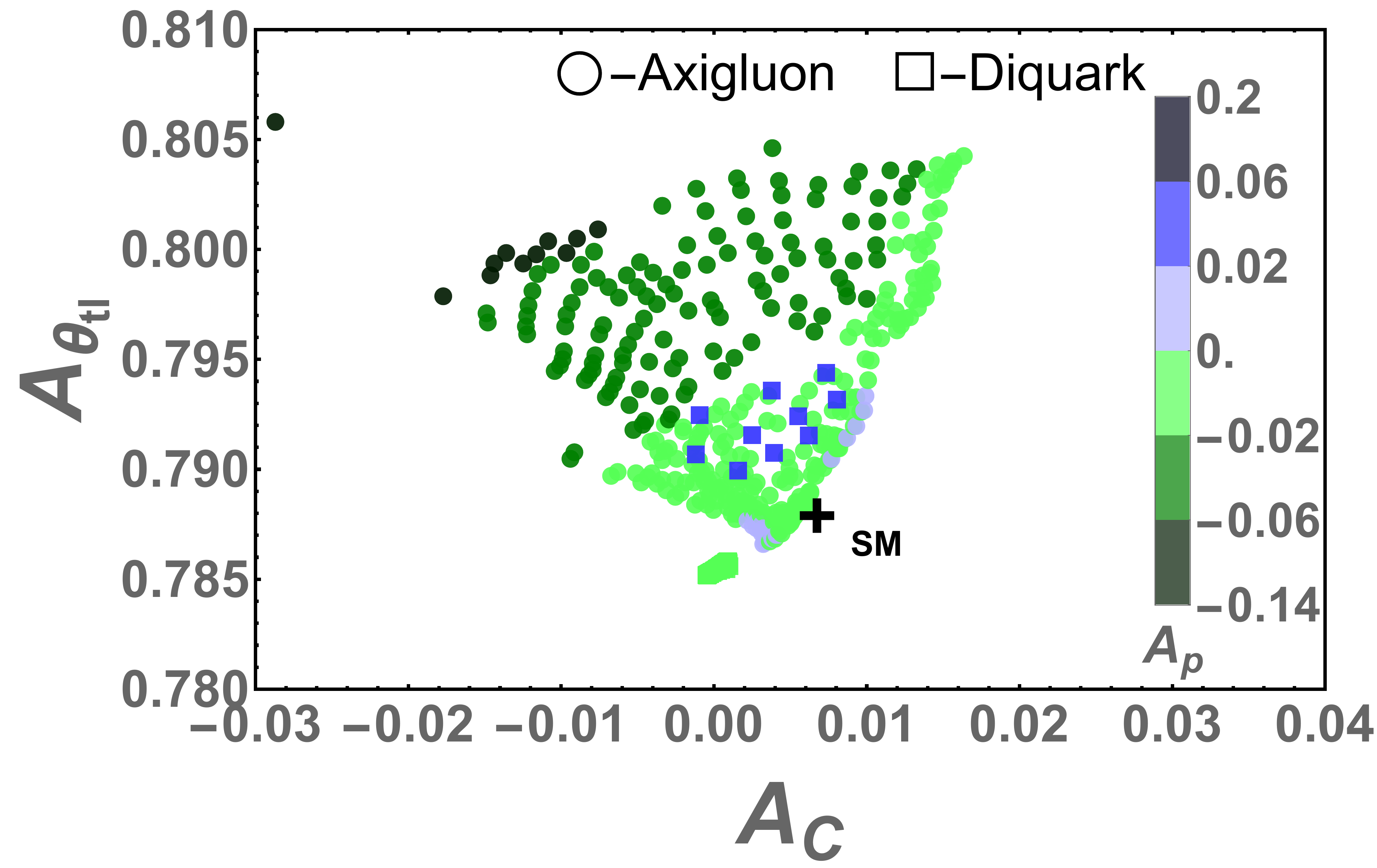}

}

\subfloat[\label{fig:Lepton-Energy-asymmetry-lhc13}Lepton Energy asymmetry
about $E_{0}=80$ GeV for the LHC $\sqrt{s}=13$ TeV]{\includegraphics[scale=0.21]{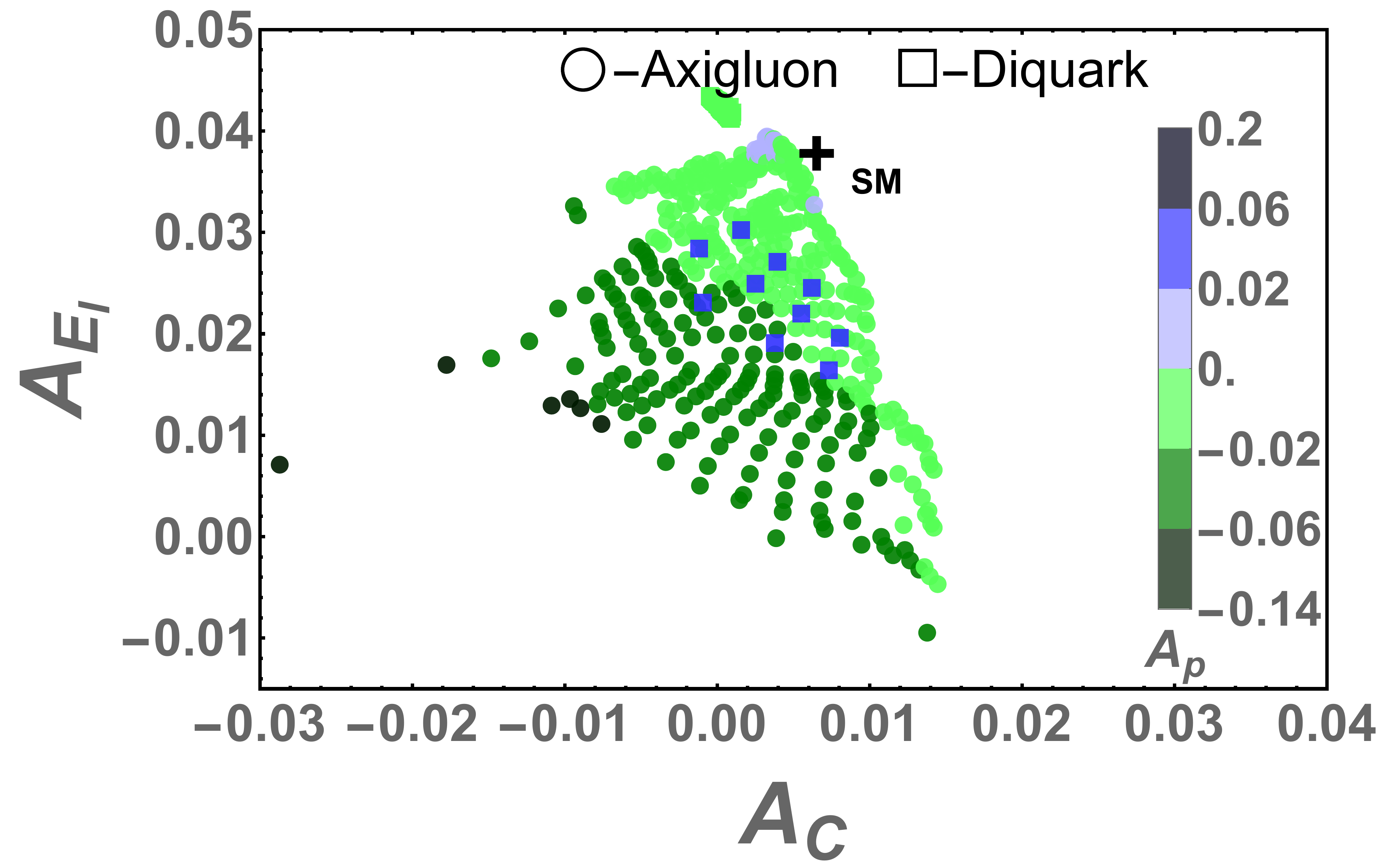}

}

\caption{Correlations between lepton and top kinematic asymmetries at the LHC-13
TeV.}
\end{figure}

The lepton-level asymmetry correlations with $t\overline{t}$ charge
asymmetry are shown in figures \ref{fig:Lepton-Azimuthal-asymmetry-lhc}-\ref{fig:Lepton-Energy-asymmetry-lhc}
for the LHC $7$ TeV run and figures \ref{fig:Lepton-Azimuthal-asymmetry-lhc13}-\ref{fig:Lepton-Energy-asymmetry-lhc13}
for the LHC $13$ TeV. The plots are made for the region of the model
parameter space constrained in sections \ref{sec:Flavor-Non-Universal-axigluon}
and \ref{sec:U-channel-Scalar-Model}. For the $\sqrt{s}=7$ TeV
calculation, we use $m_{t}=172.5$ GeV and factorization scale $Q=2m_{t}$
and $\alpha_{s}=0.108$ to remain consistent with the ATLAS and CMS
reconstruction of $A_{C}$. For $\sqrt{s}=13$ TeV, , the mass of
top quark is chosen at the updated central value $m_{t}=173.2GeV$
and $Q=2m_{t}$ with $\alpha_{s}=0.108$ and CTEQ6l pdf.

A large portion of the parameter space predicts a negative polarization
at 7 TeV LHC for the axigluon model. The diquark model predicts a
small negative polarization for small couplings with quarks. When
heavier diquark models are considered, larger couplings are allowed
leading to a large positive contribution to the polarization. Observed
values of polarization from CMS and ATLAS are compatible with $-0.03<A_{p}<0.07$,
which covers a large region of parameter space for both axigluon and
diquark models. As in the case of the Tevatron, polarization is an
important discriminant between models for the LHC as well, especially
when combined with decay-lepton asymmetries. It is able to distinguish
overlapping parameter space regions between the two models. This is
more true when the couplings are small and the bSM effects are more
difficult to detect as the $s$-channel and $u$-channel exchanges
predict small polarization, but with opposite signs in this region.
The energy asymmetry becomes smaller for the LHC at 13 TeV due to
the effect of the overall boost. The values of azimuthal and polar
asymmetries do not change significantly for higher energy and so remain
good observables for the study of top quark dynamics.

\section{Asymmetry correlations and top transverse polarization\label{sec:Off-Diagonal-Density-Matrix}}

As remarked earlier, keeping full spin correlations between the production
and decay of the top quark in a coherent manner requires the spin
density matrix formalism. In this formalism, the top polarization,
which played a significant role in the above analysis, corresponds
to the difference in the diagonal elements of the density
matrix, as seen from eqn (\ref{eq:polarization}). The off-diagonal
elements of the density matrix can also be significant in practice,
and they would contribute to the transverse polarization of the top
quark, corresponding to a spin quantization axis transverse to the
momentum. In SM, these terms arise at loop level and have been studied in the literature along-with transverse polarization and observables have been suggested to measure their contribution \cite{Bernreuther:1995cx}. We examine in this section what role these off-diagonal
matrix elements and transverse top polarization play in the two models
considered in this study.

Following the formalism developed in \cite{Godbole:2006ldistandtpol},
the spin density matrix integrated over a suitable final-state phase
space can be written as $\sigma(\lambda,\lambda')=\sigma_{tot}\mathcal{P}_{t}(\lambda,\lambda')$,
where $\sigma_{tot}$ represents the unpolarized cross section. The
matrix $\mathcal{P}_{t}(\lambda,\lambda')$ can be written as 
\begin{equation}
\mathcal{P}_{t}(\lambda,\lambda')=\left(\begin{array}{cc}
1+\eta_{3} & \eta_{1}-i\eta_{2}\\
\eta_{1}+i\eta_{2} & 1-\eta_{3}
\end{array}\right).
\end{equation}
Here $\eta_{3}$ is the longitudinal polarization, $\eta_{1}$ and
$\eta_{2}$ are polarizations along two transverse directions. The expressions for the $\eta_{i}$ in terms of the top-quark density matrix $\sigma\left(\lambda_{t},\lambda_{t}^{'}\right)$ can be written as,
\begin{eqnarray}
\eta_{3} & = & \frac{\left(\sigma\left(++\right)-\sigma\left(--\right)\right)}{\sigma_{tot}}\\
\eta_{1} & = & \frac{\left(\sigma\left(+-\right)+\sigma\left(-+\right)\right)}{\sigma_{tot}}\\
i\eta_{2} & = & \frac{\left(\sigma\left(+-\right)-\sigma\left(-+\right)\right)}{\sigma_{tot}}
\end{eqnarray}
Splitting
the top density matrix as shown in eqn (\ref{eq:total_amp sq}) under
the narrow-width approximation, the helicity-dependent decay density
matrix in the rest frame of top quark separates into a simple functions
of the decay angle: 
\begin{equation}
d\Gamma(\lambda,\lambda')=c\times A(\lambda,\lambda')d\Omega_{l}
\end{equation}
where 
\begin{equation}
A=\begin{array}{c|cc}
\lambda\downarrow,\lambda'\to & + & -\\
\hline + & 1+\cos\left(\theta_{l}\right) & \sin\left(\theta_{l}\right)e^{i\phi_{l}}\\
- & \sin\left(\theta_{l}\right)e^{-i\phi_{l}} & 1-\cos\left(\theta_{l}\right).
\end{array}
\end{equation}
$\Omega_{l}$ is the solid angle in which the lepton is emitted and
$c$ is the integrated contribution of the rest of the decay kinematic
variables. The resulting lepton angular distribution in the lab frame
is ,
\begin{equation}
\frac{d\sigma}{d\cos\left(\theta_{l}\right)d\phi_{l}}=c\sigma_{tot}\left(1+\eta_{3}\cos\left(\theta_{l}\right)+\eta_{1}\sin\left(\theta_{l}\right)\cos\left(\phi_{l}\right)+\eta_{2}\sin\left(\theta_{l}\right)\sin\left(\phi_{l}\right)\right)\label{angdist_eta_i}
\end{equation}
The off-diagonal elements in the top-quark production density matrix
do not contribute to the total cross-section due to an overall factor
of $\sin\left(\theta_{l}\right)$ which integrates to 0. They do contribute
instead to the kinematic distributions of the decay particle, although this effect
is quite small for most observables.

In this study, we find that the lepton polar angle asymmetry defined
in the lab frame is sensitive to the off-diagonal terms in the top
quark density matrix eqn (\ref{eq:top-density-matrix}). The transverse
polarization originating from these off-diagonal terms contains further
information about the dynamics of top-quark interaction. This relation
has been pointed out before in the context of a wide-width colour
octet bSM particle \cite{Aguilar-Saavedra:2014yea,Baumgart:2013yra}.

\begin{figure}[t]
\subfloat[axigluon]{\includegraphics[scale=0.15]{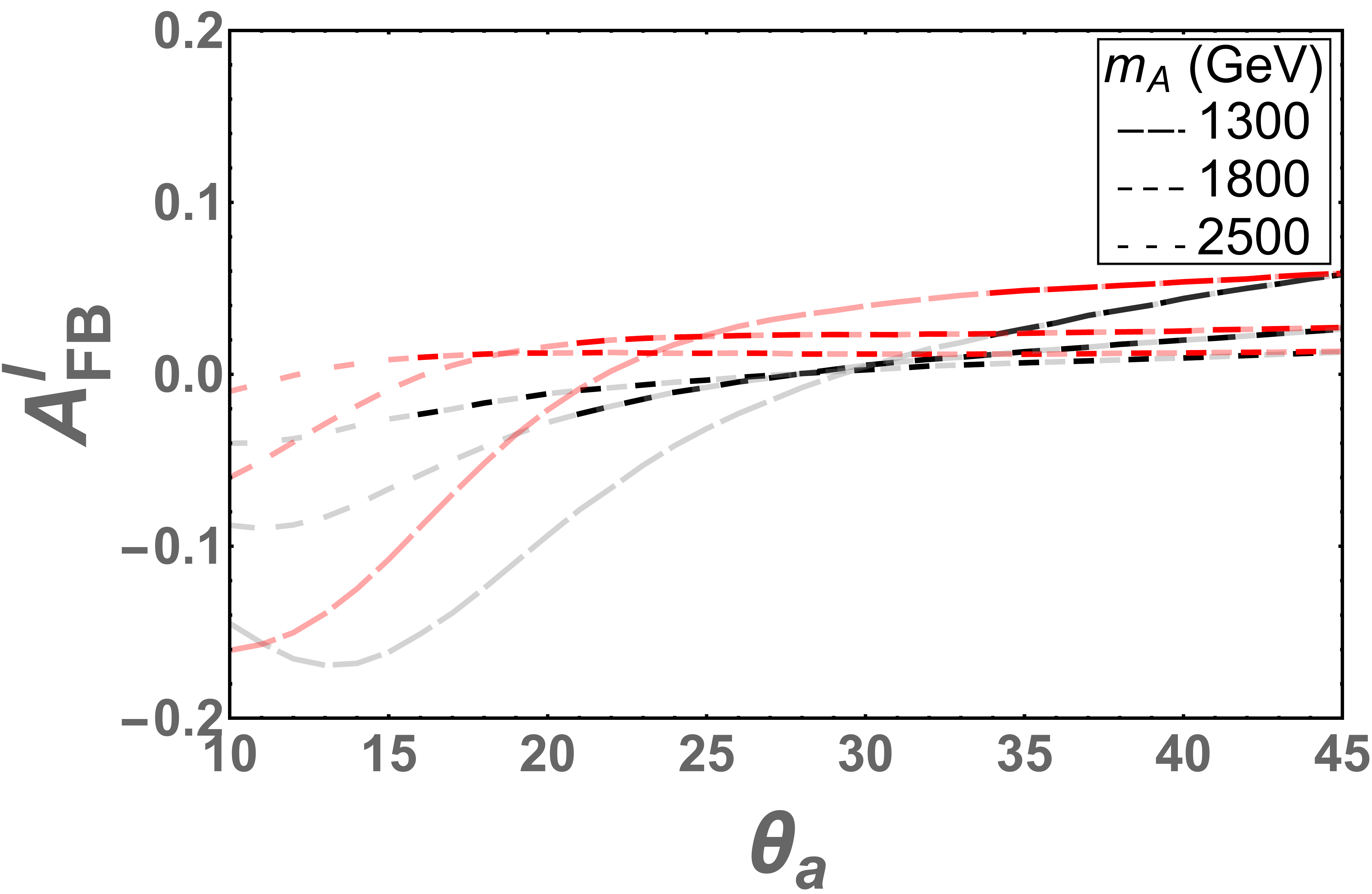}

}\hfill{}\subfloat[diquark]{\includegraphics[scale=0.15]{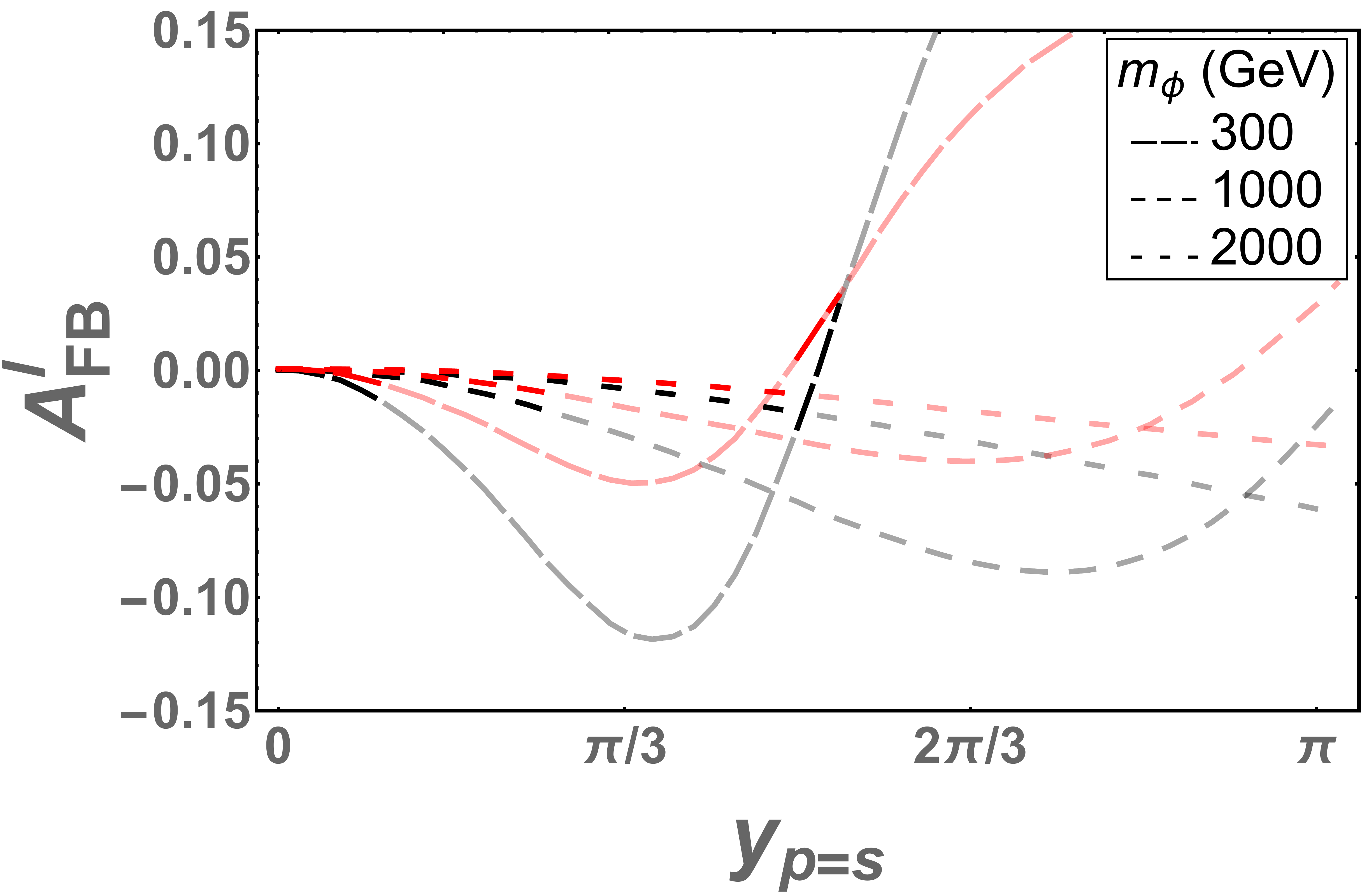}

}

\caption{\label{fig:Lepton-asymmetry-Transverse-pol}Contribution to Lepton
asymmetry from the off-diagonal terms of top density matrix calculated
in the lab frame for the Tevatron $\sqrt{s}=1.96$ TeV for the axigluon and diquark models. The
red lines represent the asymmetry for diagonal density matrix and
the black line represents the distribution for the case of total density
matrix. The darker lines represent allowed regions of parameter space.}
\end{figure}

In figure \ref{fig:Lepton-asymmetry-Transverse-pol} we study the contribution
of off-diagonal terms to the lepton distributions and present the
distributions for a few sample masses of bSM particles. 

It can be seen that the contribution of the off-diagonal density matrix
elements can be significant, and is particularly important for the
diquark model. These can in turn lead to significant transverse polarization of the top for appropriate range of parameters which could be measured experimentally.

\section{Conclusions\label{sec:Conclusions}}

The forward backward asymmetry of the top quark in top-pair production process at the Tevatron collider was, for a long time, anomalously large and a persistent effect observed independently by both D0 and CDF detectors. It has  been demonstrated only relatively recently that NNLO contributions give rise to an $A_{FB}^t$ of the right order of magnitude and seems to be in agreement with the values measured experimentally. Previously, many bSM models had been proposed with parity breaking interactions to explain the observed $A_{FB}^t$. Many of these models predicted a charge asymmetry at the LHC. Since LHC has a gluon dominated initial state as opposed to the Tevatron where $q\bar{q}$ was the primary initial state, the asymmetries predicted for LHC coming from the bSM couplings to the quarks get diluted. The data gives values for $A_C$ consistent with the SM and so far there has been no evidence for the new particles predicted in the different bSM models. Under these circumstances there is a need to construct measures which can distinguish between different sources of the $A_{FB}^t$: either SM or bSM.

One such measure is provided by polarization of the top quark which has a non-zero value in the presence of a parity breaking interaction. Within SM, top polarization is close to 0. Observables that correlate with top polarization can be used to distinguish between various SM and bSM contributions. In continuation to a previous work where correlations between polarization and forward backward asymmetry were used to constraint bSM \cite{DC:2010_W'z'Axdiq}, we have introduced the correlations between lepton polar, azimuthal and energy asymmetries and top charge asymmetry and showed how they can be used together with top longitudinal polarization to distinguish between SM and bSM.

In the reference \cite{Krohn:2011tw} the authors have constructed dilepton central charge and azimuthal asymmetries and studied it along with top quark polarization, forward backward asymmetry and $t\bar{t}$ spin correlations for benchmark models of G' and W'. Subsequently, in reference \cite{Berger:2012_assym_corr} the authors have shown that the lepton polar asymmetry and top forward backward asymmetry and lepton charge asymmetry vs top charge asymmetry correlations can be useful in the study of W' and G' models. Our work adds multiple new observables to the analysis of new physics in $t\bar{t}$ pair production which include single-lepton azimuthal angle, energy and polar angle (wrt top quark) in the lab frame which show signatures from parity breaking in top interactions and help isolate and constrain the interactions of bSM particles.

We demonstrate the efficacy of the correlations between forward backward asymmetry and the lepton asymmetries at Tevatron and charge asymmetry and lepton angluar,energy asymmetries at LHC, by utilizing a representative s-channel model, axigluon and an u-channel model, diquark. Constraints on these models are obtained based on measured values of $t\bar{t}$ cross-section at Tevatron and LHC 7TeV, $A_{FB}^t$ and $A_{C}^t$ and resonance searches in dijet, four jet cross sections. The parameter space of axigluon allowed within 2-$\sigma$ of the measured values of the stated observables includes a lower bound on the mass of axigluon at 1.5 TeV with a corresponding coupling $\theta_A>27^\circ$. The allowed mass of the diquark is bounded from below by 300 GeV and masses above are allowed with the coupling of the model bounded from above by a value of 0.2 for smaller masses which rises to $y_s<2$ corresponding to $m_\phi=3$ TeV. Another sliver of parameter space is allowed for larger couplings of the diquark due to destructive interference effects.

For the first time we have presented the complete density matrix of the top quark, including the off-diagonal elements for top quark pair production process, in the axigluon and diquark models to aid further studies. We use these to show that the lepton polar asymmetry in the lab frame shows a correlation with the transverse polarization of the top quark for axigluon model and even more significantly for the diquark model. The lepton asymmetry usually considered in studies of top polarization is calculated from lepton polar angle with respect to the top quark and it does not show this correlation with transverse polarization. The correlation of transverse polarization with lepton azimuthal or energy asymmetry is also very small. 

Finally, we extend our analysis to 13 TeV  LHC  where, even though the values of the asymmetries get diluted, the correlations between accurate measurements of charge asymmetry and lepton asymmetries still separate out bSM and the SM in the 2 dimensional space. Taken together, these correlations can indeed be used to improve significance of the constraints on bSM from LHC data even in the initial stages of low luminosity.

\acknowledgments{
The authors are pleased to acknowledge conversations with Ritesh K. Singh, Pratishruti Saha and Arunprasath V. RMG wishes to acknowledge support from the Department
of Science and Technology, India, under Grant No. SR/S2/JCB-64/2007.
SDR acknowledges support from the Department of Science and Technology,
India, under the Grant No.
SR/SB/JCB-42/2009. The research of GM was supported by CSIR, India via SPM Grant No. 07/079(0095)/2011-EMR-I.
}
\appendix
\addcontentsline{toc}{section}{Appendices}

\section{AC at $13$ TeV LHC\label{sec:Charge-Asymmetry-AtLHC13}}

Since the AC calculated at NLO for 13TeV LHC was unavailable at the
time of submission of this work, we note that the available charge
asymmetry values \cite{SM_LHC_Ac} form a smooth function of the beam
energy and fit them to a polynomial to find the AC as
a function of the beam energy. We obtain a fit to a polynomial presented
in eqn (\ref{eq:Ac-fit}) with goodness of fit parameter $r^{2}=0.9995$.

\begin{equation}
Ac\left(\sqrt{s}\right)=3.12\times10^{-2}-4.37\times10^{-3}\sqrt{s}+2.6269\times10^{-4}s-5.683\times10^{-6}s^{\frac{3}{2}}\label{eq:Ac-fit}
\end{equation}

This gives a value of Ac(13  TeV)=0.0063.

\section{$t\overline{t}$ Production density matrices}

\subsection{axigluon density matrices\label{sub:axigluon-Density-Matrices}}

With $C_{\theta}=\cos(\theta_{t})$, $S_{\theta}=\sin(\theta_{t})$,
$\beta=\sqrt{1-\frac{4m_{t}^{2}}{\hat{s}}}$ and $\beta_{A}=\sqrt{1-\frac{4m_{t}^{2}}{m_{A}^{2}}}$.

\begin{eqnarray}
\rho_{bSM}^{++} & = & \frac{1}{\Gamma_{A}{}^{2}m_{A}{}^{2}+\left(m_{A}{}^{2}-\hat{s}\right){}^{2}}\hat{s}^{2}\{\frac{1}{18}(g_{A}{}^{2}+g_{V}{}^{2})(6g_{\text{At}}g_{V}\beta+3g_{\text{At}}{}^{2}\beta^{2}-g_{V}{}^{2}\left(-4+\beta^{2}\right))\nonumber \\
 & & +\frac{4}{9}g_{A}g_{V}\left(g_{V}+g_{\text{A}}^{t}\beta\right){}^{2}C_{\theta}+\frac{1}{18}\left(g_{A}{}^{2}+g_{V}{}^{2}\right)\beta\left(2g_{\text{A}}^{t}g_{V}+g_{\text{A}}^{t}{}^{2}\beta+g_{V}{}^{2}\beta\right)C_{2\theta}\}\\
\rho_{bSM}^{+-} & = & \frac{1}{\Gamma_{A}{}^{2}m_{A}{}^{2}+\left(m_{A}{}^{2}-\hat{s}\right){}^{2}}\hat{s}^{2}\left(-\frac{8g_{A}g_{V}{}^{3}m_{t}S_{\theta}}{9\sqrt{\hat{s}}}-\frac{4g_{\text{A}}^{t}g_{V}\left(g_{A}{}^{2}+g_{V}{}^{2}\right)m_{t}\beta C_{\theta}S_{\theta}}{9\sqrt{\hat{s}}}\right)\\
\rho_{bSM}^{-+} & = & (\rho_{bSM}^{+-})^{\star}\\
\rho_{bSM}^{--} & = & \frac{1}{\Gamma_{A}{}^{2}m_{A}{}^{2}+\left(m_{A}{}^{2}-\hat{s}\right){}^{2}}\hat{s}^{2}\{\frac{1}{18}(g_{A}{}^{2}+g_{V}{}^{2})
(-6g_{\text{A}}^{t}g_{V}\beta+3g_{\text{A}}^{t}{}^{2}\beta^{2}-g_{V}{}^{2}\left(-4+\beta^{2}\right))\nonumber \\
 & & -\frac{4}{9}g_{A}g_{V}\left(g_{V}-g_{\text{A}}\beta\right){}^{2}C_{\theta}+\frac{1}{18}\left(g_{A}{}^{2}+g_{V}{}^{2}\right)\beta\left(-2g_{\text{A}}^{t}g_{V}+g_{\text{A}}^{t}{}^{2}\beta+g_{V}{}^{2}\beta\right)C_{2\theta}\}\nonumber\\
\end{eqnarray}
\begin{eqnarray}
\rho_{Interference}^{++} & = & \frac{g_{s}^{2}}{9\left(\Gamma_{A}{}^{2}m_{A}{}^{2}+\left(m_{A}{}^{2}-\hat{s}\right){}^{2}\right)}\hat{s}\left(-m_{A}{}^{2}+\hat{s}\right)\nonumber \\
 & & \times\left(4g_{A}\left(g_{V}+g_{\text{A}}^{t}\beta\right)C_{\theta}+g_{V}\left(4g_{v}+3g_{\text{A}}^{t}\beta-g_{V}\beta^{2}+\beta\left(g_{\text{A}}^{t}+g_{V}\beta\right)C_{2\theta}\right)\right)\nonumber\\
   \\
\rho_{Interference}^{+-} & = & \frac{g_{s}^{2}}{9\left(\Gamma_{A}{}^{2}m_{A}{}^{2}+\left(m_{A}{}^{2}-\hat{s}\right){}^{2}\right)}4m_{t}\sqrt{\hat{s}}S_{\theta}\nonumber \\
 & & \times\left(g_{A}\left(2g_{v}\left(m_{A}{}^{2}-\hat{s}\right)+ig_{\text{At}}\Gamma_{A}m_{A}\beta\right)+g_{\text{A}}^{t}g_{V}\left(m_{A}{}^{2}-\hat{s}\right)\beta C_{\theta}\right)\\
\rho_{Interference}^{-+} & = & (\rho_{Interference}^{+-})^{\star}\\
\rho_{Interference}^{--} & = & \frac{g_{s}^{2}}{9\left(\Gamma_{A}{}^{2}m_{A}{}^{2}+\left(m_{A}{}^{2}-\hat{s}\right){}^{2}\right)}\left(m_{A}{}^{2}-\hat{s}\right)\hat{s}\nonumber \\
 & & \times\left(4g_{A}\left(g_{V}-g_{\text{A}}^{t}\beta\right)C_{\theta}+g_{V}\left(3g_{\text{A}}^{t}\beta+g_{V}\left(\beta^{2}-4\right)+\beta\left(g_{\text{A}}^{t}-g_{V}\beta\right)C_{2\theta}\right)\right)\nonumber\\
\end{eqnarray}

To present the dependence on top boost and polar angle clearly the
amplitude square is written in terms of the polar angle $\theta$
in $t\overline{t}$ center of momentum frame. The off-diagonal terms in
the gluon initiated process are zero and the diagonal terms in gluon
initiated process are not dependent on the top quark polarization
therefore we have omitted these here and they can be found in many
references including \cite{DC:2010_W'z'Axdiq}. Decay width of axigluon
at tree level is given by,
\begin{eqnarray}
\Gamma_{A} & = & \frac{4\pi}{6m_{A}}\{g_{A}^{2}\left(m_{A}^{2}\left(\beta_{A}+5\right)-4m_{t}^{2}\beta_{A}\right)+ g_{V}^{2}\left(m_{A}^{2}\left(\beta_{A}+5\right)+2m_{t}^{2}\beta_{A}\right)\}\label{eq: Flavor non-universal axigluon decay width}
\end{eqnarray}

\subsection{diquark density matrices\label{sub:diquark-Density-Matrices}}

The top-quark spin density matrix for the $u$-channel exchange is
given below in $t\overline{t}$ center of momentum frame. The notation
and SM only contributions remain the same as for the case of the $s$-channel
model.

\begin{eqnarray}
\rho_{bSM}^{++} & = & \frac{2\hat{s}\left(y_{P}^{2}+y_{S}^{2}\right)}{48\left(\beta\hat{s}C_{\theta}-2m_{t}^{2}+2m_{\phi}^{2}+\hat{s}\right){}^{2}}\times\nonumber \\
 & & \{\hat{s}\left(2y_{P}y_{S}\left(\beta+\beta C_{\theta}^{2}+2C_{\theta}\right)+(y_{P}^{2}+y_{S}^{2})\left(2\beta C_{\theta}+C_{\theta}^{2}+1\right)\right)\nonumber \\
 & & -4C_{\theta}m_{t}^{2}\left(C_{\theta}\left(y_{P}^{2}+y_{S}^{2}\right)+2y_{P}y_{S}\right)\}\\
\rho_{bSM}^{+-} & = & -\frac{\hat{s}^{3/2}m_{t}y_{P}y_{S}\left(\beta C_{\theta}+1\right)\left(y_{P}^{2}+y_{S}^{2}\right)S_{\theta}}{6\left(\beta\hat{s}C_{\theta}-2m_{t}^{2}+2m_{\phi}^{2}+\hat{s}\right){}^{2}}\\
\rho_{bSM}^{-+} & = & (\rho_{bSM}^{+-})^{\star}\\
\rho_{bSM}^{--} & = & \frac{\hat{s}\left(y_{P}^{2}+y_{S}^{2}\right)}{24\left(\beta\hat{s}C_{\theta}-2m_{t}^{2}+2m_{\phi}^{2}+\hat{s}\right){}^{2}}\nonumber \\
 & & \{\hat{s}\left(-2y_{P}y_{S}\left(\beta+\beta C_{\theta}^{2}+2C_{\theta}\right)+(y_{P}^{2}+y_{S}^{2})\left(2\beta C_{\theta}+C_{\theta}^{2}+1\right)\right)\nonumber \\
 & & -4C_{\theta}m_{t}^{2}\left(C_{\theta}\left(y_{P}^{2}+y_{S}^{2}\right)-2y_{P}y_{S}\right)\}
\end{eqnarray}
\begin{eqnarray}
\rho_{Interference}^{++} & = & \frac{g_{s}^{2}}{18\left(\beta\hat{s}C_{\theta}-2m_{t}^{2}+2m_{\phi}^{2}+\hat{s}\right)}\times\{4\left(C_{\theta}^{2}-1\right)m_{t}^{2}\left(y_{P}^{2}+y_{S}^{2}\right)\nonumber \\
 & & -\hat{s}\left(2y_{P}y_{S}\left(\beta+\beta C_{\theta}^{2}+2C_{\theta}\right)+(y_{P}^{2}+y_{S}^{2})\left(2\beta C_{\theta}+C_{\theta}^{2}+1\right)\right)\}\\
\rho_{Interference}^{+-} & = & \frac{g_{s}^{2}2\sqrt{\hat{s}}m_{t}y_{P}S_{\theta}y_{S}\left(\beta C_{\theta}+2\right)}{9\left(\beta\hat{s}C_{\theta}-2m_{t}^{2}+2m_{\phi}^{2}+\hat{s}\right)}\nonumber \\
\rho_{Interference}^{-+} & = & (\rho_{Interference}^{+-})^{\star}\\
\rho_{Interference}^{--} & = & \frac{g_{s}^{2}}{18\left(\beta\hat{s}C_{\theta}-2m_{t}^{2}+2m_{\phi}^{2}+\hat{s}\right)}\times\{4\left(C_{\theta}^{2}-1\right)m_{t}^{2}\left(y_{P}^{2}+y_{S}^{2}\right)\nonumber \\
 & & -\hat{s}\left(-2y_{P}y_{S}\left(\beta+\beta C_{\theta}^{2}+2C_{\theta}\right)+(y_{P}^{2}+y_{S}^{2})\left(2\beta C_{\theta}+C_{\theta}^{2}+1\right)\right)\}\nonumber\\
\end{eqnarray}

\bibliographystyle{apsrev4-1}

\begin{thebibliography}{10}

\bibitem{Lees:2012ju}
{\bfseries BaBar} Collaboration, J.~Lees {\em et~al.}, ``{Evidence of $B^+ \to
  \tau^+\nu$ decays with hadronic B tags},''
  \href{http://dx.doi.org/10.1103/PhysRevD.88.031102}{{\em Phys.Rev.}
  {\bfseries D88} no.~3, (2013) 031102},
\href{http://arxiv.org/abs/1207.0698}{{\ttfamily arXiv:1207.0698 [hep-ex]}}.

\bibitem{CMS-PAS-HIG-14-005}
{\bfseries CMS} Collaboration,
``{Search for Lepton Flavour Violating Decays of the Higgs Boson},''.
CMS-PAS-HIG-14-005.

\bibitem{Abazov:2014ysa}
{\bfseries D0} Collaboration, V.~M. Abazov {\em et~al.}, ``{Measurement of the
  Forward-Backward Asymmetry in the Production of $B^{\pm}$ Mesons in
  $p\bar{p}$ Collisions at $\sqrt{s}$ = 1.96 TeV},''
  \href{http://dx.doi.org/10.1103/PhysRevLett.114.051803}{{\em Phys. Rev.
  Lett.} {\bfseries 114} (2015) 051803},
\href{http://arxiv.org/abs/1411.3021}{{\ttfamily arXiv:1411.3021 [hep-ex]}}.

\bibitem{ALEPH:2005ab}
{\bfseries SLD Electroweak Group, DELPHI, ALEPH, SLD, SLD Heavy Flavour Group,
  OPAL, LEP Electroweak Working Group, L3} Collaboration, S.~Schael {\em
  et~al.}, ``{Precision electroweak measurements on the $Z$ resonance},''
  \href{http://dx.doi.org/10.1016/j.physrep.2005.12.006}{{\em Phys. Rept.}
  {\bfseries 427} (2006) 257--454},
\href{http://arxiv.org/abs/hep-ex/0509008}{{\ttfamily arXiv:hep-ex/0509008
  [hep-ex]}}.

\bibitem{Murphy:2015cha}
C.~W. Murphy, ``{Bottom-Quark Forward-Backward and Charge Asymmetries at Hadron
  Colliders},''
\href{http://arxiv.org/abs/1504.02493}{{\ttfamily arXiv:1504.02493 [hep-ph]}}.

\bibitem{d02008afb}
{\bfseries D0} Collaboration, V.~Abazov {\em et~al.}, ``First measurement of
  the forward-backward charge asymmetry in top quark pair production,''
  \href{http://dx.doi.org/10.1103/PhysRevLett.100.142002}{{\em
  {Phys.Rev.Lett.}} {\bfseries 100} (2008) 142002},
\href{http://arxiv.org/abs/0712.0851}{{\ttfamily arXiv:0712.0851}}.

\bibitem{cdf2008afb}
{\bfseries CDF} Collaboration, T.~Aaltonen {\em et~al.}, ``Forward-backward
  asymmetry in top quark production in $p\bar{p}$ collisions at $\sqrt{s}=1.96$
  tev,'' \href{http://dx.doi.org/10.1103/PhysRevLett.101.202001}{{\em
  {Phys.Rev.Lett.}} {\bfseries 101} (2008) 202001},
\href{http://arxiv.org/abs/0806.2472}{{\ttfamily arXiv:0806.2472}}.

\bibitem{Abazov:2014cca}
{\bfseries D0} Collaboration, V.~M. Abazov {\em et~al.}, ``{Measurement of the
  forward-backward asymmetry in top quark-antiquark production in ppbar
  collisions using the lepton+jets channel},''
  \href{http://dx.doi.org/10.1103/PhysRevD.90.072011}{{\em Phys.Rev.}
  {\bfseries D90} no.~7, (2014) 072011},
\href{http://arxiv.org/abs/1405.0421}{{\ttfamily arXiv:1405.0421 [hep-ex]}}.

\bibitem{Kidonakis:2015ona}
N.~Kidonakis, ``{Top quark forward-backward asymmetry at approximate
  $N^3LO$},'' \href{http://dx.doi.org/10.1103/PhysRevD.91.071502}{{\em
  Phys.Rev.} {\bfseries D91} no.~7, (2015) 071502},
\href{http://arxiv.org/abs/1501.01581}{{\ttfamily arXiv:1501.01581 [hep-ph]}}.

\bibitem{DC:2007AfbAx}
D.~Choudhury, R.~M. Godbole, R.~K. Singh, and K.~Wagh, ``Top production at the
  tevatron/lhc and nonstandard, strongly interacting spin one particles,''
  \href{http://dx.doi.org/10.1016/j.physletb.2007.09.057}{{\em {Phys.Lett.}}
  {\bfseries B657} (2007) 69--76},
\href{http://arxiv.org/abs/0705.1499}{{\ttfamily arXiv:0705.1499}}.

\bibitem{Antunano:2007da}
O.~Antunano, J.~H. Kuhn, and G.~Rodrigo, ``{Top quarks, axigluons and charge
  asymmetries at hadron colliders},''
  \href{http://dx.doi.org/10.1103/PhysRevD.77.014003}{{\em Phys.Rev.}
  {\bfseries D77} (2008) 014003},
\href{http://arxiv.org/abs/0709.1652}{{\ttfamily arXiv:0709.1652 [hep-ph]}}.

\bibitem{Ferrario:2009ax}
P.~Ferrario and G.~Rodrigo, ``Constraining heavy colored resonances from
  top-antitop quark events,''
  \href{http://dx.doi.org/10.1103/PhysRevD.80.051701}{{\em {Phys.Rev.}}
  {\bfseries D80} (2009) 051701},
\href{http://arxiv.org/abs/0906.5541}{{\ttfamily arXiv:0906.5541}}.

\bibitem{framptonshu2010}
P.~H. Frampton, J.~Shu, and K.~Wang, ``axigluon as possible explanation for
  $p\bar{p} \to t \bar{t}$ forward-backward asymmetry,''
  \href{http://dx.doi.org/10.1016/j.physletb.2009.12.043}{{\em {Phys.Lett.}}
  {\bfseries B683} (2010) 294--297},
\href{http://arxiv.org/abs/0911.2955}{{\ttfamily arXiv:0911.2955}}.

\bibitem{DC:2010_W'z'Axdiq}
D.~Choudhury, R.~M. Godbole, S.~D. Rindani, and P.~Saha, ``Top polarization,
  forward-backward asymmetry and new physics,''
  \href{http://dx.doi.org/10.1103/PhysRevD.84.014023}{{\em {Phys.Rev.}}
  {\bfseries D84} (2011) 014023},
\href{http://arxiv.org/abs/1012.4750}{{\ttfamily arXiv:1012.4750}}.

\bibitem{Dutta:2012ai}
S.~Dutta, A.~Goyal, and M.~Kumar, ``{Top quark physics in the vector
  color-octet model},''
  \href{http://dx.doi.org/10.1103/PhysRevD.87.094016}{{\em Phys.Rev.}
  {\bfseries D87} no.~9, (2013) 094016},
\href{http://arxiv.org/abs/1209.3636}{{\ttfamily arXiv:1209.3636 [hep-ph]}}.

\bibitem{Fajfer:2012si}
S.~Fajfer, J.~F. Kamenik, and B.~Melic, ``{Discerning New Physics in
  Top-Antitop Production using Top Spin Observables at Hadron Colliders},''
  \href{http://dx.doi.org/10.1007/JHEP08(2012)114}{{\em JHEP} {\bfseries 1208}
  (2012) 114},
\href{http://arxiv.org/abs/1205.0264}{{\ttfamily arXiv:1205.0264 [hep-ph]}}.

\bibitem{AguilarSaavedra:2011ug}
J.~A. Aguilar-Saavedra and M.~Perez-Victoria, ``{Simple models for the top
  asymmetry: Constraints and predictions},''
  \href{http://dx.doi.org/10.1007/JHEP09(2011)097}{{\em JHEP} {\bfseries 09}
  (2011) 097},
\href{http://arxiv.org/abs/1107.0841}{{\ttfamily arXiv:1107.0841 [hep-ph]}}.

\bibitem{Gresham:2012kv}
M.~Gresham, J.~Shelton, and K.~M. Zurek, ``{Open windows for a light axigluon
  explanation of the top forward-backward asymmetry},''
  \href{http://dx.doi.org/10.1007/JHEP03(2013)008}{{\em JHEP} {\bfseries 1303}
  (2013) 008},
\href{http://arxiv.org/abs/1212.1718}{{\ttfamily arXiv:1212.1718 [hep-ph]}}.

\bibitem{Jung:2009_Z'}
S.~Jung, H.~Murayama, A.~Pierce, and J.~D. Wells, ``Top quark forward-backward
  asymmetry from new t-channel physics,''
  \href{http://dx.doi.org/10.1103/PhysRevD.81.015004}{{\em {Phys.Rev.}}
  {\bfseries D81} (2010) 015004},
\href{http://arxiv.org/abs/0907.4112}{{\ttfamily arXiv:0907.4112}}.

\bibitem{Papaefstathiou:2011kd}
A.~Papaefstathiou and K.~Sakurai, ``{Determining the Helicity Structure of
  Third Generation Resonances},''
  \href{http://dx.doi.org/10.1007/JHEP06(2012)069}{{\em JHEP} {\bfseries 1206}
  (2012) 069},
\href{http://arxiv.org/abs/1112.3956}{{\ttfamily arXiv:1112.3956 [hep-ph]}}.

\bibitem{Shu_Tait_2009diquark}
J.~Shu, T.~M. Tait, and K.~Wang, ``Explorations of the top quark
  forward-backward asymmetry at the tevatron,''
  \href{http://dx.doi.org/10.1103/PhysRevD.81.034012}{{\em {Phys.Rev.}}
  {\bfseries D81} (2010) 034012},
\href{http://arxiv.org/abs/0911.3237}{{\ttfamily arXiv:0911.3237}}.

\bibitem{Barger:2011W'Z'}
V.~Barger, W.-Y. Keung, and C.-T. Yu, ``Tevatron asymmetry of tops in a w',z'
  model,'' \href{http://dx.doi.org/10.1016/j.physletb.2011.03.010}{{\em
  {Phys.Lett.}} {\bfseries B698} (2011) 243--250},
\href{http://arxiv.org/abs/1102.0279}{{\ttfamily arXiv:1102.0279}}.

\bibitem{Rajaraman:2011rw}
A.~Rajaraman, Z.~Surujon, and T.~M. Tait, ``{Asymmetric Leptons for Asymmetric
  Tops},''
\href{http://arxiv.org/abs/1104.0947}{{\ttfamily arXiv:1104.0947 [hep-ph]}}.

\bibitem{Patel:2011eh}
K.~M. Patel and P.~Sharma, ``{Forward-backward asymmetry in top quark
  production from light colored scalars in SO(10) model},''
  \href{http://dx.doi.org/10.1007/JHEP04(2011)085}{{\em JHEP} {\bfseries 1104}
  (2011) 085},
\href{http://arxiv.org/abs/1102.4736}{{\ttfamily arXiv:1102.4736 [hep-ph]}}.

\bibitem{Berger:2013ysa}
E.~L. Berger, Z.~Sullivan, and H.~Zhang, ``{LHC and Tevatron constraints on a
  W' model interpretation of the top quark forward-backward asymmetry},''
  \href{http://dx.doi.org/10.1103/PhysRevD.88.114026}{{\em Phys. Rev.}
  {\bfseries D88} no.~11, (2013) 114026},
\href{http://arxiv.org/abs/1309.7110}{{\ttfamily arXiv:1309.7110 [hep-ph]}}.

\bibitem{effLag_Ko}
D.-W. Jung, P.~Ko, J.~S. Lee, and S.-h. Nam, ``Model independent analysis of
  the forward-backward asymmetry of top quark production at the tevatron,''
  \href{http://dx.doi.org/10.1016/j.physletb.2010.06.040}{{\em {Phys.Lett.}}
  {\bfseries B691} (2010) 238--242},
\href{http://arxiv.org/abs/0912.1105}{{\ttfamily arXiv:0912.1105}}.

\bibitem{Franzosi:2015osa}
D.~Buarque~Franzosi and C.~Zhang, ``{Probing the top-quark chromomagnetic
  dipole moment at next-to-leading order in QCD},''
  \href{http://dx.doi.org/10.1103/PhysRevD.91.114010}{{\em Phys.Rev.}
  {\bfseries D91} no.~11, (2015) 114010},
\href{http://arxiv.org/abs/1503.08841}{{\ttfamily arXiv:1503.08841 [hep-ph]}}.

\bibitem{Biswal:2012mr}
S.~S. Biswal, S.~Mitra, R.~Santos, P.~Sharma, R.~K. Singh, {\em et~al.}, ``{New
  physics contributions to the forward-backward asymmetry at the Tevatron},''
  \href{http://dx.doi.org/10.1103/PhysRevD.86.014016}{{\em Phys.Rev.}
  {\bfseries D86} (2012) 014016},
\href{http://arxiv.org/abs/1201.3668}{{\ttfamily arXiv:1201.3668 [hep-ph]}}.

\bibitem{Shu:2011au}
J.~Shu, K.~Wang, and G.~Zhu, ``{A Revisit to Top Quark Forward-Backward
  Asymmetry},'' \href{http://dx.doi.org/10.1103/PhysRevD.85.034008}{{\em
  Phys.Rev.} {\bfseries D85} (2012) 034008},
\href{http://arxiv.org/abs/1104.0083}{{\ttfamily arXiv:1104.0083 [hep-ph]}}.

\bibitem{Westhoff:2013ixa}
S.~Westhoff, ``{Top Charge Asymmetry -- Theory Status Fall 2013},''
\href{http://arxiv.org/abs/1311.1127}{{\ttfamily arXiv:1311.1127 [hep-ph]}}.

\bibitem{Jezabek:1988topdecayleptondist}
M.~Jezabek and J.~H. Kuhn, ``Lepton spectra from heavy quark decay,''
  \href{http://dx.doi.org/10.1016/0550-3213(89)90209-5}{{\em {Nucl.Phys.}}
  {\bfseries B320} (1989) 20}.

\bibitem{Godbole:2006ldistandtpol}
R.~M. Godbole, S.~D. Rindani, and R.~K. Singh, ``Lepton distribution as a probe
  of new physics in production and decay of the t quark and its polarization,''
  \href{http://dx.doi.org/10.1088/1126-6708/2006/12/021}{{\em {JHEP}}
  {\bfseries 0612} (2006) 021},
\href{http://arxiv.org/abs/hep-ph/0605100}{{\ttfamily arXiv:hep-ph/0605100}}.

\bibitem{Bernreuther:2010ny}
W.~Bernreuther and Z.-G. Si, ``{Distributions and correlations for top quark
  pair production and decay at the Tevatron and LHC.},''
  \href{http://dx.doi.org/10.1016/j.nuclphysb.2010.05.001}{{\em Nucl.Phys.}
  {\bfseries B837} (2010) 90--121},
\href{http://arxiv.org/abs/1003.3926}{{\ttfamily arXiv:1003.3926 [hep-ph]}}.

\bibitem{AguilarSaavedra:2012rx}
J.~Aguilar-Saavedra, W.~Bernreuther, and Z.~Si, ``{Collider-independent top
  quark forward-backward asymmetries: standard model predictions},''
  \href{http://dx.doi.org/10.1103/PhysRevD.86.115020}{{\em Phys.Rev.}
  {\bfseries D86} (2012) 115020},
\href{http://arxiv.org/abs/1209.6352}{{\ttfamily arXiv:1209.6352 [hep-ph]}}.

\bibitem{Bernreuther:2012sx}
W.~Bernreuther and Z.-G. Si, ``{Top quark and leptonic charge asymmetries for
  the Tevatron and LHC},''
  \href{http://dx.doi.org/10.1103/PhysRevD.86.034026}{{\em Phys.Rev.}
  {\bfseries D86} (2012) 034026},
\href{http://arxiv.org/abs/1205.6580}{{\ttfamily arXiv:1205.6580 [hep-ph]}}.

\bibitem{AguilarSaavedra:2012xe}
J.~Aguilar-Saavedra and R.~Herrero-Hahn, ``{Model-independent measurement of
  the top quark polarisation},''
  \href{http://dx.doi.org/10.1016/j.physletb.2012.11.031}{{\em Phys.Lett.}
  {\bfseries B718} (2013) 983--987},
\href{http://arxiv.org/abs/1208.6006}{{\ttfamily arXiv:1208.6006 [hep-ph]}}.

\bibitem{Hewett:2011wz}
J.~L. Hewett, J.~Shelton, M.~Spannowsky, T.~M.~P. Tait, and M.~Takeuchi,
  ``{$A^t\_{FB}$ Meets LHC},''
  \href{http://dx.doi.org/10.1103/PhysRevD.84.054005}{{\em Phys. Rev.}
  {\bfseries D84} (2011) 054005},
\href{http://arxiv.org/abs/1103.4618}{{\ttfamily arXiv:1103.4618 [hep-ph]}}.

\bibitem{Aaltonen:2013wca}
{\bfseries CDF, D0} Collaboration, T.~A. Aaltonen {\em et~al.}, ``{Combination
  of measurements of the top-quark pair production cross section from the
  Tevatron Collider},''
  \href{http://dx.doi.org/10.1103/PhysRevD.89.072001}{{\em Phys.Rev.}
  {\bfseries D89} no.~7, (2014) 072001},
\href{http://arxiv.org/abs/1309.7570}{{\ttfamily arXiv:1309.7570 [hep-ex]}}.

\bibitem{CMS:2013sca}
{\bfseries CMS} Collaboration,
``{Combination of ATLAS and CMS top-quark pair cross section measurements using
  proton-proton collisions at sqrt(s) = 7 TeV},''.
CMS-PAS-TOP-12-003.

\bibitem{ATLAS:2012dpa}
{\bfseries ATLAS} Collaboration,
``{Combination of ATLAS and CMS top-quark pair cross section measurements using
  up to 1.1 fb-1 of data at 7 TeV},''.
ATLAS-CONF-2012-134 ETC.

\bibitem{SM_tevatron_crs}
N.~Kidonakis and R.~Vogt, ``The theoretical top quark cross section at the
  tevatron and the lhc,''
  \href{http://dx.doi.org/10.1103/PhysRevD.78.074005}{{\em {Phys.Rev.}}
  {\bfseries D78} (2008) 074005},
\href{http://arxiv.org/abs/0805.3844}{{\ttfamily arXiv:0805.3844}}.

\bibitem{Czakon:2011xx}
M.~Czakon and A.~Mitov, ``{Top++: A Program for the Calculation of the Top-Pair
  Cross-Section at Hadron Colliders},''
  \href{http://dx.doi.org/10.1016/j.cpc.2014.06.021}{{\em Comput.Phys.Commun.}
  {\bfseries 185} (2014) 2930},
\href{http://arxiv.org/abs/1112.5675}{{\ttfamily arXiv:1112.5675 [hep-ph]}}.

\bibitem{Czakon:2013goa}
M.~Czakon, P.~Fiedler, and A.~Mitov, ``{Total Top-Quark Pair-Production Cross-Section at Hadron Colliders Through $O(\alpha^{4}_{S})$},''
  \href{http://dx.doi.org/10.1103/PhysRevLett.110.252004}{{\em Phys.Rev.Lett.}
  {\bfseries 110} (2013) 252004},
\href{http://arxiv.org/abs/1303.6254}{{\ttfamily arXiv:1303.6254 [hep-ph]}}.

\bibitem{k_Tevatron}
V.~Ahrens, A.~Ferroglia, M.~Neubert, B.~D. Pecjak, and L.~L. Yang, ``The
  top-pair forward-backward asymmetry beyond nlo,''
  \href{http://dx.doi.org/10.1103/PhysRevD.84.074004}{{\em {Phys.Rev.}}
  {\bfseries D84} (2011) 074004},
\href{http://arxiv.org/abs/1106.6051}{{\ttfamily arXiv:1106.6051}}.

\bibitem{FBA_mttbar2013}
{\bfseries CDF} Collaboration, T.~Aaltonen {\em et~al.}, ``Measurement of the
  top quark forward-backward production asymmetry and its dependence on event
  kinematic properties,''
  \href{http://dx.doi.org/10.1103/PhysRevD.87.092002}{{\em {Phys.Rev.}}
  {\bfseries D87} (2013) 092002},
\href{http://arxiv.org/abs/1211.1003}{{\ttfamily arXiv:1211.1003}}.

\bibitem{Czakon:2014xsa}
M.~Czakon, P.~Fiedler, and A.~Mitov, ``{Resolving the Tevatron top quark
  forward-backward asymmetry puzzle},''
\href{http://arxiv.org/abs/1411.3007}{{\ttfamily arXiv:1411.3007 [hep-ph]}}.

\bibitem{CMS:2014jua}
{\bfseries ATLAS, CMS} Collaboration,
``{Combination of ATLAS and CMS $t\bar{t}$ charge asymmetry measurements using
  LHC proton-proton collisions at $\sqrt{s}=7$ TeV},''.
ATLAS-CONF-2014-012 ETC.

\bibitem{SM_LHC_Ac}
J.~H. Kuhn and G.~Rodrigo, ``Charge asymmetries of top quarks at hadron
  colliders revisited,'' \href{http://dx.doi.org/10.1007/JHEP01(2012)063}{{\em
  {JHEP}} {\bfseries 1201} (2012) 063},
\href{http://arxiv.org/abs/1109.6830}{{\ttfamily arXiv:1109.6830}}.

\bibitem{cdf_Alpm}
{\bfseries CDF} Collaboration, T.~A. Aaltonen {\em et~al.}, ``Measurement of
  the leptonic asymmetry in ttbar events produced in ppbar collisions at
  $\sqrt{s}=1.96$ tev,''
  \href{http://dx.doi.org/10.1103/PhysRevD.88.072003}{{\em {Phys.Rev.}}
  {\bfseries D88} (2013) 072003},
\href{http://arxiv.org/abs/1308.1120}{{\ttfamily arXiv:1308.1120}}.

\bibitem{Abazov:2014oea}
{\bfseries D0} Collaboration, V.~M. Abazov {\em et~al.}, ``{Measurement of the
  forward-backward asymmetry in the distribution of leptons in $t\bar{t}$
  events in the lepton$+$jets channel},''
  \href{http://dx.doi.org/10.1103/PhysRevD.90.072001}{{\em Phys.Rev.}
  {\bfseries D90} no.~7, (2014) 072001},
\href{http://arxiv.org/abs/1403.1294}{{\ttfamily arXiv:1403.1294 [hep-ex]}}.

\bibitem{Aaltonen:2010nz}
{\bfseries CDF} Collaboration, T.~Aaltonen {\em et~al.}, ``{Measurement of
  $t\bar{t}$ Spin Correlation in $p\bar{p}$ Collisions Using the CDF II
  Detector at the Tevatron},''
  \href{http://dx.doi.org/10.1103/PhysRevD.83.031104}{{\em Phys.Rev.}
  {\bfseries D83} (2011) 031104},
\href{http://arxiv.org/abs/1012.3093}{{\ttfamily arXiv:1012.3093 [hep-ex]}}.

\bibitem{Peters:2012wg}
{\bfseries D0} Collaboration, Y.~Peters, ``{$t\bar{t}$ Spin Correlations at
  D0},'' {\em PoS} {\bfseries ICHEP2012} (2013) 236,
\href{http://arxiv.org/abs/1210.7189}{{\ttfamily arXiv:1210.7189 [hep-ex]}}.

\bibitem{Aad:2014pwa}
{\bfseries ATLAS} Collaboration, G.~Aad {\em et~al.}, ``{Measurements of spin
  correlation in top-antitop quark events from proton-proton collisions at
  $\sqrt{s}=7$ TeV using the ATLAS detector},''
  \href{http://dx.doi.org/10.1103/PhysRevD.90.112016}{{\em Phys.Rev.}
  {\bfseries D90} no.~11, (2014) 112016},
\href{http://arxiv.org/abs/1407.4314}{{\ttfamily arXiv:1407.4314 [hep-ex]}}.

\bibitem{Chatrchyan:2013wua}
{\bfseries CMS} Collaboration, S.~Chatrchyan {\em et~al.}, ``{Measurements of
  $t\bar{t}$ spin correlations and top-quark polarization using dilepton final
  states in $pp$ collisions at $\sqrt{s}$ = 7 TeV},''
  \href{http://dx.doi.org/10.1103/PhysRevLett.112.182001}{{\em Phys.Rev.Lett.}
  {\bfseries 112} no.~18, (2014) 182001},
\href{http://arxiv.org/abs/1311.3924}{{\ttfamily arXiv:1311.3924 [hep-ex]}}.

\bibitem{Frixione:2002ik}
S.~Frixione and B.~R. Webber, ``{Matching NLO QCD computations and parton
  shower simulations},''
  \href{http://dx.doi.org/10.1088/1126-6708/2002/06/029}{{\em JHEP} {\bfseries
  0206} (2002) 029},
\href{http://arxiv.org/abs/hep-ph/0204244}{{\ttfamily arXiv:hep-ph/0204244
  [hep-ph]}}.

\bibitem{Aad:2013ksa}
{\bfseries ATLAS} Collaboration, G.~Aad {\em et~al.}, ``{Measurement of Top
  Quark Polarization in Top-Antitop Events from Proton-Proton Collisions at
  $\sqrt{s}$ = 7TeV Using the ATLAS Detector},''
  \href{http://dx.doi.org/10.1103/PhysRevLett.111.232002}{{\em Phys.Rev.Lett.}
  {\bfseries 111} no.~23, (2013) 232002},
\href{http://arxiv.org/abs/1307.6511}{{\ttfamily arXiv:1307.6511 [hep-ex]}}.

\bibitem{Frampton:1987dn}
P.~H. Frampton and S.~L. Glashow, ``Chiral color: An alternative to the
  standard model,'' \href{http://dx.doi.org/10.1016/0370-2693(87)90859-8}{{\em
  {Phys.Lett.}} {\bfseries B190} (1987) 157}.

\bibitem{Frampton:1992flv_nonunivAx}
P.~Frampton, ``Chiral dilepton model and the flavor question,''
  \href{http://dx.doi.org/10.1103/PhysRevLett.69.2889}{{\em {Phys.Rev.Lett.}}
  {\bfseries 69} (1992) 2889--2891}.

\bibitem{CTEQ6l1:Pumplin2002}
J.~Pumplin, D.~Stump, J.~Huston, H.~Lai, P.~M. Nadolsky, {\em et~al.}, ``New
  generation of parton distributions with uncertainties from global qcd
  analysis,'' \href{http://dx.doi.org/10.1088/1126-6708/2002/07/012}{{\em
  {JHEP}} {\bfseries 0207} (2002) 012},
\href{http://arxiv.org/abs/hep-ph/0201195}{{\ttfamily arXiv:hep-ph/0201195}}.

\bibitem{Khachatryan:2015sja}
{\bfseries CMS} Collaboration, V.~Khachatryan {\em et~al.}, ``{Search for
  resonances and quantum black holes using dijet mass spectra in proton-proton
  collisions at $\sqrt{s} =$ 8 TeV},''
  \href{http://dx.doi.org/10.1103/PhysRevD.91.052009}{{\em Phys.Rev.}
  {\bfseries D91} no.~5, (2015) 052009},
\href{http://arxiv.org/abs/1501.04198}{{\ttfamily arXiv:1501.04198 [hep-ex]}}.

\bibitem{Aguilar-Saavedra:2014nja}
J.~Aguilar-Saavedra, ``{Portrait of a colour octet},''
  \href{http://dx.doi.org/10.1007/JHEP08(2014)172}{{\em JHEP} {\bfseries 1408}
  (2014) 172},
\href{http://arxiv.org/abs/1405.5826}{{\ttfamily arXiv:1405.5826 [hep-ph]}}.

\bibitem{Chivukula:2010Axnotallowed}
R.~S. Chivukula, E.~H. Simmons, and C.-P. Yuan, ``axigluons cannot explain the
  observed top quark forward-backward asymmetry,''
  \href{http://dx.doi.org/10.1103/PhysRevD.82.094009}{{\em {Phys.Rev.}}
  {\bfseries D82} (2010) 094009},
\href{http://arxiv.org/abs/1007.0260}{{\ttfamily arXiv:1007.0260}}.

\bibitem{CMS:2013gqa}
{\bfseries CMS} Collaboration, C.~Collaboration,
``{Search for pair production of resonances decaying to a top quark plus a jet
  in final states with two leptons},''.

\bibitem{Chatrchyan:2013izb}
{\bfseries CMS} Collaboration, S.~Chatrchyan {\em et~al.}, ``{Search for
  pair-produced dijet resonances in four-jet final states in pp collisions at
  $\sqrt{s}$ = 7 TeV},''
  \href{http://dx.doi.org/10.1103/PhysRevLett.110.141802}{{\em Phys.Rev.Lett.}
  {\bfseries 110} (2013) 141802},
\href{http://arxiv.org/abs/1302.0531}{{\ttfamily arXiv:1302.0531 [hep-ex]}}.

\bibitem{Gresham:2012wc}
M.~I. Gresham, I.-W. Kim, S.~Tulin, and K.~M. Zurek, ``{Confronting Top AFB
  with Parity Violation Constraints},''
  \href{http://dx.doi.org/10.1103/PhysRevD.86.034029}{{\em Phys.Rev.}
  {\bfseries D86} (2012) 034029},
\href{http://arxiv.org/abs/1203.1320}{{\ttfamily arXiv:1203.1320 [hep-ph]}}.

\bibitem{Grinstein:2011yv}
B.~Grinstein, A.~L. Kagan, M.~Trott, and J.~Zupan, ``{Forward-backward
  asymmetry in $t \bar{t}$ production from flavour symmetries},''
  \href{http://dx.doi.org/10.1103/PhysRevLett.107.012002}{{\em Phys.Rev.Lett.}
  {\bfseries 107} (2011) 012002},
\href{http://arxiv.org/abs/1102.3374}{{\ttfamily arXiv:1102.3374 [hep-ph]}}.


\bibitem{Diaz:2013tfa}
B.~Diaz and A.~R. Zerwekh, ``{axigluon Phenomenology using ATLAS dijet data},''
  \href{http://dx.doi.org/10.1142/S0217751X13501339}{{\em Int.J.Mod.Phys.}
  {\bfseries A28} (2013) 1350133},
\href{http://arxiv.org/abs/1308.0166}{{\ttfamily arXiv:1308.0166 [hep-ph]}}.

\bibitem{Choudhury:2011cg}
D.~Choudhury, R.~M. Godbole, and P.~Saha, ``{Dijet resonances, widths and all
  that},'' \href{http://dx.doi.org/10.1007/JHEP01(2012)155}{{\em JHEP}
  {\bfseries 01} (2012) 155},
\href{http://arxiv.org/abs/1111.1054}{{\ttfamily arXiv:1111.1054 [hep-ph]}}.

\bibitem{Melnikov:2009dn}
K.~Melnikov and M.~Schulze, ``{NLO QCD corrections to top quark pair production
  and decay at hadron colliders},''
  \href{http://dx.doi.org/10.1088/1126-6708/2009/08/049}{{\em JHEP} {\bfseries
  0908} (2009) 049},
\href{http://arxiv.org/abs/0907.3090}{{\ttfamily arXiv:0907.3090 [hep-ph]}}.

\bibitem{Bernreuther:2014dla}
W.~Bernreuther, P.~González, and C.~Mellein, ``{Decays of polarized top quarks
  to lepton, neutrino and jets at NLO QCD},''
  \href{http://dx.doi.org/10.1140/epjc/s10052-014-2815-5}{{\em Eur.Phys.J.}
  {\bfseries C74} no.~3, (2014) 2815},
\href{http://arxiv.org/abs/1401.5930}{{\ttfamily arXiv:1401.5930 [hep-ph]}}.

\bibitem{Berger:2012_assym_corr}
E.~L. Berger, Q.-H. Cao, C.-R. Chen, and H.~Zhang, ``Interpretations and
  implications of the top quark rapidity asymmetries $a_{FB}^{t}$ and
  $a_{FB}^{E_{l}}$,'' \href{http://dx.doi.org/10.1103/PhysRevD.88.014033}{{\em
  {Phys.Rev.}} {\bfseries D88} (2013) 014033},
\href{http://arxiv.org/abs/1209.4899}{{\ttfamily arXiv:1209.4899}}.

\bibitem{Berger2013_AlandPol}
E.~L. Berger, Q.-H. Cao, J.-H. Yu, and H.~Zhang, ``Measuring top-quark
  polarization in top-pair + missing energy events,''
\href{http://arxiv.org/abs/1305.7266}{{\ttfamily arXiv:1305.7266}}.

\bibitem{Godbole:2010phidist_tpol_LHC}
R.~M. Godbole, K.~Rao, S.~D. Rindani, and R.~K. Singh, ``On measurement of top
  polarization as a probe of $t\bar{t}$ production mechanisms at the lhc,''
  \href{http://dx.doi.org/10.1007/JHEP11(2010)144}{{\em {JHEP}} {\bfseries
  1011} (2010) 144},
\href{http://arxiv.org/abs/1010.1458}{{\ttfamily arXiv:1010.1458}}.

\bibitem{Godbole:2011_AlAphiAp}
R.~M. Godbole, L.~Hartgring, I.~Niessen, and C.~D. White, ``Top polarisation
  studies in $h^{-}t$ and $wt$ production,''
  \href{http://dx.doi.org/10.1007/JHEP01(2012)011}{{\em {JHEP}} {\bfseries
  1201} (2012) 011},
\href{http://arxiv.org/abs/1111.0759}{{\ttfamily arXiv:1111.0759}}.

\bibitem{Barger:2011philcorr}
V.~Barger, W.-Y. Keung, and B.~Yencho, ``Azimuthal correlations in top pair
  decays and the effects of new heavy scalars,''
  \href{http://dx.doi.org/10.1103/PhysRevD.85.034016}{{\em {Phys.Rev.}}
  {\bfseries D85} (2012) 034016},
\href{http://arxiv.org/abs/1112.5173}{{\ttfamily arXiv:1112.5173}}.

\bibitem{Baumgart:2012ay}
M.~Baumgart and B.~Tweedie, ``{A New Twist on Top Quark Spin Correlations},''
  \href{http://dx.doi.org/10.1007/JHEP03(2013)117}{{\em JHEP} {\bfseries 1303}
  (2013) 117},
\href{http://arxiv.org/abs/1212.4888}{{\ttfamily arXiv:1212.4888 [hep-ph]}}.

\bibitem{Boudjema:2009fz}
F.~Boudjema and R.~K. Singh, ``{A Model independent spin analysis of
  fundamental particles using azimuthal asymmetries},''
  \href{http://dx.doi.org/10.1088/1126-6708/2009/07/028}{{\em JHEP} {\bfseries
  0907} (2009) 028},
\href{http://arxiv.org/abs/0903.4705}{{\ttfamily arXiv:0903.4705 [hep-ph]}}.

\bibitem{Shelton:2008topPol}
J.~Shelton, ``Polarized tops from new physics: signals and observables,''
  \href{http://dx.doi.org/10.1103/PhysRevD.79.014032}{{\em {Phys.Rev.}}
  {\bfseries D79} (2009) 014032},
\href{http://arxiv.org/abs/0811.0569}{{\ttfamily arXiv:0811.0569}}.

\bibitem{Carmona:2014gra}
A.~Carmona, M.~Chala, A.~Falkowski, S.~Khatibi, M.~M. Najafabadi, {\em et~al.},
  ``{From Tevatron's top and lepton-based asymmetries to the LHC},''
  \href{http://dx.doi.org/10.1007/JHEP07(2014)005}{{\em JHEP} {\bfseries 1407}
  (2014) 005},
\href{http://arxiv.org/abs/1401.2443}{{\ttfamily arXiv:1401.2443 [hep-ph]}}.

\bibitem{Prasath:2014mfa}
A.~Prasath, R.~M. Godbole, and S.~D. Rindani, ``{Top polarisation measurement
  and anomalous $Wtb$ coupling},''
\href{http://arxiv.org/abs/1405.1264}{{\ttfamily arXiv:1405.1264 [hep-ph]}}.

\bibitem{Drobnak:2012cz}
J.~Drobnak, J.~F. Kamenik, and J.~Zupan, ``{Flipping t tbar Asymmetries at the
  Tevatron and the LHC},''
  \href{http://dx.doi.org/10.1103/PhysRevD.86.054022}{{\em Phys.Rev.}
  {\bfseries D86} (2012) 054022},
\href{http://arxiv.org/abs/1205.4721}{{\ttfamily arXiv:1205.4721 [hep-ph]}}.

\bibitem{Bernreuther:1995cx}
W.~Bernreuther, A.~Brandenburg, and P.~Uwer, ``{Transverse polarization of top
  quark pairs at the Tevatron and the large hadron collider},''
  \href{http://dx.doi.org/10.1016/0370-2693(95)01475-6}{{\em Phys.Lett.}
  {\bfseries B368} (1996) 153--162},
\href{http://arxiv.org/abs/hep-ph/9510300}{{\ttfamily arXiv:hep-ph/9510300
  [hep-ph]}}.

\bibitem{Aguilar-Saavedra:2014yea}
J.~Aguilar-Saavedra, ``{Quantum coherence, top transverse polarisation and the
  Tevatron asymmetry $A^l_{FB}$},''
  \href{http://dx.doi.org/10.1016/j.physletb.2014.07.013}{{\em Phys.Lett.}
  {\bfseries B736} (2014) 132--136},
\href{http://arxiv.org/abs/1405.1412}{{\ttfamily arXiv:1405.1412 [hep-ph]}}.

\bibitem{Baumgart:2013yra}
M.~Baumgart and B.~Tweedie, ``{Transverse Top Quark Polarization and the ttbar
  Forward-Backward Asymmetry},''
  \href{http://dx.doi.org/10.1007/JHEP08(2013)072}{{\em JHEP} {\bfseries 1308}
  (2013) 072},
\href{http://arxiv.org/abs/1303.1200}{{\ttfamily arXiv:1303.1200 [hep-ph]}}.

\bibitem{Krohn:2011tw}
D.~Krohn, T.~Liu, J.~Shelton, and L.-T. Wang, ``{A Polarized View of the Top
  Asymmetry},'' \href{http://dx.doi.org/10.1103/PhysRevD.84.074034}{{\em Phys.
  Rev.} {\bfseries D84} (2011) 074034},
\href{http://arxiv.org/abs/1105.3743}{{\ttfamily arXiv:1105.3743 [hep-ph]}}.
\end{thebibliography}

\providecommand{\href}[2]{#2}
\begingroup
\raggedright

\endgroup
\end{document}